\documentclass[10pt]{article}
\usepackage{nexilume-report}
\usepackage{longtable}
\usepackage{multirow}
\usepackage{placeins}
\usepackage{url}
\usepackage{xurl}
\reporttitle
  {Nexus: An Execution Fabric for AI Agents Across Cloud, Edge, and Devices}
  []

\reportmeta
  {NX-SYS-2026-001}
  {v1}
  {September 2026}
  {Nexilume Research}
  {research@nexilume.com}
  {Research Manuscript}

\newcommand{\system}{\textsc{Nexus}}

\newcommand{\code}[1]{\texttt{#1}}

\hypersetup{
  pdftitle={Nexus: An Edge-Oriented Platform for Agents across Heterogeneous Environments},
  pdfauthor={Nexilume Research},
  pdfsubject={Distributed AI agent execution, capability routing, OpenWrt, and IPv6}
}

\begin{document}

\makereporttitle{%
Language-model agents are evolving into long-running services that interact with models, tools, computers, mobile devices, and distributed environments. Existing agent frameworks simplify reasoning and tool invocation. However, cloud-centric designs face three limitations: centralized execution increases failure impact, scaling pressure, and compute cost; extending agents across computers, mobile devices, and edge environments requires a unified execution abstraction with permission control; and long-running executions require consistent lifecycle management across failures, recovery, results, usage, and settlement.
We present Nexus, a cloud–edge platform for deploying and operating agent services. To reduce dependence on centralized serving, Nexus extends agent execution to distributed edge infrastructure through an OpenWrt-based runtime, where edge nodes register their available service and computation capabilities and remain accessible through a unified invocation interface. To support heterogeneous environments, Nexus separates agent reasoning from environment-specific execution and uses run-scoped delegation to bind Computer, Mobile, and other resource operations to explicitly authorized subjects, devices, and capabilities. To manage long-running executions, Nexus represents each invocation as a persistent task and jointly tracks its execution state, intermediate events, outputs, failures, recovery state, resource usage, and charging records across cloud and edge components.
We evaluate Nexus using controlled workloads, device interactions, and model-driven workflows. For a 16-MiB, 32-round hashing task arriving once per second with a two-second deadline, both capacity experiments pass all batches at a tested Cloud CPU ceiling of 0.25 vCPU with additional edge execution, compared with 2 vCPU for Cloud execution. Ten five-action Computer--Android workflows produce the expected artifacts, and all six live-revocation cases block the subsequent write while retaining an already-authorized read. Under worker loss, operation journaling reduces duplicate appends from six per task to zero, with 0.933 s mean normal-path overhead. Across 24 pairs using the same model and device backend, Nexus completes 24 tasks and Dify completes 22; among the 22 jointly successful pairs, Nexus's observed completion time is lower by a median 3.88 s. Dapr provides lower successful-call latency in a separate service workload, while Nexus completes under a smaller tested incremental-runtime memory ceiling, 16 MiB versus 64 MiB. The results show how data locality, operation-scoped authority, and persistent result identity support Agent services across cloud and edge resources, with workload-dependent resource and execution costs.\footnote[1]{https://github.com/Nexilume-AI}
}

\section{Introduction}
Language-model agents combine reasoning with tool use and external state,
supporting applications in software development, information retrieval,
and interactive task execution
\cite{yao2023react,schick2023toolformer,wu2023autogen}.
Deploying these agents as reusable services requires coordinating not
only model inference but also the distributed resources on which their
actions depend. For example, a reporting agent may process files on a
workstation, consult a hosted model to generate a report, and submit it
through a mobile application on the caller's device. Although these
steps serve a single request, they span different execution hosts,
user-owned environments, and service interfaces. Reliable service
execution therefore requires making these resources accessible,
restricting agent actions to authorized environments and operations,
and preserving the association between each request, its actions, and
their outcomes across interruptions.

Existing work provides complementary abstractions for constructing and
operating agents. ReAct and Toolformer connect language models to external
actions \cite{yao2023react,schick2023toolformer}, while AutoGen and Dify
support agent interaction and workflow orchestration
\cite{wu2023autogen,dify2026}. For distributed execution, Ray schedules
tasks and actors across cluster nodes, and Dapr provides service discovery,
invocation, and access control \cite{moritz2018ray,dapr2026}. For interaction
with external resources, the Model Context Protocol (MCP) standardizes
tool and resource interfaces, with the host coordinating connections,
permissions, and user consent \cite{mcp2025architecture}. For execution
continuity, LangGraph persists workflow state and completed node outputs,
allowing interrupted workflows to resume without rerunning successfully
recorded work \cite{langgraphPersistence}. Together, these approaches
provide foundations for distributed serving, controlled tool invocation,
and workflow recovery.

\begin{figure*}[t]
  \centering
  \includegraphics[width=\textwidth]{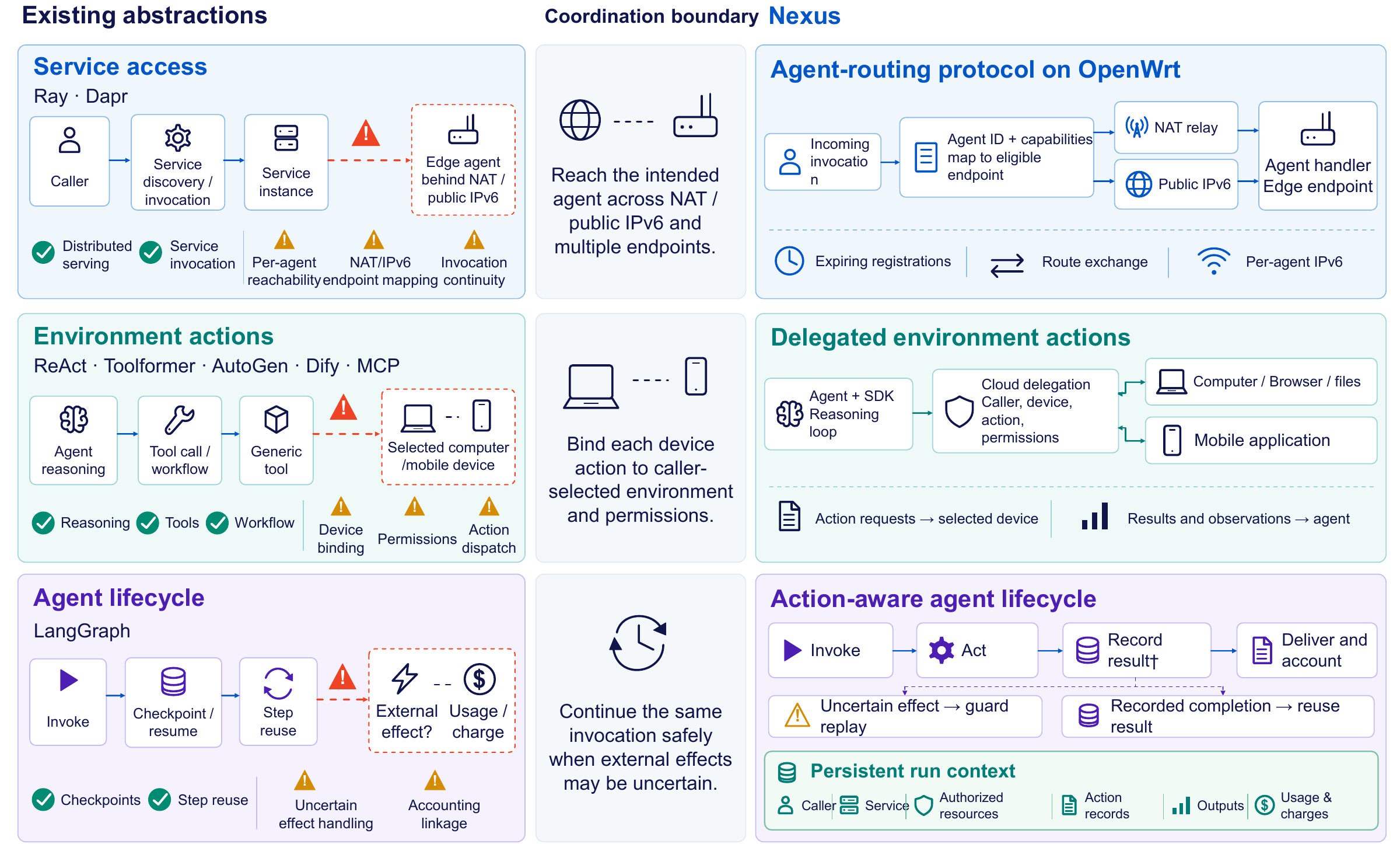}
  \caption{Existing service, workflow, and recovery abstractions and their relationship to \system{} \cite{moritz2018ray,dapr2026,yao2023react,schick2023toolformer,wu2023autogen,dify2026,mcp2025architecture,langgraphPersistence}. Nexus associates service reachability, authorized actions, and recorded outcomes with persistent invocations. Arrows show logical relationships. The Dify and Dapr experiments compare the corresponding workflow and service-invocation functions. Public IPv6 requires provisioned addresses and ingress permission; suppression of repeated external effects requires cooperating resources.}
  \label{fig:intro-comparison}
\end{figure*}

However, as shown in Figure ~\ref{fig:intro-comparison}, extending these abstractions to independently managed edge
resources requires additional coordination across network access,
environment actions, and execution lifecycles. Service discovery identifies
an endpoint, but making an edge-hosted agent callable across network
address translation (NAT) and public IPv6 also requires suitable
reachability mechanisms and endpoint deployment support. Hosting all
execution in the cloud simplifies some of these requirements but retains
dependence on centrally provisioned capacity and shared failure domains.
Likewise, permission to invoke a tool service does not by itself specify
which caller-owned computer or mobile device should perform an action,
or whether that action remains authorized when dispatched. Workflow
persistence addresses another boundary: it preserves recorded progress,
but does not make a completion record atomic with an external file write
or device action. A failure between the action and its recorded completion
can therefore leave retry safety unresolved, even when the workflow state
is recoverable. The remaining problem is to maintain the association among
an agent's service endpoint, its authorized environment actions, and their
execution outcomes throughout invocation, recovery, result delivery, and
accounting, without requiring all computation to share a hosting
environment.

Our key observation is that \emph{distributed execution need not fragment
service access, resource authority, or invocation continuity}. First,
calling a cloud model or agent does not require subsequent computation to remain in
the cloud: edge-hosted capabilities can be exposed through the same service
interface, allowing distributed resources to supplement centralized serving.
Second, the host running an agent's reasoning loop need not determine which
environments it can access: resource selection and permissions can instead
be associated with the caller's invocation, allowing the same agent to
operate in different authorized environments. Third, the lifetime of an
invocation need not coincide with that of a particular process or
connection: its identity and recorded progress can persist across execution
attempts, retaining the association among completed work, outputs, usage,
and charges. These separations suggest that agent execution can extend to
distributed resources while service access, authorization, and lifecycle
state remain jointly managed.

Realizing this opportunity requires addressing three challenges. First,
distributed serving must make edge-hosted agents uniformly addressable
and invocable despite differences in capabilities, connectivity, and
resource limits. Agents may reside behind network address translation
(NAT) or be accessible through public IPv6. Bridging these network
conditions must preserve a common service interface without requiring
centralized execution or replicating a complete cloud service stack at
every endpoint. Second, agents must be able to perform actions in
caller-selected environments without conflating connectivity with
authority. Executing commands, accessing files, navigating browsers, and
operating mobile applications involve different action semantics and
execution contexts. The system must dispatch each action to the intended
environment, return results or observations to the agent, and enforce
resource bindings and permissions throughout the interaction rather than
treating an established connection as unrestricted access. Third, agent
lifecycle management must preserve continuity between an invocation and
the actions performed on its behalf across interaction, interruption,
recovery, and completion. An action may take effect before a failure
prevents its completion from being recorded, making replay potentially
duplicative and abandonment potentially wasteful. Persistent task state
alone is therefore insufficient: the system must distinguish recorded
completion from uncertain effects, determine when actions can be safely
retried, and keep recovered progress, outputs, usage, and charges
associated with the originating invocation.

We present \system{}, a cloud--edge platform that combines agent routing, authorized environment interaction, and persistent lifecycle management.
To make distributed agents accessible across network boundaries, we design an agent-routing protocol implemented natively on the OpenWrt router
operating system. The protocol maps agent identities and advertised
capabilities---the operations they provide---to reachable endpoints through
inter-router route exchange and registrations that expire unless renewed.
It supports outbound relay connections for agents behind NAT and direct
access over public IPv6; where address provisioning and network ingress
permit, each published agent can be bound to a dedicated public IPv6
address. Shared registration and routing on OpenWrt, together with cloud
service management, allow endpoints to execute agent operations without
replicating the complete service stack. To enable agents to act in
heterogeneous environments, a shared software development kit (SDK)
exposes caller-selected computers, browsers, workspaces, and mobile
devices through action interfaces. Environment-specific adapters dispatch
commands, file operations, browser interactions, and mobile actions to
the selected environment and return results or observations to the agent,
allowing its reasoning loop to operate independently of the device
performing the action. Invocation-scoped delegation binds actions to the
caller, selected resources, and permitted operations; binding validation
and permission checks govern dispatch, with live revalidation on supported
execution paths. To manage the agent service lifecycle, the cloud layer
connects publication and invocation with persistent execution state,
user interaction, recovery, output retention, and settlement. Each
invocation establishes a run---a persistent execution context that
associates execution attempts, reported progress, outputs, and usage
with the originating request. For actions registered through the SDK's
recovery interface, stable operation identifiers link recorded outcomes
to that run. Recovery reuses recorded results instead of repeating
completed actions and suspends unsafe automatic replay when external
effects remain uncertain. Cooperating resources that enforce duplicate
suppression can additionally use these identifiers to recognize retries
without repeating previously applied effects. Outputs and usage reports
remain associated with the original invocation, while pricing terms
captured at invocation and cost reports checked against configured limits
govern settlement. The SDK exposes environment actions and lifecycle
operations alongside model and file interfaces, reducing the
deployment-specific control and recovery logic required in task handlers.

We evaluate how these mechanisms affect serving capacity, environment actions, and execution continuity (Section~\ref{sec:evaluation}). For a 16-MiB, 32-round hashing workload with fixed arrivals and a two-second deadline, edge processing passes all batches at a tested Cloud CPU ceiling of 0.25 vCPU, compared with 2 vCPU for Cloud processing, in both the initial and repeated experiments. These configurations share the business operation while differing in kernel, bridge, and additional edge capacity. A separate unconstrained experiment measures a paired median reduction of 11.102 s in Cloud CPU per twenty-task batch. Smaller inputs show a different latency tradeoff, making the benefit workload-dependent.

Across heterogeneous environments, ten five-action Computer--Android workflows produce the expected artifacts. In six live-revocation pairs, retained grants allow completion, while revoked grants prevent the next write despite an earlier authorized read. Resource adapters preserve operation identities through shared hosting and recovery infrastructure. Under controlled worker loss, journaling avoids six duplicate appends per interrupted task for 0.933 s mean healthy-path overhead. A separate crash between an external effect and its completion record identifies the additional role of resource-side deduplication.

External experiments characterize workflow and service execution. Nexus and Dify complete 24 and 22 tasks across 24 pairs sharing the same model source and device tools; Nexus has a median observed completion-time advantage of 3.88 s among jointly successful pairs. Dapr provides lower successful-call latency in a byte-echo workload, while the smallest successful tested incremental-runtime memory ceilings are 16 MiB for Nexus and 64 MiB for Dapr in the separate memory-constrained experiment. Together, these results connect the proposed service abstractions with their capacity, coordination, and recovery tradeoffs.

This work makes four contributions:
\begin{itemize}
  \item \textbf{An agent-routing protocol for distributed serving.}
  We design an agent-routing protocol that maps agent identities and
  advertised capabilities to reachable endpoints across NAT and public
  IPv6 networks. Its OpenWrt-native implementation combines leased
  registration, inter-router route exchange, relay access, and direct
  IPv6 access, including per-agent public IPv6 bindings where network
  provisioning permits. These mechanisms expose edge-hosted agents
  through a common invocation interface without requiring a complete
  service stack at every endpoint.

  \item \textbf{Authorized agent interaction with heterogeneous environments.}
  We develop a delegated execution model that allows agents to issue
  actions to caller-selected computers, browsers, workspaces, and mobile
  devices and receive execution results or observations. Environment-specific
  adapters preserve operation semantics, while invocation-scoped resource
  bindings and permission checks constrain action dispatch, with live
  revalidation on supported execution paths.

  \item \textbf{Action-aware agent lifecycle management.}
  We connect agent publication and invocation with persistent execution,
  recovery, output retention, and accounting. For actions registered
  through the recovery interface, operation records support completed-result
  reuse and prevent unsafe automatic replay of uncertain effects.
  Stable operation identifiers additionally enable duplicate suppression
  by cooperating resources, while run identities retain the association
  among execution attempts, action outcomes, usage, and settlement.

  \item \textbf{A unified implementation and empirical evaluation.}
  We integrate these mechanisms into a shared SDK and cloud--edge platform,
  separating task-specific agent logic from network access, device
  interaction, and lifecycle control. Matched deployment and workflow
  comparisons, fault-injection experiments, and accounting checks evaluate
  performance, resource requirements, authorization, recovery, and
  settlement consistency. A matched SDK comparison further quantifies
  application-side adapter-code requirements.
\end{itemize}

\section{Related Work}

\subsection{Agent reasoning, tools, and communication}

Research on language-model agents studies how models select actions and
use the resulting observations. ReAct alternates reasoning with external
actions so that observations can inform subsequent decisions
\cite{yao2023react}. Toolformer learns which tools to call, when to call
them, and how to incorporate their outputs \cite{schick2023toolformer}.
Wang et al. organize agent designs around role specification, memory,
planning, and action \cite{wang2024survey}. At the application level,
AutoGen supports programmable conversations among agents,
AgentScope provides reusable components for tool use and agent interaction,
and Dify supports visual workflow construction and model and tool
integration \cite{wu2023autogen,agentscope2025,dify2026}.
These approaches support the decision-making and application logic that
\system{} deploys and connects to distributed execution resources.

Communication protocols make independently implemented components
interoperate. The Model Context Protocol (MCP) connects a host application
to servers exposing tools and contextual resources through clients that
manage individual connections. The host coordinates permissions and user
consent \cite{mcp2025architecture}. The Agent2Agent (A2A) protocol
standardizes communication between agents, including descriptions of
available capabilities, task-status updates, result exchange, and
authentication requirements \cite{a2a2025spec}. These protocols therefore
provide more than message formats. \system{} complements them by mapping
supported protocol operations to registered agent services and coordinating
their network routes, caller-selected execution environments, and
invocation records. Here, an invocation denotes one admitted request to
an agent service, which may produce multiple actions and execution attempts.

\subsection{Distributed execution and edge service routing}

Distributed runtimes separate application logic from the machines executing
it. Ray provides remote tasks, which execute functions, and actors, which
retain state across method calls \cite{moritz2018ray}. Ray Serve builds on
this runtime to distribute requests among service replicas---multiple
running instances of a deployment---and adjust their number with demand
\cite{rayserve2026}. Service-based routing similarly separates a service
identity from a fixed host address: requests identify the desired service,
and the network directs them to an available instance at the edge
\cite{trossen2019service}.

Dapr provides service discovery and invocation through a sidecar, a helper
process associated with an application that handles communication on its
behalf \cite{dapr2026}. Istio manages a service mesh, a communication layer
that applies routing, security, and monitoring policies across services;
Envoy provides a proxy that forwards and controls service traffic
\cite{istio2026,envoy2026}. At the model-access layer, Portkey Gateway
routes inference requests among model providers and supports retries
and response caching \cite{portkey2026}. These systems address related
but distinct routing targets: distributed computations, application
services, and model requests.

\system{} focuses on making agents hosted on independently managed edge
resources callable through a common interface. Its OpenWrt-based routing
protocol maps agent identities and capabilities---the operations an agent
provides---to eligible execution endpoints. Registrations expire unless
renewed, while route exchange maintains information about remote services.
Relay paths support endpoints behind network address translation (NAT),
which translates private network addresses and can restrict incoming
connections;
direct paths support publicly reachable Internet Protocol version 6
(IPv6) endpoints. The distinction is the coordination of agent-specific
service access with device authorization and invocation continuity,
rather than assuming that service discovery or distributed execution is
itself new.

\subsection{Environment interaction and authorization}

Agent actions must be mapped to concrete execution environments.
AgentScope, for example, supports tool interaction and sandboxed
execution, in which operations run in an isolated environment
\cite{agentscope2025}. Isolation and resource authorization address
different questions: the former limits an operation's effects within an
execution environment, while the latter determines which environment and
operations a caller may use. \system{} connects agent actions to
caller-selected computers, browsers, workspaces, and mobile devices,
and returns their results or observations to the reasoning process.

This design follows established access-control principles. Zero-trust
architecture requires access decisions to depend on the requesting identity
and target resource rather than network location alone
\cite{rose2020zerotrust}. Transport Layer Security (TLS) 1.3 protects
communication against eavesdropping and tampering and supports peer
authentication \cite{rfc8446}. OAuth grants limited resource access through tokens. Its resource-indicator
extension identifies the servers for which a token is intended, enabling
access to be restricted to those recipients \cite{rfc8707}. Such channel and credential
protections must be combined with application-level rules specifying which
device an agent may operate and which actions it may perform. \system{}
applies these principles through invocation-scoped delegation: permissions
are associated with a particular caller, selected resources, and admitted
agent request. Environment-specific adapters preserve operation semantics,
while binding checks and live permission revalidation constrain action
dispatch on supported paths.

\subsection{Persistent execution and agent service lifecycles}

Agent runtimes also manage resources and execution progress. AIOS provides
shared services for scheduling agent requests, managing context and memory,
and controlling resource access through a software development kit (SDK)
\cite{aios2024}. LangGraph supports checkpoints, which save workflow state
for later resumption, and retains outputs from completed workflow steps so
that those steps need not be repeated after an interruption
\cite{langgraphPersistence}. Dapr Workflow reconstructs execution from
recorded history and reuses completed task results. Its activity functions,
which perform external work, can execute more than once; the documentation
therefore recommends idempotent activities, for which repeating a request
does not change the intended outcome \cite{daprWorkflow}.

These mechanisms motivate a distinction between recovering recorded
progress and determining whether an external action can safely be repeated.
For example, a device may apply an action before a failure prevents its
completion record from being stored. Restoring workflow state alone does
not resolve that uncertainty. \system{} addresses this boundary within the
agent service lifecycle by recording the identities and outcomes of actions registered through its
recovery interface and associating them with the originating invocation. Recovery reuses completed results and
withholds unsafe automatic replay of uncertain actions. Where an external
resource supports duplicate suppression, stable operation identifiers allow
it to recognize retries without repeating an already applied effect.
Invocation identity also connects recovered results and usage to the
applicable service charge. The contribution is the coordination of these
mechanisms with routed edge execution and delegated device actions,
not persistence or an SDK in isolation.

\paragraph{Scope of empirical comparisons.}
Table~\ref{tab:baselines} distinguishes related systems from the components
evaluated in this work. The Dify comparison uses the same model and shared
Nexus device tools to compare agent workflow execution
(Section~\ref{sec:dify-agent-measured}). The Dapr comparison evaluates
matched service invocation and memory-budget feasibility, rather than its
workflow engine (Section~\ref{sec:external-service-measured}). The resulting
latency--memory tradeoff and workflow measurements characterize the tested
configurations, not an overall ranking of the platforms. Component
ablations separately assess the effects of mechanisms within \system{}.

\begin{table}[t]
\centering
\small
\caption{Related system roles and the scope of empirical comparison in this
work. Systems without a matched experiment provide architectural context.}
\label{tab:baselines}
\begin{tabularx}{\linewidth}{@{}
  >{\raggedright\arraybackslash}p{0.20\linewidth}
  >{\raggedright\arraybackslash}X
  >{\raggedright\arraybackslash}X@{}}
\toprule
System & Relevant functionality & Comparison in this work \\
\midrule
Dify \cite{dify2026} &
Agent workflows and tool integration &
Matched workflows using shared model and device tools. \\
Ray Serve \cite{rayserve2026} &
Distributed application serving and replica management &
Architectural context; not experimentally compared. \\
Istio / Envoy \cite{istio2026,envoy2026} &
Service-traffic routing, security, and monitoring &
Architectural context; not experimentally compared. \\
Dapr \cite{dapr2026} &
Service discovery and invocation &
Matched invocation and constrained-memory experiments. \\
Portkey Gateway \cite{portkey2026} &
Model-provider routing, retries, and caching &
Architectural context; not experimentally compared. \\
\bottomrule
\end{tabularx}
\end{table}

\FloatBarrier

\section{System Overview}
\label{sec:overview}

\begin{figure}[htbp]
\centering
\includegraphics[width=\linewidth]{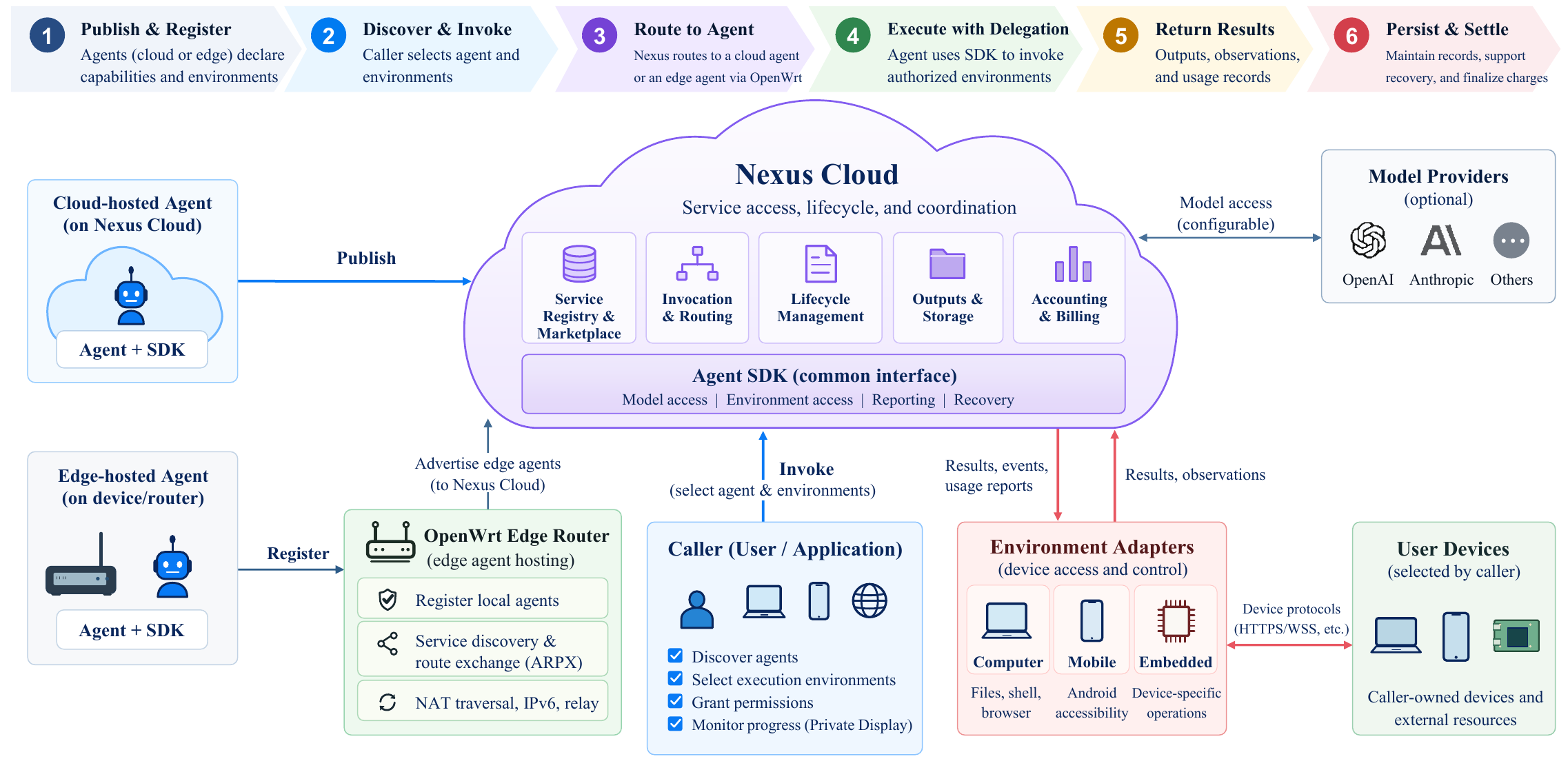}
\caption{System architecture and invocation lifecycle of \system{}.
Cloud hosts agent deployments and manages service publication,
invocation state, outputs, and accounting. OpenWrt makes edge-hosted
agents discoverable and reachable. The SDK runs within each agent process
and exposes interfaces for caller-authorized environment actions and
execution reporting. Results and usage remain associated with the
originating run. Agent-service routing and device-action dispatch are
separate paths: OpenWrt reaches agent handlers rather than routing their
commands directly to environment adapters. Arrows represent logical
exchanges, not mandatory physical paths.}
\label{fig:architecture}
\end{figure}

\system{} connects distributed agent services and caller-selected execution
environments through a shared invocation lifecycle.
Figure~\ref{fig:architecture} shows its four components.
The OpenWrt routing layer makes edge-hosted agents discoverable and
reachable; environment adapters execute authorized device actions and
return results or observations; a shared software development kit (SDK)
exposes these facilities to agent code; and the cloud layer manages
service publication, invocation state, output retention, and accounting.
The SDK runs within the agent process, which may be hosted in a cloud
container or deployed on an edge endpoint. This organization separates
where an agent executes from how its service is accessed, its actions
are authorized, and its lifecycle is managed.

An owner first deploys an agent handler and declares its capabilities,
the operations that the service accepts. Edge-hosted agents register
these capabilities with OpenWrt using leases, which are time-limited
registrations that must be renewed. Route exchange makes registered
services known to other participating routers, while cloud publication
makes them available to authorized callers. The routing layer maps agent
requests to eligible endpoints through local, relayed, or direct IPv6
paths. Relay access supports endpoints behind network address translation
(NAT); public IPv6 access supports direct entry points, including
dedicated per-agent addresses where network provisioning permits.
Model access is configured separately, so invoking a hosted model does
not require the agent handler to execute in the cloud.

When invoking a published agent, the caller selects the required
execution environments and grants permitted operations. Admission
checks access and establishes a run---a persistent execution context
associating the request with its caller, agent deployment, selected
resources, and applicable pricing terms. The platform invokes the
handler through the selected service path and supplies this context.
Through the SDK, the agent can issue computer commands, access workspace
files, navigate a browser, or operate a mobile application.
Environment-specific adapters dispatch actions to the selected resources
and return results or observations for subsequent reasoning.
Resource-binding and permission checks constrain dispatch, with live
revalidation on supported paths. A reachable device is therefore not
automatically an authorized execution environment.

The run also connects agent execution to its service lifecycle.
Progress reports, user interactions, outputs, and usage records refer
to the same invocation across execution attempts. For actions registered
through the SDK's recovery interface, operation records retain stable
identifiers and completed results. Recovery reuses those results rather
than repeating completed actions and withholds unsafe automatic replay
when an external effect remains uncertain. Cooperating resources can
use the same operation identifiers to suppress duplicate effects.
When charging applies, the pricing terms captured at admission and
cost reports checked against configured limits determine settlement.
Route leases, deployment status, device connections, and accounting
retain their respective state transitions; the run associates them
without requiring one shared state machine or an atomic transaction
across external devices.

\section{System Design}

\subsection{OpenWrt-native agent routing}
\label{sec:edge}

\begin{figure}[t]
\centering
\includegraphics[width=\linewidth]{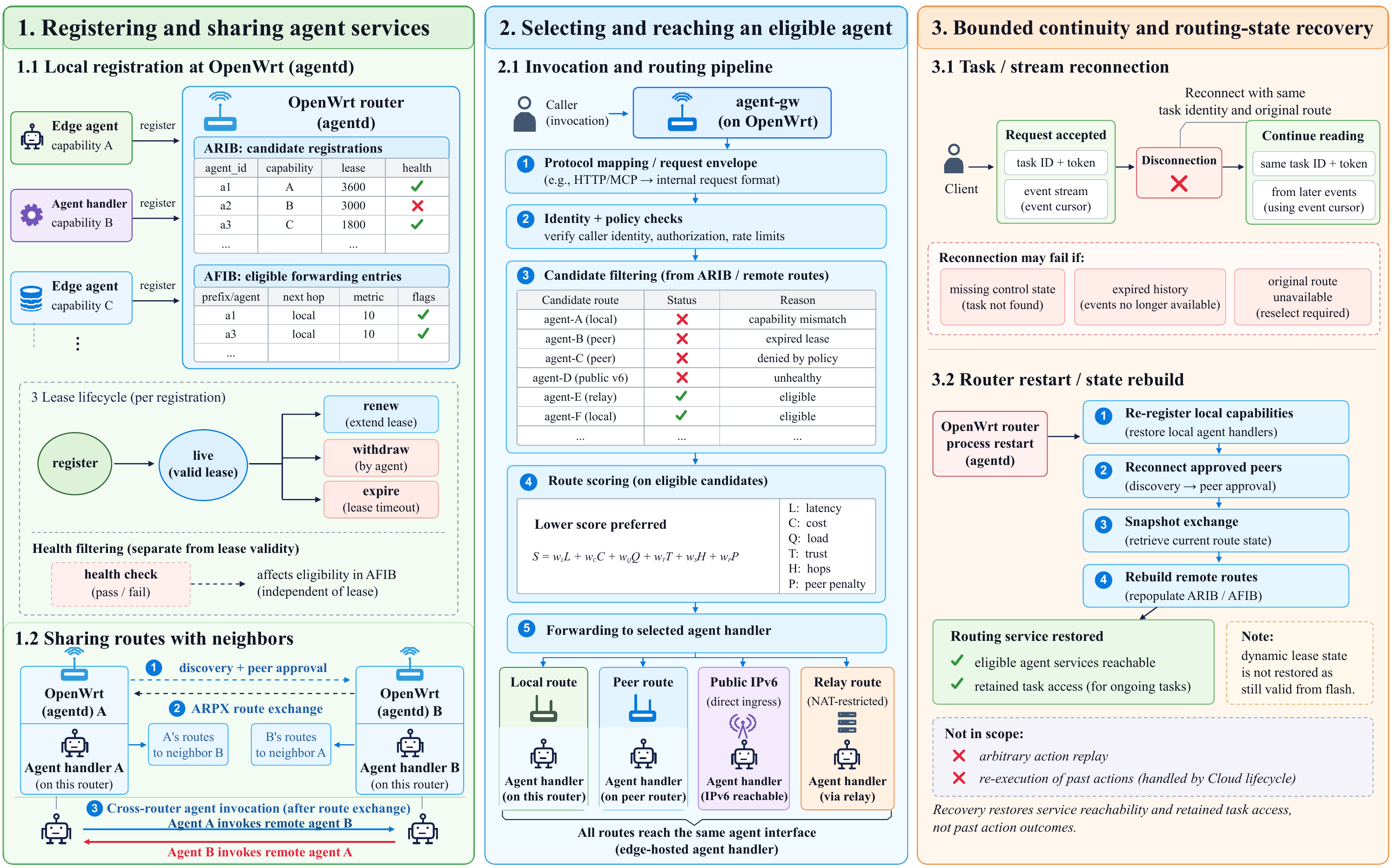}
\caption{OpenWrt-native agent routing.
(1) Time-limited registrations and route exchange maintain candidate
endpoints and enable authorized calls between agents on different
routers. (2) Identity and policy checks precede candidate filtering;
only eligible routes are ranked and used to reach handlers through
local, peer, public IPv6, or relay paths. (3) Reconnection uses retained
task state and the original route, while router restart reconstructs
service availability. These mechanisms recover access to services or
retained executions, not arbitrary completed actions. Public IPv6 entry
points require address provisioning and permitted incoming access.}
\label{fig:openwrt-path}
\end{figure}

The OpenWrt component implements the routing layer in
Section~\ref{sec:overview}. An agent exposes named operations, called
\emph{capabilities}, through an \emph{endpoint}, the network entry point
receiving requests. The routing layer selects an eligible endpoint and
delivers an \emph{invocation}, a request to execute an operation, to its
\emph{handler}, the code implementing it. OpenWrt is a Linux-based router
operating system: ordinary Internet Protocol (IP) forwarding reaches a
network address, whereas agent routing selects which service may perform
the requested work. The routing service, \code{agentd}, maintains service
records and policies; the invocation gateway, \code{agent-gw}, checks
requests and forwards them using these decisions
(Figure~\ref{fig:openwrt-path}). Handlers may run on the router or nearby
servers without hosting model inference there. Cloud manages publication
and persistent invocation records, while environment adapters separately authorize
the handlers' device actions.

\subsubsection{Registering and sharing agent services}

A registration specifies an operation name and interface version, origin,
endpoint, access restrictions, and a \emph{lease}, the period for which
the registration remains valid without renewal. The software development
kit (SDK) renews registrations, re-registers missing records, and withdraws
them when a managed server stops. Expiry immediately removes eligibility;
health checks separately exclude temporarily unhealthy endpoints.
Registration therefore establishes a candidate route, not guaranteed
availability or permission for every request.

Discovery identifies candidate routers; approval makes them \emph{peers}
permitted to exchange routes. Explicit approval is tied to the discovered
record's version so that an earlier approval does not authorize a changed
endpoint. Configured local policies can instead approve matching candidates
automatically. Approved peers exchange advertisements through ARPX, the
inter-router route-exchange protocol. Advertisements include ownership,
update sequence, remaining lease, and the routers already traversed. This
traversed-router list, or \emph{path vector}, supports loop rejection,
while ownership checks reject conflicting claims. Disconnected peers'
routes become ineligible immediately. Reconnection transfers a
\emph{snapshot}, a versioned collection of current routes, and removes
records absent from the completed collection. Transfer restarts when the
source version changes, so completion requires a sufficiently stable
version. A configured \emph{route reflector} redistributes learned routes
without sending them back toward their source, reducing pairwise router
configuration.

The optional Cloud connector links edge registration to publication.
After pairing establishes router ownership and identity, it publishes
compatible, healthy local registrations.
It renews Cloud registrations and versions synchronization updates across
restarts. Router presence remains separate from individual agent readiness,
and synchronization failure is not interpreted as service withdrawal.
Sections~\ref{sec:discovery-measured},
\ref{sec:peer-propagation-measured}, and~\ref{sec:peer-invoke-measured}
evaluate discovery, route exchange, and invocation separately.

\subsubsection{Selecting and reaching an agent}

Protocol adapters translate supported external calls into a common request
containing the operation and routing context. The Model Context Protocol
(MCP) exposes tools and resources; Agent2Agent (A2A) supports communication
between agents. Their supported entry points are mapped by protocol,
addressed service, and configured operation, not inferred from prompts.
Native and translated requests then share the same eligibility checks.

Selection separates permission from preference. In authenticated
configurations, the gateway verifies the requesting agent and its
\emph{tenant}, the administrative group governing access and resource
limits. Rules match tenant, source agent, and operation; highest priority
wins, with rule identifiers breaking ties. A selected deny rule rejects
the request rather than falling through to a weaker allow rule. Allowed
requests filter candidates by capability, liveness, and constraints on
region, latency, cost, load, trust, and permitted paths. Lookup by route identifier or operation
name reduces search work without bypassing checks. Limits on request rates
and simultaneous execution additionally bound request admission.

For an allow policy $p$, eligible route $r$ has advertised latency $L_r$
in milliseconds, cost $C_r$ in millionths of a monetary unit, load $U_r$
in parts per thousand, policy trust score $T_r$, and hop count $H_r$, the number
of routers traversed. Configured weights $w_L,w_U,w_T,w_H$ and divisor
$d_C$ scale these terms; $P_p(r)$ penalizes a nonpreferred peer.
$\operatorname{sat}_{64}$ caps arithmetic at $2^{64}-1$ to avoid overflow,
and the floor operation rounds down. Lower scores are preferred:
\begin{equation}
\begin{aligned}
 S_p(r)=\operatorname{sat}_{64}\!\bigl(
 &w_L L_r+\lfloor C_r/d_C\rfloor+w_U U_r\\
 &+w_T(100-T_r)+w_H H_r+P_p(r)\bigr).
\end{aligned}
\label{eq:openwrt-score}
\end{equation}
Route identifiers break score ties; the configured default ranking applies when
no policy matches. Ranking cannot override eligibility, and advertised
properties do not establish model quality or future prices.

Selected handlers are reached locally, through peer routers, or through
public entry points. Where address allocation and incoming access permit,
an agent can be bound to a dedicated public Internet Protocol version 6
(IPv6) address, a \code{/128} denoting one address, through the SDK or
router's public entry point. For endpoints behind network address translation (NAT),
which translates private addresses and can restrict incoming connections,
the router opens an outbound relay connection carrying remote invocations.
Multiple request streams share that connection, with cancellation and
limits on outstanding data. Relay access preserves edge execution but
consumes relay resources and does not itself conceal payloads from the
relay operator. Required address bindings and protected transport are not
silently replaced by other endpoints or unencrypted connections.

Across administrative boundaries, optional short-lived signed
\emph{forwarding assertions} convey request-specific authority instead of
the caller's reusable credential. With both routers' checks enabled, the
destination verifies the signature, caller and routing context, request-body
digest---a value detecting content changes---and markers identifying
repeated submissions. Exhausted replay-tracking capacity causes rejection.
These checks authorize service forwarding, not device actions, and depend
on trusted signers rather than protecting against their compromise.

\subsubsection{Recovering connections and routing state}

Connection recovery should contact an existing task rather than start a
new execution. For asynchronous calls, which acknowledge acceptance before
completion, the destination returns a task identifier and access token.
Later queries and control requests retain that identity and the original
routing constraints, allowing a surviving execution process with
retained state to be contacted without resubmitting the work. A lost acceptance reply remains
ambiguous, and process termination or record expiry can remove control state.
The evaluated capacity-limited implementation preserves unexpired records
during lookups and rejects new tasks when full
(Sections~\ref{sec:peer-task-recovery} and~\ref{sec:task-retention}).

Streaming delivers events before execution completes. A caller resumes
reading with the same task identity, original route, unchanged request
content, and an \emph{event cursor}, the identifier of its last complete
event. The retained history supplies later events without restarting the
handler. Unavailable routes, changed content, expired history, or exhausted
capacity cause explicit failure. Task controls and stream history have
bounded, process-local retention; they do not provide general restart
recovery. Retained-cursor replay uses the explicit Nexus stream interface,
not standard MCP resumption
(Sections~\ref{sec:mcp-client} and~\ref{sec:mcp-correction}).

Trying another candidate route is separate from stream resumption. This
applies only to synchronous requests, which wait for a response, with
caller retry permission, a nonempty key identifying the logical request,
and an explicitly \emph{idempotent} operation, whose repetition produces
no additional intended effect. Retry uses one ordered snapshot of
eligible routes and a shared time budget. This does not authorize replay of arbitrary
device actions or relocation of a partially delivered stream.

Router restart reconstructs service availability rather than agent
progress. OpenWrt restarts routing processes; local services re-register
and peers exchange fresh snapshots instead of restoring expired leases.
Configuration changes are validated before replacing active tables:
invalid updates retain the last valid state, while invalid startup tables
remain empty and invalid policy denies requests until repaired
(Section~\ref{sec:router-recovery}). The routing layer thus restores paths
to eligible services or retained tasks; persistent action outcomes,
outputs, usage, and settlement remain lifecycle responsibilities
(Table~\ref{tab:openwrt-functions}).

\begin{table}[t]
\centering
\small
\setlength{\tabcolsep}{3pt}
\caption{Routing responsibilities and their relationship to the other
components in the system overview.}
\label{tab:openwrt-functions}
\begin{tabularx}{\linewidth}{@{}
  >{\raggedright\arraybackslash}p{0.20\linewidth}
  >{\raggedright\arraybackslash}X
  >{\raggedright\arraybackslash}X@{}}
\toprule
Stage & Routing responsibility & Relationship to other components \\
\midrule
Maintain service records &
Register local services, expire old records, and exchange remote routes. &
Cloud uses compatible registrations for service publication. \\
\addlinespace
Forward invocations &
Apply access rules and reach eligible handlers through configured paths. &
Handlers use environment adapters to request separately authorized
 device actions. \\
\addlinespace
Reconnect and rebuild &
Reach retained tasks, resume available event histories, and reconstruct
 routing state. &
Persistent action recovery, output retention, and settlement remain
 lifecycle responsibilities. \\
\bottomrule
\end{tabularx}
\end{table}

\FloatBarrier

\subsection{Delegated agent interaction with execution environments}
\label{sec:environments}

An \emph{execution environment} is the device-side context where an
agent's action, such as a file read or application interaction, takes
effect. Following Section~\ref{sec:overview}, \system{} separates this
environment from the reasoning host through \emph{delegation}, permission
to act on the caller's behalf during an invocation, and adapters that
translate requests into device-specific operations. Each action is bound
to an authorized environment, executed there, and followed by a result
or observation returned to the agent.

\subsubsection{Binding actions to authorized environments}

A \emph{run} is the persistent context of an admitted agent invocation.
Its environment context identifies the selected resources, permitted
operations, and access credentials. \emph{Pairing} establishes device
identity; delegation authorizes specified operations for a run. A paired
computer may therefore permit an agent to read, but not write, its
\emph{workspace}, the directory configured for file access. Workspace
grants are restricted to agent-declared capabilities and associated with
the caller and project; Mobile retains its own device-binding and grant
records. Device identity, permission, and readiness are checked separately.

\begin{figure}[t]
\centering
\includegraphics[width=\linewidth]{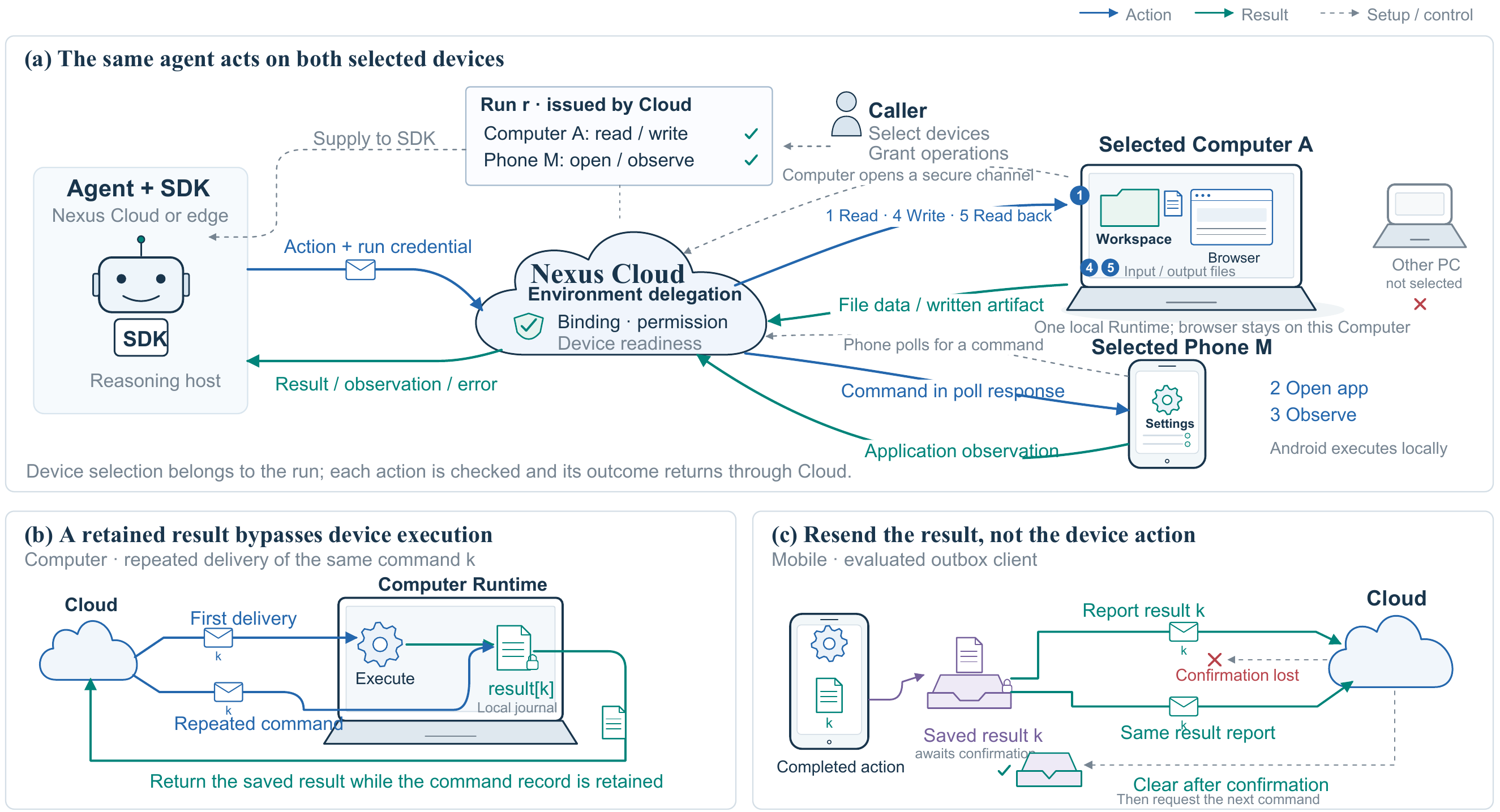}
\caption{Delegated agent interaction with caller-selected environments.
(a) The same agent accesses a Computer workspace and a Mobile application
through Cloud, which checks the applicable bindings and permissions.
Computer establishes an outbound channel; Mobile retrieves commands
through polling. (b) Repeated Computer commands reuse retained results
without re-executing completed actions. (c) The evaluated Mobile client
stores and retransmits completed results until acknowledgment, provided
Cloud recognizes repeated reports as the same completion. The bottom
strip identifies failures between an action taking effect and result
retention, when completion can remain uncertain. Result replay requires
retained state and valid authority; it does not replace run-level recovery.}
\label{fig:environment-flow}
\end{figure}

As shown in Figure~\ref{fig:environment-flow}, the agent uses the shared
software development kit (SDK) to send an action and run-scoped credential
to Cloud. Cloud checks the applicable resource binding and permissions,
delivers the command to the selected device, and returns its result,
observation, or error. This path controls device actions rather than
service routing: a Computer connection need not traverse the OpenWrt
agent gateway in Section~\ref{sec:edge}. Production connections use
protected web requests (HTTPS) or a protected, persistent two-way channel
(secure WebSocket, WSS). Cloud nevertheless receives delegated commands
and results; these connections do not conceal that data from Cloud.

\subsubsection{Executing actions through device-specific adapters}

The adapters share an action-and-result interface without making device
operations interchangeable. The \emph{Computer Runtime}, a device-side
execution program, opens an outbound command channel; Browser reuses it,
while Android periodically requests queued commands
(Table~\ref{tab:environment-protocols}).

\begin{table}[t]
\centering
\small
\renewcommand{\arraystretch}{1.12}
\setlength{\tabcolsep}{3pt}
\caption{Environment-specific action paths. Connection direction identifies
who establishes access, not the direction of subsequent messages.
Embedded support is an adapter extension rather than a universal device
implementation.}
\label{tab:environment-protocols}
\begin{tabularx}{\linewidth}{@{}
  >{\raggedright\arraybackslash}p{.18\linewidth}
  >{\raggedright\arraybackslash}X
  >{\raggedright\arraybackslash}X@{}}
\toprule
Target & Action delivery & Execution and authority boundary \\
\midrule
Computer & Device opens HTTPS setup and WSS command connections to Cloud. &
Runs under a local user account; file operations use the configured workspace. \\
\addlinespace
Browser & Actions use the selected Computer Runtime connection. &
Actions stay on the selected Computer and are checked against stored observations. \\
\addlinespace
Mobile & Android sends HTTPS requests for commands and reports results. &
Caller/device grants, configured approval, and Android permissions govern actions. \\
\addlinespace
Embedded & An explicit adapter exposes supported actions through a routed endpoint. &
Each adapter defines its supported operations, transport, and errors. \\
\bottomrule
\end{tabularx}
\end{table}

\emph{Computer.}
The Runtime runs under a local operating-system account. It signs a
Cloud-issued challenge with its private key; Cloud checks the signature
using the public key registered during pairing, verifying device identity.
Cloud then issues a single-use ticket for the outbound channel and assigns
a connection version to distinguish it from replaced connections. This supports devices without
public addresses or incoming remote-login connections. Commands carry
identifiers, required permissions, inputs, and deadlines. The Runtime
checks the operation--permission mapping before dispatch; connection
versions and message sequence numbers reject stale or repeated messages.
Receipt acknowledgment remains separate from an execution result.
Workspace file operations resolve paths within the configured directory,
but shell commands retain the local account's privileges: a workspace is
not a security boundary for arbitrary command execution.

\emph{Browser.}
The adapter executes browser actions on the caller-selected Computer,
checking the active run, device binding, unexpired credential, and saved
browser-control grant. It returns page descriptions and screenshot
references as observations. Each stored observation has a revision number;
an action supplying an expected revision older than the current one is
rejected. This prevents acting on a superseded observation, but does not
detect page changes that occur without a new observation being stored
(Section~\ref{sec:attached-browser-measured}). A missing browser, offline
Computer, or lost browser session---the retained browser-control
context---produces an error rather than execution on a substitute device.

\emph{Mobile.}
The Android client uses a paired device credential to request commands and
report results over HTTPS. It executes supported actions through Android
Accessibility, the operating-system facility for observing and interacting
with applications under user-granted permissions. Agent access separately
requires an active run, the caller-selected device, an unexpired delegation
credential, current grants, and device readiness. Cloud applies the
configured approval policy and claims an approved command while marking
it running in the same database transaction, so these updates occur
together before delivery. Reusing a request identifier can suppress
duplicate command creation, but does not guarantee that a device action
occurs only once. Android permissions and protected screens continue to
restrict execution, and revocation cannot undo an already dispatched action.

Embedded devices use explicitly supported operations and device-specific
adapters. This is an extension point, not a universal embedded-device
driver; each adapter retains its own transport and error behavior.

\subsubsection{Command-result retention and recovery}

A restored connection does not establish that an earlier action completed.
Cloud may redeliver an unexpired Computer command, so the Runtime combines
duplicate requests for the same still-running command rather than starting
separate executions. Completed results are retained in a bounded encrypted
local \emph{journal}, a stored record indexed by command identifier.
A repeated command with a retained result returns that result without
re-executing the action (Section~\ref{sec:computer-ack}).

Unlike the baseline Android client, the evaluated candidate retains one
completed result in an encrypted, pairing-scoped \emph{outbox}, storage
for a result awaiting Cloud acknowledgment. It retries that report before
requesting another command and removes the stored result after acknowledgment.
This repeats result delivery rather than device execution, provided Cloud
treats repeated reports as the same completion. The baseline and candidate
are compared in Section~\ref{sec:mobile-result-recovery}; earlier baseline
observations remain in Section~\ref{sec:mobile-ack}.

Neither mechanism closes the failure window between applying an action
and recording its result. Loss of a retained result can likewise leave
completion uncertain. Interactive shells and browser sessions also depend
on live process state that a command journal cannot recreate. Reconnection
does not renew an expired run credential, and Mobile stops its service on
authentication failure rather than treating it as a transient network loss.
The environment layer therefore preserves available command outcomes and
leaves uncertain effects unresolved; the lifecycle component in
Section~\ref{sec:marketplace} uses the run to associate action records,
recovery decisions, outputs, usage, and settlement.

\subsection{A shared SDK for agent services}
\label{sec:sdk}

The software development kit (SDK) provides the application-facing interface
for the routing, environment access, and invocation lifecycle described in
Section~\ref{sec:overview}. The developer implements a \emph{handler}, the
function performing an agent operation, while the SDK manages registration,
request transport, resource access, interaction, and reporting. The handler
retains responsibility for reasoning and action selection. Cloud, router
connections, model credentials, environment grants, and pricing remain
operator-configured rather than being created by the SDK.

\subsubsection{Service registration and scoped resource access}

The Python SDK registers named operations as agent capabilities and manages
their registration renewal and request handling. Listing~\ref{lst:serve}
shows a minimal capability; its router connection and credentials must
already be configured.

\begin{lstlisting}[language=Python,float=htbp,caption={Registering an agent capability through the SDK.},label={lst:serve}]
from nexus_agent import NexusAgent

agent = NexusAgent(router="auto", tenant="demo",
                   agent_id="echo-server")

@agent.capability("demo.echo")
def echo(payload):
    return {"echo": payload}

if __name__ == "__main__":
    agent.run()
\end{lstlisting}

For Cloud-admitted invocations, \code{NexusRunContext} supplies identifiers,
selected resources, and access credentials restricted to the invocation.
A \emph{run} associates an interaction with its execution, outputs, and
accounting; a \emph{turn} is one admitted handler invocation within that run.
Access to file, workspace, Computer, Mobile, interaction, and reporting
operations depends on the supplied interfaces and grants. The context
neither grants unrestricted device access nor replaces the checks in
Section~\ref{sec:environments}.

Resource interfaces retain distinct meanings. A workspace exposes granted
operations on a selected computer directory; run files are inputs and
outputs transferred through a private file service. Downloads verify size
and content integrity before releasing the completed file. Uploads return
file references, and supported interrupted transfers can resume using a
retained identifier within the same active run, with an unchanged source.
Model calls use a separately configured endpoint and the selected execution
configuration. The run context provides access to these resources, not
administrative control over their models, credentials, or permissions.

\subsubsection{Interaction during agent execution}
\label{sec:sdk-chat}
\label{sec:sdk-async}
\label{sec:sdk-inbox}

A hosted chat agent uses \code{NexusMCPServer} to expose its handler through
the Model Context Protocol (MCP), an interface for invoking tools.
Declaring \code{chat=True} enables task execution and caller interaction;
\code{continuable=True} separately permits later turns in the same run.
Cloud uses these declarations to provide the Web chat entry point.
The SDK supplies an invocation-specific context and closes it when the
handler returns. \code{chat.say} reports a message and \code{chat.ask}
requests a caller response; neither performs model inference.

An asynchronous handler can suspend while waiting for a model, device,
file transfer, or caller, allowing independent operations to proceed.
For example, history retrieval can overlap with a confirmation request,
but an action requiring consent must await approval. The handler must
finish or join its background work before returning; asynchronous execution
does not make that work persistent across process failure. Similarly,
cancelling a local wait does not establish that a remote action stopped.

Table~\ref{tab:sdk-interaction} distinguishes three forms of input during
execution. A reply answers an agent question, steering changes future work
in the current turn, and queueing requests a later turn. Steering is
\emph{cooperative}: the handler processes instructions at safe points,
such as between actions, rather than interrupting an ongoing device action
or undoing its effects.

\begin{table}[htbp]
\centering\small
\caption{User input and the execution it affects.}
\label{tab:sdk-interaction}
\begin{tabularx}{\linewidth}{@{}>{\raggedright\arraybackslash}p{18mm}
  >{\raggedright\arraybackslash}X@{}}
\toprule
Input & Effect on execution \\
\midrule
Reply & Answers a pending \code{chat.ask} and resumes the same handler
in the current turn. \\
Steering & The handler accepts or rejects a caller instruction at a safe
point; accepted changes affect subsequent actions in the current turn. \\
Queued request & Cloud admits a later turn after successful completion
of the current turn and renewed access and pricing checks. \\
\bottomrule
\end{tabularx}
\end{table}

\code{chat.ask} associates a question with a turn-specific identifier and
waits for a reply within a bounded timeout. An unavailable channel,
expired question, or cancellation is not approval. For steering, the
handler explicitly enables an \emph{inbox}, a channel for unsolicited
caller instructions, and acknowledges each instruction after applying or
rejecting it. Unacknowledged instructions may be delivered again.
Applications must therefore record relevant instruction state and action
identities before acknowledgment when restart recovery or nonrepeatable
effects are involved; a local variable update is not persistent recovery.
Cloud stores queued requests independently of the browser connection and
rechecks authority, execution context, and funding or pricing before the
next turn. Queue acceptance does not guarantee execution or fix its charge,
and unused steering instructions do not become queued turns.

\subsubsection{Outputs, usage, and settlement reporting}
\label{sec:sdk-display}

\emph{Private Display} is the caller-authorized page showing a run's
messages, progress, observations, and outputs. SDK helpers publish events
and protected file references rather than modifying the page directly.
Their role is reporting: a shell log does not execute a command, a browser
frame reports an existing observation, and a readiness notification does
not create or verify an output. Run-file uploads and writes to a selected
output workspace use different interfaces and require their corresponding
permissions. The runtime's credential for writing events is separate from
the caller's permission to read them. Caller-visible events remain within
the authorized run rather than being published to the Internet.

Usage and service-cost reporting are also separate. After an actual model
call, \code{usage.report} records its model and measured token counts;
missing usage remains incomplete rather than being replaced with zero.
Each logical model request has a stable report identifier so retries can
be associated with the same request. Under reported pricing, the handler
aggregates the applicable service cost and submits one final report per
turn using a stable billing key, the identifier retained across report
retries. The charge follows the configured pricing terms and authorized
limit, not necessarily the model provider's cost.

File transfer, event reporting, handler completion, and settlement have
separate outcomes. Some reporting failures allow execution to continue;
an optional bounded \emph{outbox}, storage for pending events, supports
report retries but cannot recreate interrupted computation or undo device
actions. Usage and cost reports are separate requests, so partial success
must retain which report was accepted. An accepted cost report is likewise
not proof of completed settlement. These distinctions preserve the
SDK's role as a shared interface while leaving persistent action recovery
and final accounting to the lifecycle mechanisms in
Section~\ref{sec:marketplace}.


\subsection{Persistent agent service lifecycle and bounded accounting}
\label{sec:marketplace}

An agent service may continue after its caller disconnects, wait for user
input, or require another execution attempt after a worker failure.
Its lifecycle therefore cannot be represented by a single network
connection. The platform must preserve which work was authorized, which
action outcomes have been recorded, and which pricing terms govern the
resulting charge. The Cloud component addresses this requirement through
persistent execution records that connect publication, invocation,
interaction, recovery, output delivery, and accounting. Following
Section~\ref{sec:overview}, OpenWrt determines how requests reach agents,
environment adapters control device actions, and the software development
kit (SDK) supplies access and reporting interfaces. Cloud maintains the
association among these activities without treating request acceptance,
action completion, output availability, and settlement as interchangeable
outcomes.

\subsubsection{Admitting invocations and preserving execution identity}

Publication exposes an agent's callable operations and selected
\emph{deployment}, the configured server running its \emph{handler},
the code implementing an operation. The Model Context Protocol (MCP)
provides interfaces for discovering and invoking these operations
\cite{mcp2025transport,mcp2025lifecycle}. Cloud combines the declared input
requirements and interaction behavior with access policy, allowing the
same service to support Web and programmatic callers without an
agent-specific Web interface. Publication makes an operation available
for selection; permission to execute it is checked during admission.

\emph{Admission} validates the requested operation, inputs, execution
configuration, and environment permissions before accepting work.
Cloud distinguishes the identities and records in
Table~\ref{tab:run-identities} to separate the caller's continuing
interaction from individual invocations and execution attempts.
A \emph{run} is a persistent record associating an agent with its caller,
project, selected environments, and consumer \emph{tenant}, the
administrative group governing access and resources. Within a run, a
\emph{turn} identifies one admitted tool invocation and its execution
configuration, usage, and applicable charge. One \emph{execution task}
tracks the run's current work and is updated for later turns. Separate
attempt records identify the \emph{worker}, a process executing queued
work, and its ownership of each execution attempt. A replacement worker
can therefore attempt the same turn without redefining the caller's run
or admitting a new turn.

\begin{table}[htbp]
\centering\small
\caption{Execution identities and records at different lifecycle scopes.
A run can contain multiple turns, and a turn can involve multiple worker
attempts; one execution task tracks the run's current work.}
\label{tab:run-identities}
\begin{tabularx}{\linewidth}{@{}
  >{\raggedright\arraybackslash}p{.20\linewidth}
  >{\raggedright\arraybackslash}X@{}}
\toprule
Identity / record & Responsibility and relationship \\
\midrule
Run & Preserves caller scope and associates selected resources,
conversation history, outputs, and usage across turns. \\
Turn & Identifies one admitted tool invocation, including its execution
selection, usage, and applicable pricing terms. \\
Execution task & Tracks the run's current execution, waiting, and recovery
state; updated for subsequent turns rather than recreated for each message. \\
Worker attempt & Identifies an execution attempt and its assigned worker,
separately from the persistent task and caller-visible status. \\
\bottomrule
\end{tabularx}
\end{table}

\begin{figure}[htbp]
\centering
\includegraphics[width=\linewidth]{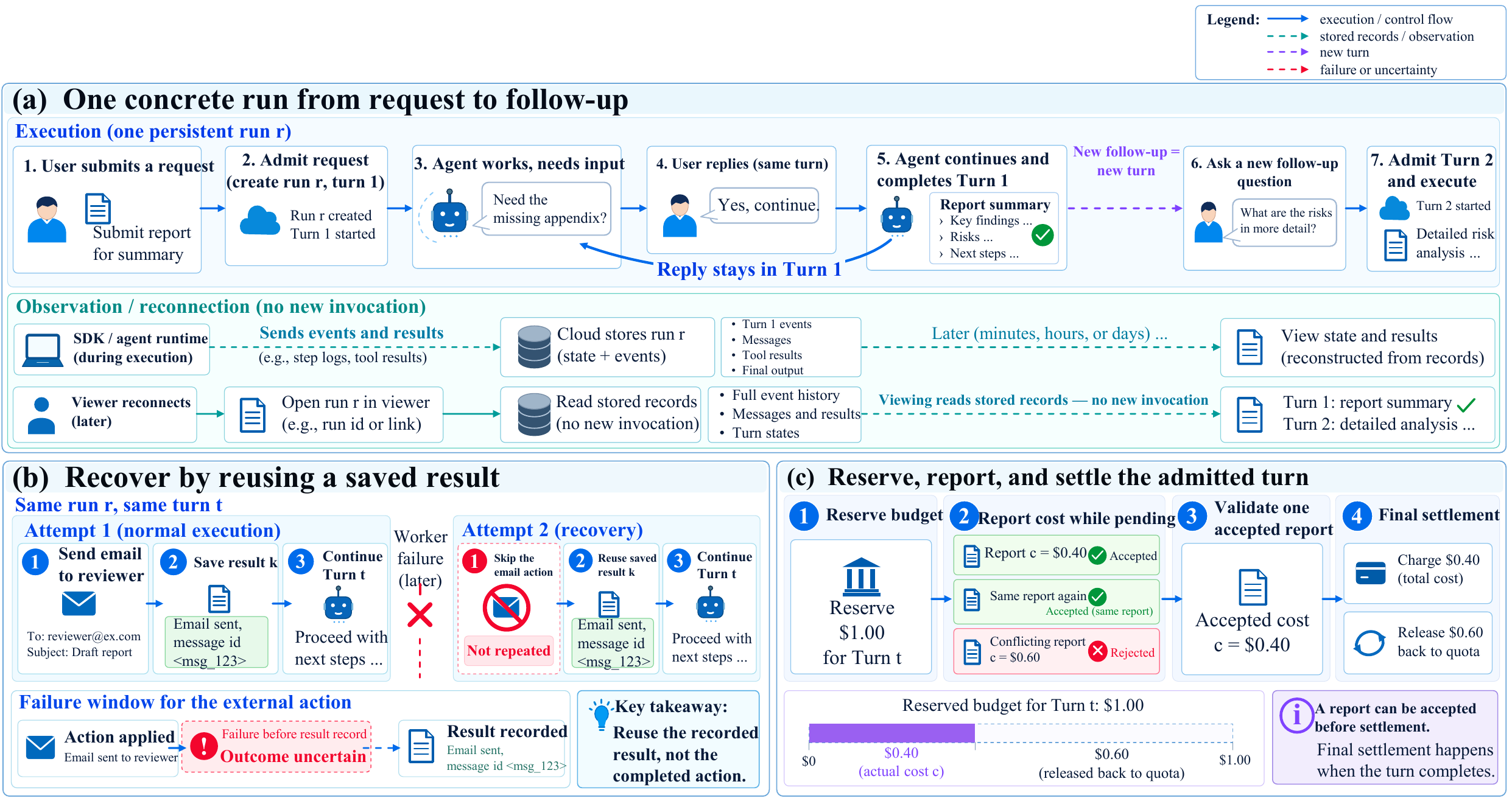}
\caption{Persistent agent service lifecycle and bounded accounting.
(a) Replies continue the current turn; follow-ups require renewed admission,
while reconnecting viewers read stored records without reinvoking the
handler. (b) Recovery reuses completed results of SDK-recorded actions;
a failure between an external effect and its completion record can leave
the outcome uncertain. (c) In the illustrated successful reported-price
turn, identical report retries are recognized while pending. Settlement
separately finalizes the charge within the admitted limit and releases
unused reserved funds. The email action and monetary amounts are
illustrative.}
\label{fig:web-invocation-flow}
\end{figure}

Accepted requests enter a queue, and Cloud returns the run identifier
before execution completes. A worker subsequently invokes the selected
handler with trusted run and turn identifiers, permitted resource
interfaces, and restricted reporting credentials. Cloud replaces
caller-supplied values in reserved execution-context fields before
supplying this information to the SDK. The handler thus receives the
context required for authorized execution and reporting without the
caller's browser login credentials. MCP sessions, which retain negotiated
communication context, remain separate from persistent run history;
programmatic session mappings are restricted to the caller and selected
service context. Figure~\ref{fig:web-invocation-flow}(a) illustrates this
separation: an approval reply continues the current turn, an authorized
follow-up starts a later turn in the same run, and a page refresh only
retrieves existing records.

\subsubsection{Observing execution and continuing a run}

Observation is separated from execution so that a caller can monitor or
revisit an invocation without starting its work again. \emph{Private
Display}, the caller-authorized interface for progress, replies, and
outputs, combines a \emph{snapshot}, a saved summary of current run state,
with \emph{events}, individual progress and action reports. Event types
follow the AG-UI vocabulary for agent--interface communication
\cite{agui2026events}. The SDK submits events using a run-specific write
credential. Cloud validates that credential and, for managed tasks, the
active \emph{execution lease}, a time-limited assignment authorizing a
worker to execute the task, before accepting reports.

Event identifiers and sequence numbers serve different purposes.
A producer retains an event identifier across delivery retries; identical
content and visibility under that identifier return the existing record,
whereas conflicting reuse is rejected. Cloud assigns accepted events
increasing sequence numbers to support ordered retrieval. A reconnecting
viewer reads the snapshot and events after its \emph{cursor}, the sequence
number of the last retrieved event. This combination supports repeated
report delivery and incremental observation without another handler
invocation. It prevents duplicate presentation records, not repeated
execution of the actions being reported.

Observation remains subject to caller authorization. Display reads check
the authenticated caller, tenant, project, run visibility, and a signed
display credential. This read authority is separate from runtime event
writing, device delegation, and billing; internal reporting credentials
are not exposed to callers. Private Display therefore isolates callers
from one another, not from Cloud, which processes the stored information.
Output availability is checked separately as well: a downloadable artifact
requires a ready stored copy, rather than merely an output-ready event.
Because reporting can fail while execution continues, accepted display
history is not necessarily a complete record of performed actions
(Sections~\ref{sec:sdk-async}--\ref{sec:sdk-display}).

Continuing execution requires more than reading these records. A reply to
a pending interaction resumes the current turn only after checks of the
caller, pending state, expiry, and permitted choices. A later turn requires
a tool that permits continuation, no prior work still executing or
finalizing, and renewed checks of the tool, deployment, and current
authority. Cloud then advances the turn index, refreshes delegated context,
and retains earlier messages. Thus, supplying requested input does not
admit another invocation, while retained conversation history does not
replace authorization for new work. Section~\ref{sec:sdk-inbox} further
distinguishes instructions affecting future actions in the current turn
from requests queued for a later turn.

\subsubsection{Recovering actions and settling bounded charges}

Persistent identity establishes which invocation should continue; action
records determine which completed work can be reused. Worker ownership
and attempt records distinguish active execution from work awaiting an
agent server or an explicit recovery decision. For actions registered
through the SDK's recovery interface, an \emph{operation journal} stores
completed outcomes under stable operation identifiers. A recovery attempt
can reuse these outcomes within the original run and turn instead of
repeating the corresponding actions, as illustrated in
Figure~\ref{fig:web-invocation-flow}(b). Section~\ref{sec:worker-recovery}
evaluates this recorded-result reuse under worker failure.

The journal's effectiveness depends on whether completion was recorded.
If a file write takes effect and its result is saved before a worker fails,
recovery can reuse the recorded result. If the failure occurs after the
write but before result recording, the external effect remains uncertain:
restoring task state cannot establish whether replay would repeat an
already completed action. Such cases require an explicit recovery decision
rather than an assumption that missing evidence means no execution.
Cancellation preserves a related distinction. A cancellation request does
not prove that an ongoing action has stopped; cancellation may not be
recorded as complete until the tool call returns. Closing a browser tab
neither cancels execution nor reverses completed effects
(Section~\ref{sec:measured-cancellation}). Likewise, restoring a route or
stream in Section~\ref{sec:edge} restores access, not the outcome of an
uncertain action.

The distinction between logical work and execution attempts also determines
the accounting unit. Several worker attempts or repeated reports can belong
to one admitted turn, so charging follows that turn rather than display
reads or transport retries. Where commercial policy applies, supported
pricing includes free service, a fixed charge, and a final cost reported
by the agent. Before execution, admission checks funding, retains a
\emph{pricing snapshot}---the currency, pricing mode, and authorized amount
for the turn---and reserves eligible funds for subsequent charging.
Retaining these terms prevents later price changes from redefining an
already admitted invocation.

For reported pricing, let $a_{r,t}$ denote the authorized amount for run
$r$, turn $t$, and $c_{r,t}$ the reported cost. If the report provides an
itemized breakdown, let $\ell_{r,t,j}$ denote the amount of item $j$.
Accepted reports satisfy
\begin{equation}
\begin{aligned}
  0 \leq c_{r,t} &\leq a_{r,t},\\
  \sum_j \ell_{r,t,j} &= c_{r,t}
  \quad\text{if itemized}.
\end{aligned}
\label{eq:bounded-turn-cost}
\end{equation}
The first condition enforces the amount authorized at admission; the second
checks consistency between an optional breakdown and its total. Reports
use six-decimal amounts and must match the turn, currency, and billing
credential. An \emph{idempotency key}, an identifier retained across retries
of the same logical report, associates repeated submissions. While the
invocation remains pending, identical reports are recognized as repeats
and conflicting reuse is rejected. This rule governs report acceptance
within that state; it neither extends to unrestricted submissions after
settlement nor prevents repetition of external actions.

Report validation and final charging complete different parts of the
lifecycle. An accepted report establishes a permitted cost attributed to
the turn, whereas \emph{settlement} finalizes the charge and disposition
of reserved funds according to the invocation outcome
(Figure~\ref{fig:web-invocation-flow}(c)). Accounting checks reconcile the
pricing snapshot, accepted report, reserved and released funds, charged
amount, and \emph{ledger entries}, the recorded accounting transactions.
Device actions may contribute to service usage or reported cost without
each command having an independent tariff. These charging rules are
implemented by Cloud's commercial modules, not the standalone SDK or every
deployment. Together, run and turn identities preserve the association
among authorized work, recorded outcomes, recovery, and settlement, while
the corresponding checks determine what has actually completed at each
stage.

\section{Evaluation}
\label{sec:evaluation}
\subsection{Experimental Design}
\label{sec:eval-setup}

We evaluate how Nexus distributes computation, controls access to user
devices, and continues work after interruptions. Three research questions
(RQs) guide the experiments. RQ1 asks how much Cloud processing capacity
is needed to complete a fixed workload on time, and how efficiently
requests can find a suitable service. RQ2 asks whether a common programming
interface preserves the requested file and device operations while
enforcing permissions that may change during execution. RQ3 asks whether
interrupted tasks can reuse completed work, recover results under their
original request identities, and avoid repeated actions or charges.
Table~\ref{tab:rqmap} summarizes the comparisons and measurements.

\begin{table}[htbp]
\centering\small
\caption{Research questions, experimental comparisons, and primary
measurements.}
\label{tab:rqmap}
\begin{tabularx}{\linewidth}{@{}p{9mm}
>{\raggedright\arraybackslash}X
>{\raggedright\arraybackslash}X@{}}
\toprule
RQ & Experimental comparison & Primary measurements \\
\midrule
RQ1 &
The same task and request-arrival schedule in Cloud and at the edge;
service lookup using an index that narrows the search versus checking
every entry. &
Smallest tested Cloud processor limit meeting the deadline; processor
time and transferred bytes; lookup time and whether the selected
service remains unchanged. \\
\addlinespace
RQ2 &
Shared device-access functions versus explicitly constructed requests
for the same operations; permissions retained versus withdrawn. &
Correct files and device responses; association of commands with the
original request; permission enforcement before subsequent actions;
device-access code size. \\
\addlinespace
RQ3 &
Recovery with and without stored operation records; resending stored
results; preventing repeated actions at the service performing them. &
Repeated actions; recovery of the original task and its results;
additional time when no failure occurs; final charging outcomes. \\
\bottomrule
\end{tabularx}
\end{table}

\paragraph{Platforms and workloads.}
Cloud runs on an Intel Core i7-12700H machine with 14 cores, 20 logical
processors, and approximately 32~GiB RAM. OpenWrt runs in a 64-bit x86
virtual machine (VM) managed by Hyper-V, with two virtual processors
(vCPUs) and 1~GiB RAM. Its software versions are OpenWrt 25.12.4 and
Linux 6.12.87. The Computer Runtime, which executes operations on the
caller-selected computer, and an Android emulator configured for
application programming interface (API) level 34 run on the same machine.
The emulator polls for commands through an Android Debug Bridge (ADB)
reverse connection, which forwards its connections to the host-side
service.

Each experiment specifies its processor and memory limits. The experiment
comparing the request-forwarding program, referred to as the
\emph{bridge}, under the same software environment also restricts both
configurations to the same selected processor cores. Other experiments
share processor resources with background activity.

We use fixed computations and scripted device operations to study
execution, communication, and recovery without variation in model
decisions. The hashing workloads map input bytes to a fixed-length
digest, which is checked against an independently computed result.
Model-driven workflows instead issue fresh requests to Qwen/Qwen3-8B
during each execution. We check their generated files, recorded device
commands, and model-usage records. The twelve-task workload includes
expense summaries, inventory checks, and Android checklists, combining
model and device operations with independently checkable outputs.

\paragraph{Experimental controls.}
A paired trial executes the same task in each of two configurations.
The pair shares input data, expected results, and the request-arrival
schedule specified for that comparison. In mechanism ablations, we
enable or disable a mechanism while retaining the other stated execution
conditions. Deployment comparisons instead vary the complete execution
configuration, including its operating-system kernel, request path,
and available edge resources.

The device-access comparison uses functions supplied by the software
development kit (SDK) or explicitly constructed Hypertext Transfer
Protocol (HTTP) requests. Both implementations retain the same task
execution, authentication, and recovery services. Recovery experiments
use an \emph{operation journal}, a stored record of operation status
and results that can be consulted after an interruption. Separate tests
vary whether the service performing an action recognizes a repeated
operation identifier and returns its stored result instead of applying
the action again.

Dify provides the comparison for coordinating model and device operations
into a workflow; Dapr provides the comparison for calling a target
service. Table~\ref{tab:eval-comparability} identifies the shared
conditions and the differences in each comparison.

\begin{table}[htbp]
\centering\small
\caption{Shared conditions and varying factors in the principal
comparisons.}
\label{tab:eval-comparability}
\begin{tabularx}{\linewidth}{@{}
>{\raggedright\arraybackslash}p{27mm}
>{\raggedright\arraybackslash}X
>{\raggedright\arraybackslash}X@{}}
\toprule
Comparison & Shared conditions & Factors that vary \\
\midrule
Edge / Cloud execution &
Compiled task program, input data, output checks, and completion
reporting; request-arrival schedule and Cloud processor limit in the
capacity experiment. &
Execution location, operating-system kernel, bridge version,
connection path, and additional edge resources. \\
\addlinespace
Bridge outside / inside Docker &
Kernel, system image, programs and libraries, selected processor
cores, requests, and resource-measurement method. &
How the bridge is started and isolated from other processes. \\
\addlinespace
Nexus / Dify workflows &
Tasks, model source, action-selection instructions, maximum number
of model decisions, device tools, and output checks. &
Coordination of workflow steps, actions actually generated by the
model, and handling of the remaining execution time. \\
\addlinespace
Nexus / Dapr service calls &
VM, target program returning the same bytes, and request schedule;
equal total processor and memory limits in the memory experiment. &
Call interfaces, how services are located and authenticated, and
supporting processes. Latency and memory limits are tested separately. \\
\addlinespace
SDK / explicit HTTP &
Task code, selected devices, requests, permissions, and shared
execution and recovery services. &
Device-access code: SDK functions versus explicitly constructed
HTTP requests. \\
\addlinespace
Journal / no journal &
Task steps, their execution pace, the point at which the execution
process is stopped, the wait before task reassignment, and checks
of outputs and actual actions. &
Saving and reusing operation records. Handling repeated requests at
the service performing an action is varied in a separate experiment. \\
\bottomrule
\end{tabularx}
\end{table}

\paragraph{Measurements and analysis.}
Correct completion requires the expected file, device result, or
response. Timely completion additionally requires the result to arrive
within the specified deadline, measured from the scheduled request
arrival. We separately record whether an unauthorized action is rejected,
whether a result is recovered for the original command, and whether
an action is repeated. These measurements distinguish outcomes that
final-output correctness alone can hide: writing the same content twice,
for example, can leave the expected file while repeating the write.
Scheduled attempts include startup and execution failures. Preliminary
setup runs are reported separately, and tasks used to refine model
instructions are identified as calibration tasks.

Timing summaries distinguish the time until any attempt ends from
successful-completion time. Comparisons restricted to pairs in which
both executions succeed identify that condition explicitly. Paired
differences or ratios are calculated within each pair before being
summarized across trials. To estimate uncertainty, bootstrap resampling
repeatedly selects complete pairs or scheduled groups of trials,
allowing the same group to be selected more than once, and recalculates
the statistic. Calls from one workflow remain together because they
share execution conditions. Small fixed failure experiments report
individual outcomes and counts. Repeated executions are distinguished
from distinct tasks, while calibration tasks and component-version
groups are reported separately.

Resource measurements specify which processes they cover. Process
resident set size (RSS) measures memory resident in physical RAM for
a process, including shared pages mapped by that process. Memory
charged to a Linux control group (cgroup), which groups processes for
resource accounting and limits, is reported as a separate measure.
Other Cloud measurements sample the designated service processes
together.

Transfer counters measure request or response body bytes, excluding
protocol headers and lower-level network overhead. Model usage records
count executed requests and tokens, the text units processed or generated
by the model. Charging experiments use test account balances to verify
final charges and the release of unused reserved funds; commercial
invoices are not measured. Additional settings and complete result
tables appear in the corresponding appendices.

\subsection{Cloud Capacity and Data Locality}
\label{sec:eval-distributed}
\paragraph{Capacity under a completion deadline.}
Moving computation beside its input can reduce the work performed by the Cloud, but the operational benefit depends on whether a smaller Cloud CPU ceiling still permits the same offered demand to meet its deadline. Both configurations execute the same static capability binary and 32-round SHA-256 computation. Edge execution reads the local source, whereas Cloud execution fetches that source before computation. Each batch offers twenty tasks at one-second intervals using up to eight caller workers. Completion is measured from planned arrival, including any wait for a worker, with a two-second deadline.

The Cloud CPU ceiling varies over 0.25, 0.5, 1, and 2 vCPU at a fixed 512-MiB memory ceiling. Both configurations receive the same Cloud ceiling at each grid point; edge resources are additional. Four paired batches per cell use balanced arm order and no application retry. A setting passes when all twenty outputs and receipts in every batch are correct and timely.

\begin{table}[htbp]
\centering\small
\caption{Cloud-capacity feasibility for the 16-MiB, 32-round workload. Each
entry is passing batches out of four; each batch offers 20 tasks. Initial
and repeated campaigns are shown separately, including all deadline misses.}
\label{tab:eval-capacity}
\begin{tabular}{@{}r rrrr@{}}
\toprule
Cloud CPU & \multicolumn{2}{c}{Initial} & \multicolumn{2}{c}{Repeat} \\
ceiling (vCPU) & Edge & Cloud & Edge & Cloud \\
\midrule
0.25 & 4/4 & 0/4 & 4/4 & 0/4 \\
0.5  & 4/4 & 1/4 & 4/4 & 3/4 \\
1    & 4/4 & 3/4 & 4/4 & 3/4 \\
2    & 4/4 & 4/4 & 4/4 & 4/4 \\
\bottomrule
\end{tabular}
\end{table}
\begin{figure}[htbp]
\centering\includegraphics[width=.94\linewidth]{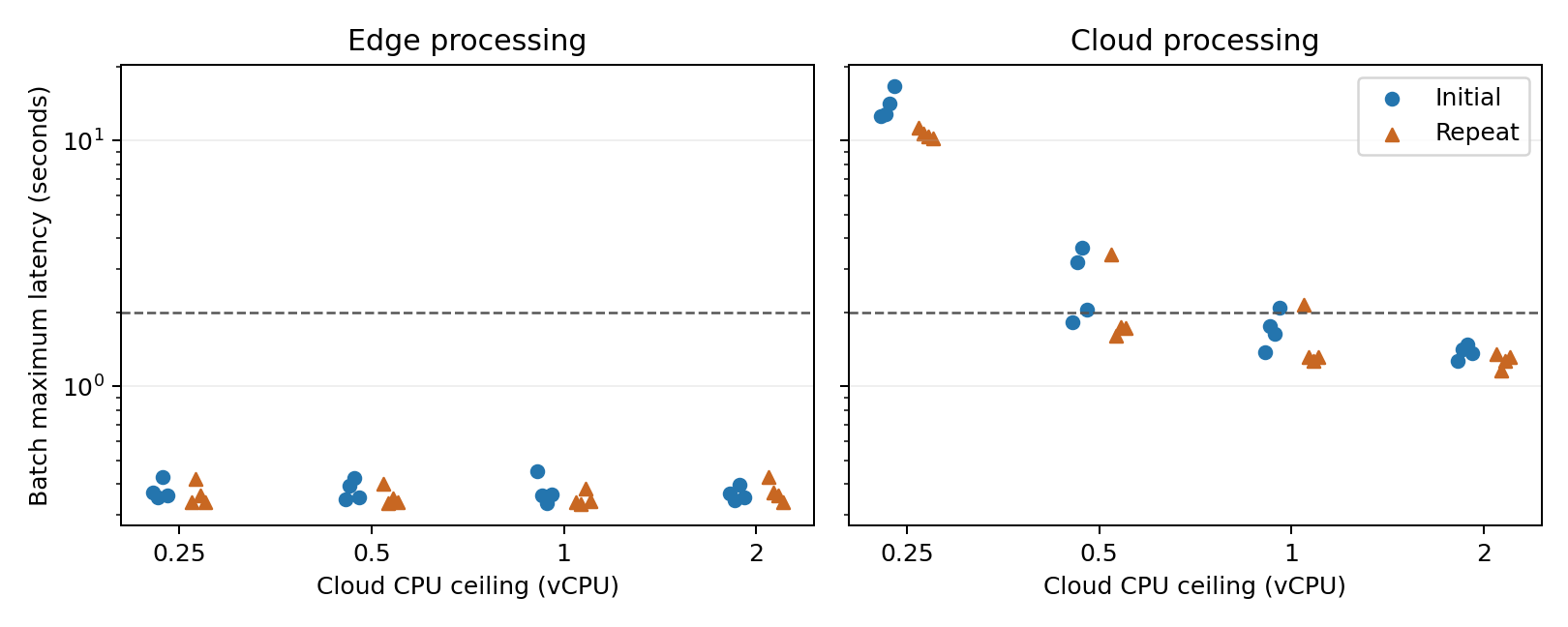}
\caption{Maximum planned-arrival-to-receipt latency of each large-input batch. The dashed line marks the two-second deadline, and horizontal jitter separates batches at the same CPU ceiling.}
\label{fig:eval-capacity}
\end{figure}

For the 16-MiB input, edge processing passes every batch at 0.25 vCPU, while Cloud processing first passes every batch at 2 vCPU in both experiments (Table~\ref{tab:eval-capacity}). The 1-KiB control passes at 0.25 vCPU in either configuration. The benefit therefore depends on the input and computation intensity. The initial experiment returns 1,280 exact outputs, of which 1,170 meet the deadline. The large-input repetition returns 640 exact outputs, with 555 timely completions. This repetition was selected after inspecting the initial results and scheduled before execution; its batches are shown separately in Figure~\ref{fig:eval-capacity}.

The capacity result characterizes the two deployments in Table~\ref{tab:eval-comparability}: kernels and bridge builds differ, and edge processing contributes resources beyond the Cloud ceiling. The measured benefit is lower required Cloud-side capacity at the specified load; the total Cloud-plus-edge budget is not held equal. Sections~\ref{sec:rightsizing-measured} and~\ref{sec:rightsizing-replication} provide the full grid. Section~\ref{sec:cost-evidence-bridge} relates this capacity relief to provisioned allocations and expenditure.

\paragraph{Where computation and data move.}
A second experiment crosses 1-KiB, 1-MiB, and 16-MiB inputs with one and 32 digest rounds. Ten paired batches per cell, each with twenty serial tasks, yield 2,399 exact results from 2,400 attempts. The edge completes all 1,200 tasks; one Cloud fetch fails. For the 16-MiB, 32-round workload, the median paired Cloud CPU reduction is 11.102 s per batch, with a 95\% interval of [10.911, 11.368] s. Edge CPU increases by 5.470 s. These counters include the dedicated endpoint, bridge children, and observation work, while shared services and VM overhead lie outside the interval.

The corresponding marginal medians of batch-level median attempt duration are 984.2 ms in Cloud and 334.8 ms at the edge. For 1-KiB inputs, the direction reverses: edge execution adds paired median times of 12.025 and 11.135 ms for one and 32 rounds. A complete 16-MiB Cloud batch fetches 320 MiB of source data, while the measured edge-execution interval sends only summary receipts to Cloud. Warmup has already delivered the data to both endpoints; the reduction therefore concerns repeated normal-path transfers. Together, the CPU and byte results identify data locality as a useful deployment property for the tested large-input workload (Section~\ref{sec:measured-placement}).

\paragraph{Complete model-and-tool workflows.}
The portability experiment moves an entire Agent loop while retaining its source, Python executable, SDK, device bindings, and one-CPU/512-MiB process ceilings. Across two experiments, all 32 workflows in sixteen pairs complete, each executing two fresh model requests and actual Computer and Android actions. Kernels and hosting paths differ. The mean native-minus-Docker non-model residual is +0.559, -0.029, -0.495, and +0.131 s across the four experiment/input cells. This arithmetic decomposition has no stable direction.

A further twelve-task, three-family experiment completes all 24 workflows, with independently realized action sequences at each placement. Across these workloads, the same Agent operates at either location, while the faster configuration varies with task family and model waiting. Data-local capability execution and whole-Agent placement thus offer different deployment choices: the former reduces central processing for large inputs, whereas the latter also changes the path coordinating models and devices (Sections~\ref{sec:full-agent-placement}, \ref{sec:full-agent-resources}, and~\ref{sec:diverse-agent}).
\FloatBarrier

\subsection{Workflow and Service Baselines}
\label{sec:eval-external}
\paragraph{Workflow execution over shared tools.}
We compare a direct Nexus SDK loop with an official Dify 1.17.1 workflow using native loop, LLM, code, and HTTP nodes and its signed provider version 0.0.68. Both execute the same policy through the same Qwen/Qwen3-8B source, gateway, Computer/Android tool backend, and output validator. The tool adapter performs one requested action without planning or invoking the model. The experiment therefore compares two orchestration implementations over common services.

Twelve synthetic tasks, four per family, have two paired repetitions in opposite product orders, giving 24 pairs and 48 serial executions. Each arm uses fresh paths and resets Android to Home. Temperature zero, a 256-token response limit, disabled thinking, and a twelve-decision budget are common. Invalid actions consume a decision; inference and transport are not automatically retried. Three task identifiers used for prompt calibration contribute six pairs, while the nine remaining tasks contribute eighteen.

\begin{table}[htbp]
\centering\small
\caption{Paired workflow outcomes for Nexus and Dify. Calibration status is defined by task identity.}
\label{tab:eval-dify-outcomes}
\begin{tabular}{@{}lrrr@{}}
\toprule
Task stratum & Both succeed & Nexus only succeeds & Total pairs \\
\midrule
Calibration & 6 & 0 & 6 \\
Non-calibration & 16 & 2 & 18 \\
All tasks & 22 & 2 & 24 \\
\bottomrule
\end{tabular}
\end{table}
\begin{figure}[htbp]
\centering\includegraphics[width=.94\linewidth]{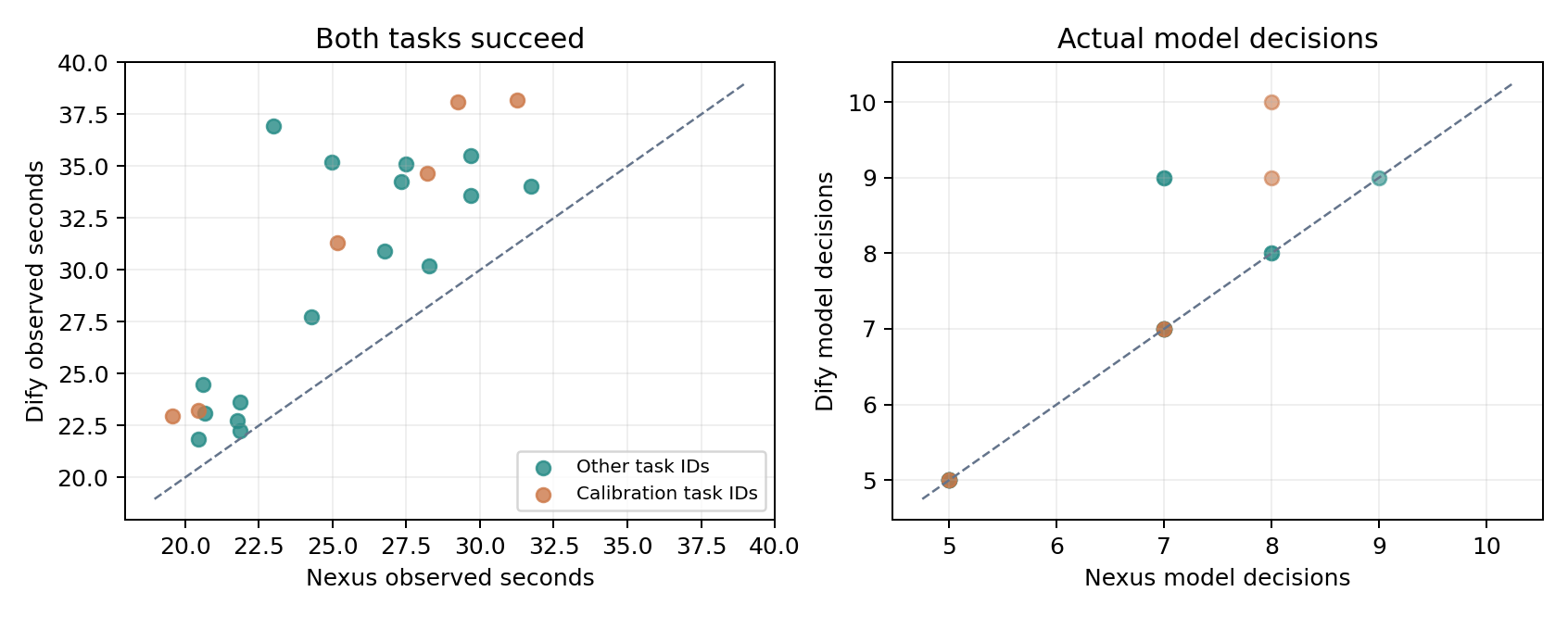}
\caption{Completion time and model decisions in the 22 jointly successful pairs. The diagonal denotes equality, and markers distinguish calibration tasks. Table~\ref{tab:eval-dify-outcomes} includes the two other pairs.}
\label{fig:dify-agent-paired}
\end{figure}

Nexus completes 24/24 executions and Dify 22/24. Among the 22 jointly successful pairs, the median Dify-minus-Nexus completion-time difference is 3.88 s and its mean is 4.80 s. The sixteen jointly successful non-calibration pairs have a median difference of 3.67 s. Table~\ref{tab:eval-dify-outcomes} includes all paired outcomes, while Figure~\ref{fig:dify-agent-paired} shows completion times conditional on joint success. The detailed all-attempt medians of 25.06 and 31.10 s include unsuccessful termination and use a different population.

Nexus and Dify make 159 and 168 model decisions, including nine and fifteen invalid actions. The observed time difference therefore combines orchestration, realized model trajectories, and upstream waiting. Of the two unsuccessful Dify executions, one stops at the shared Nexus confirmation tool for an undetermined transient reason; the other exhausts its model-action budget. These incident locations distinguish shared-service and model-control outcomes from a failure attributed to Dify's core. The nominal 540-second budget is cooperative, with different remaining-budget and cancellation handling. We consequently compare task outcomes and jointly successful completion times under this fixed policy; the experiment does not isolate orchestration overhead or match deadline enforcement. Section~\ref{sec:dify-agent-measured} gives the token-attribution and resource-measurement details.

\paragraph{Service latency and incremental memory.}
Dapr 1.18.4 and Nexus run on the same two-vCPU, 1-GiB OpenWrt VM and invoke the same byte-echo backends. Each provides discovery and mutual TLS through its own service interface. Entry APIs, name resolution, certificate issuance, and supporting processes differ. Plaintext direct HTTP is included only as a transport reference. Two order-balanced experiments contain 648 requests across the three paths.

\begin{table}[htbp]
\centering\small
\caption{Service-invocation results across both experiments. Latency is conditional on successful completion; RSS sums cover sampled product processes.}
\label{tab:eval-dapr}
\begin{tabular}{@{}lrrrr@{}}
\toprule
Configuration & Exact/scheduled & Success p50 (ms) & Success p95 (ms) & RSS sum (MiB) \\
\midrule
Dapr & 207/216 & 1.239 & 2.655 & 292.26 \\
Nexus & 207/216 & 99.169 & 148.770 & 9.73 \\
\bottomrule
\end{tabular}
\end{table}

Dapr has lower successful-call latency, with a 1.239-ms median compared with 99.169 ms for Nexus. Nexus has the smaller sampled product-process RSS sum, 9.73 MiB compared with 292.26 MiB. These RSS sums can include repeated shared mappings and measure sampled residency. Each product completes 207 of 216 requests. The nine Dapr failures occur during discovery/interface failure, while the nine Nexus failures follow a backend bind failure. Latency describes the successful requests from these separately identified incidents and includes Nexus's configured 100-ms peer tick together with the rest of its invocation path; the tick is not independently varied.

A separate grid tests incremental deployment under common aggregate limits: one CPU, disabled swap, memory ceilings of 16, 32, 64, 128, and 256 MiB, and a 180-second observation bound. Each arm includes the runtime, backend, measurement processes, and issuer where needed, with eighteen measured calls and two warmups. Two pairs per ceiling use opposite orders. Nexus completes at every tested ceiling. Dapr encounters out-of-memory termination at 16 and 32 MiB and completes at the three higher settings.

The smallest successful tested ceilings are thus 16 MiB for Nexus and 64 MiB for Dapr. Executable files were pre-staged in host memory, so some shared pages can be charged outside the experimental cgroups. The grid identifies feasible incremental runtimes under fixed resource limits; the earlier latency experiment characterizes unconstrained invocation. These are complementary operating conditions, with cold whole-device memory and equal-service-level latency--memory behavior left unmeasured (Section~\ref{sec:external-service-measured}).
\FloatBarrier

\subsection{Cross-Environment Execution and Live Authorization}
\label{sec:eval-environments}
\paragraph{A common interface with stable operation identities.}
The resource comparison holds the hosted handler, caller, permissions, device bindings, and recovery infrastructure fixed while replacing SDK helpers with explicit HTTP operations. Both adapters execute the same Computer reads, Android actions, confirmation, and file write/readback with matched payloads and polling settings. The HTTP arm retains the shared hosting and journal, isolating the application-side resource adapter.

All sixteen Runs in the balanced, eight-pair active-journal experiment produce the expected artifacts, 48 Computer commands, 32 Mobile commands, and sixteen confirmations. All 128 expected journal operations are recorded. Device request identities remain associated with their stable operation keys, connecting heterogeneous execution with recovery metadata. The SDK adapter contains fourteen authored lines compared with 99 for explicit HTTP under the same counting rule, excluding imports and shared infrastructure. This is a reduction in local protocol code rather than a measurement of programmer time.

Both adapters preserve the requested outputs and operation identities, while their whole-workflow median ordering changes with input size. The earlier inactive-journal study uses a different deployment context and imbalanced order, so its durations are analyzed separately. In the model-driven comparison, fresh inference contributes its own waiting and action time to the shared resource workflow (Sections~\ref{sec:sdk-active} and~\ref{sec:priority-real-composition}).

\paragraph{Composing actions and accounting for delegation.}
A five-action Computer--Android workflow completes all ten scheduled outputs, with marginal end-to-end medians of 11.366 and 11.462 s for 1-KiB and 64-KiB inputs. It demonstrates composition through one invocation; its completion time is descriptive because there is no alternative implementation arm (Section~\ref{sec:measured-composition}).

A separate paired experiment keeps the Android executor, Settings action, and output validator constant while adding hosted admission, worker scheduling, MCP, and run-scoped delegation. Both paths complete ten healthy workflows and ten workflows under an approximately ten-second outage. The healthy median added time is 10.503 s, with a 95\% interval of [9.534, 11.705] s; during interruption it is 0.797 s [0.271, 1.361] s. The network gate lasts 10.165--10.332 s and begins at submission, so delegated queueing can overlap disconnection. This overlap can account for the smaller interrupted difference; queueing and reconnection are not independently varied. Delegation supplies additional service semantics at a measurable experienced cost (Section~\ref{sec:matched-mobile}).

\paragraph{Permissions at the action boundary.}
Six pairs retain the same file task and Runtime interruption. Once an authorized read is queued, the treatment removes workspace grants and the control reaffirms them. Both restart the same Runtime, while the handler retains its originally captured full-scope context. All six controls finish. Every revoked-grant Run completes the already-queued read but stops before creating a write command, and no output artifact appears.

Because the Runtime interruption is shared, the contrast identifies live permission revalidation rather than an incidental connection failure. Authority governs subsequent dispatch without retroactively retracting already-authorized work (Section~\ref{sec:measured-authority}). Browser experiments further examine whether an action matches its observed snapshot, and terminal experiments examine foreground signals, resizing, and session retention. The corresponding device observations, session behavior, and experiment-specific outcomes appear in Sections~\ref{sec:attached-browser-measured} and~\ref{sec:posix-terminal-measured}.
\FloatBarrier

\subsection{Recovery across Execution and Result Boundaries}
\label{sec:eval-lifecycle}
An interruption can leave several kinds of state available: a committed operation result, an external effect whose completion has not been recorded, or a device receipt awaiting delivery. We vary the failure point to determine which state allows work to be reused and when recovery needs an additional decision or a cooperating resource.

\paragraph{Reusing committed work.}
Six blocks cross journal use with healthy or failed workers, producing 24 Runs. The handler executes six paced stages and appends actual records without resource-side deduplication. Failure terminates the Cloud worker supervisor and child after stage three is acknowledged; in the journal arm that acknowledgment follows commit. The Agent process, Cloud application, and database remain live. Replacement follows the configured 90-second lease and fifteen-second maintenance interval.

\begin{table}[htbp]
\centering\small
\caption{Worker-loss ablation after committed-operation acknowledgment. Each row contains six Runs. Duplicate appends and completion times are per-Run means; all Runs retain their original identity and single settlement.}
\label{tab:eval-worker}
\begin{tabular}{@{}llrrr@{}}
\toprule
Worker & Journal & Exact outputs & Duplicate appends & Completion (s) \\
\midrule
Healthy & Off & 6/6 & 0.0 & 7.260 \\
Healthy & On & 6/6 & 0.0 & 8.193 \\
Lost & Off & 6/6 & 6.0 & 108.414 \\
Lost & On & 6/6 & 0.0 & 108.618 \\
\bottomrule
\end{tabular}
\end{table}

All Runs complete with correct outputs. Under worker loss, the journal reduces duplicate appends from six per Run to zero (Table~\ref{tab:eval-worker}). Its mean paired normal-path overhead is 0.933 s, with an exploratory six-block bootstrap interval of [0.305, 1.874] s. Fault-to-terminal time instead ranges from 101.096 to 103.223 s, reflecting lease detection and recovery. Every successful Run retains one invocation, usage record, and settlement despite replacement of its execution attempt. The experiment isolates reuse after committed-operation acknowledgment while the Agent remains alive (Section~\ref{sec:worker-recovery}).

\paragraph{External effects before journal commit.}
A second experiment terminates the Agent after an external service durably applies stage three but before its result reaches the journal. Four configurations share the six-stage handler, service, and crash boundary: no journal, a guarded journal followed by a caller retry, a reconciliation declaration alone, and that declaration with transactional resource-side deduplication. Each configuration has six Runs.

\begin{table}[htbp]
\centering\small
\caption{Agent failure after an external effect but before its completion record. Duplicate counts are totals over six Runs per arm. The guarded configuration includes an explicit caller retry.}
\label{tab:eval-effects}
\begin{tabularx}{\linewidth}{@{}>{\raggedright\arraybackslash}Xrrr@{}}
\toprule
Recovery configuration & Final outputs & Caller retries & Duplicate effects \\
\midrule
No journal & 6/6 & 0 & 18 \\
Guarded journal + caller retry & 6/6 & 6 & 6 \\
Reconciliation declaration only & 6/6 & 0 & 6 \\
Journal + resource deduplication & 6/6 & 0 & 0 \\
\bottomrule
\end{tabularx}
\end{table}

All configurations eventually produce the final outputs, but their repeated effects differ (Table~\ref{tab:eval-effects}). A recorded journal entry supports reuse; an external effect with no corresponding entry leaves its completion uncertain. The guarded configuration pauses for at least thirty seconds before an explicit caller retry, whereas the other configurations continue autonomously. Adding resource-side deduplication to automatic reconciliation reduces duplicate effects from six to zero across six Runs. This contrast identifies the resource's contribution alongside journal reuse, while the guarded configuration demonstrates caller-assisted handling of uncertainty (Section~\ref{sec:ambiguous-effect}).

\paragraph{Reconciling a published device result.}
Eight Android pairs compare clients with and without a durable receipt outbox against the same idempotent Cloud receipt server. Both use the same debug activity and an independent click journal that does not suppress repeats. Healthy transport, result loss, acknowledgment loss, and result loss followed by process exit each have two opposite-order pairs. The fault gate is twelve seconds, and process replacement follows receipt publication.

\begin{table}[htbp]
\centering\small
\caption{Original Android command completion and subsequent-command delay. Each condition contains two pairs. Subsequent-observation times are marginal medians in seconds.}
\label{tab:eval-mobile-result}
\begin{tabular}{@{}lrrrr@{}}
\toprule
 & \multicolumn{2}{c}{Original completed} & \multicolumn{2}{c}{Next observation (s)} \\
Condition & No outbox & Outbox & No outbox & Outbox \\
\midrule
Healthy & 2/2 & 2/2 & 0.82 & 1.19 \\
Result loss & 0/2 & 2/2 & 1.72 & 6.86 \\
Acknowledgment loss & 2/2 & 2/2 & 0.66 & 7.43 \\
Result loss + process exit & 0/2 & 2/2 & 4.27 & 5.74 \\
\bottomrule
\end{tabular}
\end{table}

The outbox client completes all eight original commands. The no-outbox client leaves the original command unresolved in both result-loss conditions, although every arm produces exactly one click. Receipt replay therefore reconciles an existing action rather than executing a replacement. Waiting for acknowledgment also delays subsequent commands (Table~\ref{tab:eval-mobile-result}). In these tests, faults and process replacement follow durable receipt publication with pairing authority unchanged. A crash between the action and receipt persistence is not tested here (Section~\ref{sec:mobile-result-recovery}).

\paragraph{Interaction and settlement continuity.}
Four queued-turn pairs use identical business turns but intentionally differ in the time of second-turn submission. During caller absence, all four queued results become ready, while none of the manual-continuation results is ready before the caller returns; all eight workflows ultimately complete. Queueing therefore removes a later client-submission dependency rather than accelerating identical arrival schedules (Section~\ref{sec:queued-turn}). Ten Web/API price-matched pairs separately produce fourteen successful and six intentionally failed executions. All twenty yield the expected reservation and settlement, using synthetic amounts to test accounting semantics (Section~\ref{sec:web-billing-measured}).
\FloatBarrier

\subsection{Mechanism Ablation and Sensitivity}
\label{sec:eval-ablation}

We examine three sources of overhead: finding a service endpoint,
executing the request bridge, and redelivering output after a
disconnection. Each experiment varies the relevant mechanism or
execution configuration while preserving the requested operation
and expected result.

\paragraph{Faster route lookup without changing the decision.}
A route index should reduce the work needed to find an eligible
endpoint while preserving which endpoint is selected. We compare
indexed lookup with a full scan of all routes, using identical route
entries, eligibility rules, ranking scores, and query order. For each
query, we check that both implementations either return no route or
select the same route with the same score. These checks run outside
the timing intervals. Eleven workload settings on each of two platforms
use ten paired trials of 4,000 queries per setting, with no observed
selection mismatch.

\begin{figure}[htbp]
\centering
\includegraphics[width=.96\linewidth]{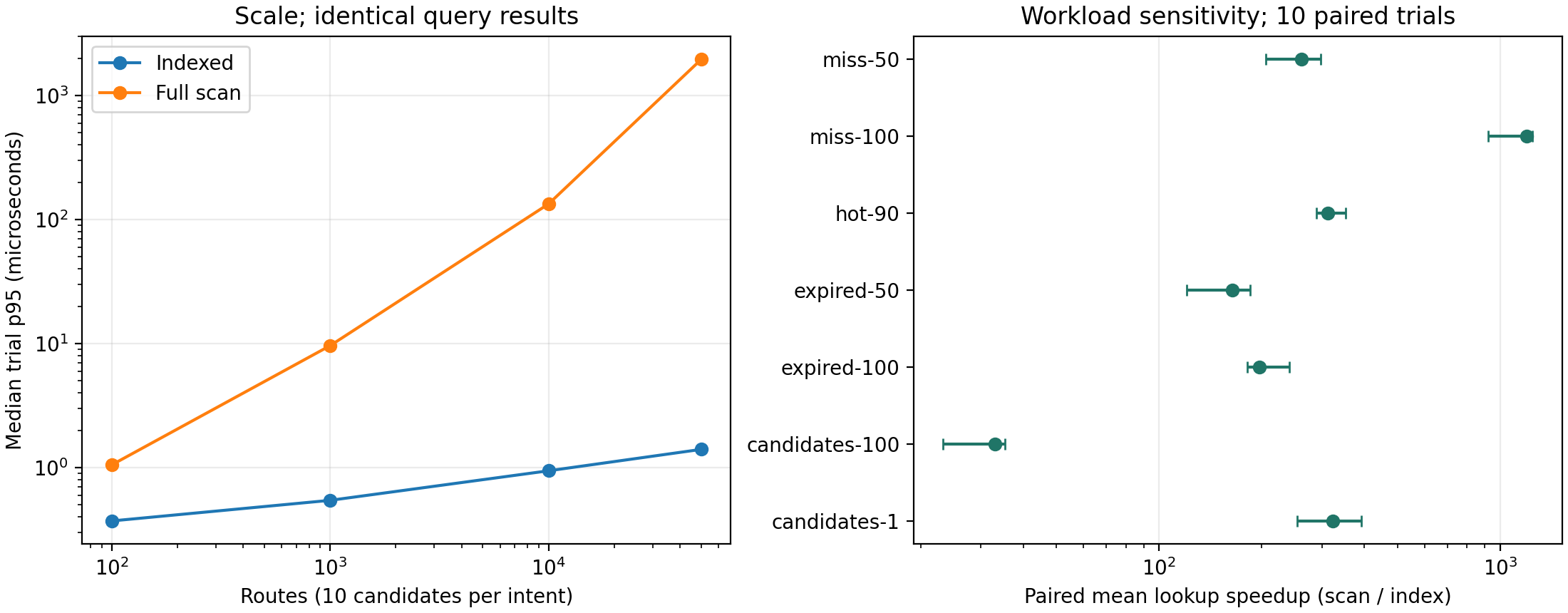}
\caption{Route-index ablation on OpenWrt. Left: median across trials
of the 95th-percentile lookup latency as the route count increases.
Right: median paired speedup based on mean lookup times, with 95\%
trial-bootstrap intervals. Workloads vary the number of candidates
per requested capability, missing queries, expired routes, and query
concentration. Both implementations use the same selection policy.}
\label{fig:route-matched}
\end{figure}

For each paired trial, speedup is the full-scan mean lookup time
divided by the indexed mean lookup time. At 10,000 routes on OpenWrt,
the median speedup is 158.4$\times$, with a 95\% interval of
[117.9, 187.0], using 262,144 additional bytes (256~KiB) of index
storage. The median trial-level 95th-percentile latencies are
0.942~$\mu$s for indexed lookup and 133.618~$\mu$s for scanning.

The index saves time by avoiding unrelated routes, but matching
candidates still require eligibility checks and scoring. Its advantage
therefore decreases as more routes match the same requested capability.
Queries with no eligible route retain the same no-result behavior,
including missing and expired-route cases. The experiment identifies
a storage-for-search-time tradeoff: the index accelerates the existing
decision rather than replacing it with a different policy
(Section~\ref{sec:matched-route}).

\paragraph{Execution overhead under the same software and kernel.}
The earlier deployment comparisons change both execution location
and the underlying system. To examine execution configuration
separately, we run the request bridge from the same filesystem image
either in a native chroot, which sets the process's filesystem root,
or in Docker. Both configurations share the same kernel boot, Python
interpreter, libraries, bridge code, workload driver, CPU affinity,
and resource counters. Three payload sizes and concurrency levels
of one and eight form six settings, each with ten order-balanced pairs.

All 240,000 measured requests return the correct bytes and digests.
Across the six settings, the median paired increase in latency under
Docker ranges from 0.026 to 1.187~ms; four exploratory intervals exclude
zero. For peak resident memory (RSS), every interval for the difference
includes zero, leaving the memory difference inconclusive in these
measurements. Thus, the latency effect remains observable in several
settings even after matching the software and kernel. Resource
measurements cover the bridge processes; the Docker daemon, backend,
workload driver, and VM are accounted for separately from this
comparison (Section~\ref{sec:native-isolation-measured}).

\paragraph{Resuming output delivery without repeating execution.}
After a connection breaks, the caller may need only the remaining
output, not another execution of the task. We compare two recovery
policies: full-history replay, in which the caller discards events
already processed, and cursor-based continuation, which requests only
events after the last processed event. Both reconnect to the same
output-producing task and must reconstruct the same result. Eight
settings vary event size, interruption point, and direct or bridge
access, with ten balanced pairs per setting. All 160 tasks reconstruct
their 32-event outputs with one producer execution.

For 4,096-byte events interrupted after event sixteen, cursor-based
continuation reduces the combined initial and reconnect server-sent
event (SSE) response bodies from 203,701 to 135,815 bytes, a 33.3\%
reduction. Both clients still wait for the same producer to generate
events at its configured pace, so the reduced retransmission does not
consistently shorten total completion time.

The history-loss and process-loss tests reveal why retained state
matters. A request for the remaining events is rejected when the original
task is unavailable, leaving its output incomplete. Restarting from
event zero instead creates a new producer and reconstructs the output,
but repeats the associated effects. Retaining the original task and
its output therefore separates two operations that would otherwise
be conflated: continuing delivery and repeating execution
(Section~\ref{sec:measured-stream}).
\FloatBarrier

\subsection{Workload Dependence and Deployment Conditions}
\label{sec:eval-validity}
The experiments use controlled workloads on one host with virtualized OpenWrt, an Android emulator, and fresh model requests. Balanced order reduces temporal drift, while host activity and upstream waiting contribute variation. Task families and resource grids retain separate estimates and the outcome populations defined in Section~\ref{sec:eval-setup}. Shared components are fixed within each pair; intentional configuration changes, software revisions, and process-accounting changes distinguish experiment groups. These configurations are evaluated separately, without an end-to-end experiment combining all mechanisms in one runtime release.

Topology determines how the results transfer to other deployments. Logical peers share a physical failure domain, and the IPv6/Relay experiment adds a laboratory bridge only to Relay while both paths traverse Cloud (Sections~\ref{sec:priority-transport}, \ref{sec:relay-followup}, and~\ref{sec:transport-concurrency-measured}). The capacity grid fixes arrival rate and deadline while adding edge resources. It measures central execution relief, with admission, authorization, result delivery, and settlement still using Cloud. Independent-node scaling, whole-platform resource use, and complete Cloud loss require measurements in those deployment conditions.
\FloatBarrier

\FloatBarrier

\section{Discussion and Conclusion}
\label{sec:discussion}
\subsection{Locality, Authority, and Recovery as Deployment Choices}
The results connect three deployment choices. First, a common invocation interface allows computation to move toward its input while retaining the same business operation. For the large-input hashing workload, the tested Cloud CPU ceiling meeting the fixed deadline changes from 2 to 0.25 vCPU when additional edge resources execute the computation. Small inputs already meet that deadline at 0.25 vCPU. Complete Agent workflows add model waiting and different local coordination costs, so their faster placement varies by workload. The useful choice is therefore where to perform each operation given its data, computation, and service dependencies.

Second, a shared environment abstraction separates the reasoning host from the device and authority used for an action. The resource experiments preserve output and operation identities through both SDK and explicit HTTP adapters, while the SDK reduces application-side protocol code. Live revocation prevents a subsequent write without retracting an already-authorized read. The additional admission, delegation, and result handling have measurable time costs; their purpose is to preserve the requested service and access semantics across execution locations.

Third, the state available after interruption determines how an invocation can continue. A committed journal result can be reused after worker loss. An external effect lacking a completion record instead calls for resource-supported reconciliation or a caller decision about retrying. A durably recorded device receipt can be retransmitted under the original command identity, with acknowledgment waiting delaying subsequent work. These experiments relate each recovery mechanism to the state it preserves and the execution decision it enables.

\subsection{Implications for Agent Service Design}
The external comparisons place these choices in concrete service settings. Over the same model and device tools, Nexus's observed completion time is lower by a median 3.88 s among the 22 jointly successful Dify pairs, with independently realized model trajectories included in each duration. For byte-echo invocation, Dapr provides lower successful-call latency. In the separate constrained-memory experiment, Nexus operates under a smaller tested incremental-runtime ceiling. Selecting between such configurations requires both the required service behavior and the resource constraints under which it will run.

More generally, execution location, action authority, and recovery state can vary independently while referring to the same logical invocation. This separation permits device-specific actions and edge-hosted computation to participate in a persistent service. It also leaves explicit dependencies: Cloud admission and settlement remain centrally managed, and external-effect uniqueness requires support at the resource that performs the effect. Larger physical deployments will need to account for these dependencies alongside node capacity, network conditions, and device behavior.

\subsection{Conclusion}
This work reveals a gap between task completion and reliable agent
operation: correct final outputs can coexist with repeated external
effects. Extending agents beyond the cloud therefore requires more than
relocating computation; it requires preserving the relationship between
authorization, execution, recovery, and settlement. Nexus addresses this
problem through OpenWrt-native routing, invocation-scoped delegation,
and persistent lifecycle management. Experiments demonstrate lower Cloud
capacity requirements for a data-intensive workload, live enforcement
of device permissions, and reuse of committed work after worker loss.
The recovery results further distinguish result persistence from
duplicate-effect suppression, which requires cooperating resources in
the evaluated design. These findings support a broader design principle:
agent execution may span heterogeneous environments, but its actions,
outcomes, and charges must remain associated with the same authorized
invocation. Reliable agent infrastructure must preserve this continuity,
not merely deliver a successful final response.
\clearpage

\section*{Authors and Contributors} \textbf{Cary Chang} served as the contributor to this work, leading the conceptualization,  system design, and manuscript preparation. \textbf{Jialin Zhou} contributed to project support. We also thank members of the broader research and developer community for discussions, feedback, and ideas that helped refine the research questions, system design, and evaluation presented in this report.

\section*{Software and Documentation}

The Nexus implementation and accompanying documentation are maintained
by the Nexilume-AI organization at
\url{https://github.com/Nexilume-AI}. The repositories follow the
architectural separation of Cloud service management, edge routing,
agent development, and device interaction described in this paper.

The \href{https://github.com/Nexilume-AI/nexus-cloud}
{\texttt{nexus-cloud}} repository contains the community
implementation of the Cloud service-management component.
The \href{https://github.com/Nexilume-AI/nexus-openwrt}
{\texttt{nexus-openwrt}} repository provides the OpenWrt runtime
for service registration, inter-router route exchange, and request
forwarding. The
\href{https://github.com/Nexilume-AI/nexus-agent-sdk-python}
{\texttt{nexus-agent-sdk-python}} repository provides the Python
software development kit for implementing agents, accessing models
and authorized environments, and reporting execution progress
and results. The
\href{https://github.com/Nexilume-AI/nexus-mobile}
{\texttt{nexus-mobile}} repository contains the Android client
for executing authorized device actions and returning observations
and command results. Supporting system documentation is maintained
in \href{https://github.com/Nexilume-AI/nexus-docs}
{\texttt{nexus-docs}}.
\begingroup\small
\setlength{\bibsep}{2pt}
\bibliographystyle{unsrtnat}
\bibliography{references}

@inproceedings{yao2023react,
  title={ReAct: Synergizing Reasoning and Acting in Language Models},
  author={Yao, Shunyu and Zhao, Jeffrey and Yu, Dian and Du, Nan and Shafran, Izhak and Narasimhan, Karthik and Cao, Yuan},
  booktitle={International Conference on Learning Representations}, year={2023}}

@inproceedings{schick2023toolformer,
  title={Toolformer: Language Models Can Teach Themselves to Use Tools},
  author={Schick, Timo and Dwivedi-Yu, Jane and Dessi, Roberto and Raileanu, Roberta and Lomeli, Maria and Hambro, Eric and Zettlemoyer, Luke and Cancedda, Nicola and Scialom, Thomas},
  booktitle={Advances in Neural Information Processing Systems}, volume={36}, year={2023}}

@article{wu2023autogen,
  title={AutoGen: Enabling Next-Gen LLM Applications via Multi-Agent Conversation},
  author={Wu, Qingyun and Bansal, Gagan and Zhang, Jieyu and Wu, Yiran and Li, Beibin and Zhu, Erkang and others},
  journal={arXiv preprint arXiv:2308.08155}, year={2023}, doi={10.48550/arXiv.2308.08155}}

@article{wang2024survey,
  title={A Survey on Large Language Model Based Autonomous Agents},
  author={Wang, Lei and Ma, Chen and Feng, Xueyang and Zhang, Zeyu and Yang, Hao and others},
  journal={Frontiers of Computer Science}, volume={18}, number={6}, pages={186345}, year={2024}, doi={10.1007/s11704-024-40231-1}}

@inproceedings{moritz2018ray,
  title={Ray: A Distributed Framework for Emerging AI Applications},
  author={Moritz, Philipp and Nishihara, Robert and Wang, Stephanie and Tumanov, Alexey and Liaw, Richard and others},
  booktitle={13th USENIX Symposium on Operating Systems Design and Implementation}, pages={561--577}, year={2018}}

@misc{daprWorkflow,
  author       = {{Dapr Authors}},
  title        = {{Dapr Workflow}: Features and Concepts},
  year         = {2026},
  howpublished = {Official documentation},
  url          = {https://docs.dapr.io/developing-applications/building-blocks/workflow/workflow-features-concepts/},
  note         = {Accessed September 27, 2026}
}

@article{trossen2019service,
  author  = {Trossen, Dirk and Robitzsch, Sebastian and Hergenhan, Scott and
             Riihijarvi, Janne and Reed, Martin and Al-Naday, Mays},
  title   = {Service-based Routing at the Edge},
  journal = {arXiv preprint arXiv:1907.01293},
  year    = {2019},
  doi     = {10.48550/arXiv.1907.01293}
}

@techreport{rose2020zerotrust,
  title={Zero Trust Architecture}, author={Rose, Scott and Borchert, Oliver and Mitchell, Stu and Connelly, Sean},
  institution={National Institute of Standards and Technology}, number={NIST SP 800-207}, year={2020}, doi={10.6028/NIST.SP.800-207}}

@techreport{rfc8446,
  title={The Transport Layer Security Protocol Version 1.3}, author={Rescorla, Eric},
  institution={Internet Engineering Task Force}, number={RFC 8446}, year={2018}, doi={10.17487/RFC8446}}

@techreport{rfc8707,
  author      = {Campbell, Brian and Bradley, John and Tschofenig, Hannes},
  title       = {Resource Indicators for {OAuth} 2.0},
  institution = {Internet Engineering Task Force},
  number      = {RFC 8707},
  year        = {2020},
  doi         = {10.17487/RFC8707}
}

@misc{mcp2025architecture,
  title={Model Context Protocol: Architecture Overview}, author={{Model Context Protocol Contributors}},
  year={2025}, howpublished={Specification, revision 2025-06-18}, url={https://modelcontextprotocol.io/specification/2025-06-18/architecture}}

@misc{a2a2025spec,
  title={Agent2Agent Protocol Specification}, author={{Linux Foundation}}, year={2025},
  howpublished={Open protocol specification}, url={https://a2a-protocol.org/latest/specification/}}

@misc{dify2026,
  title={Dify: Open-Source LLM Application Development Platform}, author={{LangGenius}}, year={2026},
  note={Software release 1.17.1}, url={https://github.com/langgenius/dify}}

@misc{portkey2026,
  title={Portkey AI Gateway}, author={{Portkey AI}}, year={2026},
  note={Software release 1.15.2}, url={https://github.com/Portkey-AI/gateway}}

@misc{dapr2026,
  title={Dapr: Distributed Application Runtime}, author={{Dapr Authors}}, year={2026},
  note={Software release 1.18.4}, url={https://github.com/dapr/dapr}}

@misc{envoy2026,
  title={Envoy Proxy}, author={{Envoy Project Authors}}, year={2026},
  note={Software release 1.39.1}, url={https://github.com/envoyproxy/envoy}}

@misc{istio2026,
  title={Istio Service Mesh}, author={{Istio Authors}}, year={2026},
  note={Software release 1.31.0}, url={https://github.com/istio/istio}}

@misc{rayserve2026,
  title={Ray Serve Architecture and Performance Guidance}, author={{Ray Project}}, year={2026},
  note={Software release 2.58.0}, url={https://docs.ray.io/en/latest/serve/architecture.html}}

@misc{aios2024,
  title={AIOS: LLM Agent Operating System},
  author={Kai Mei and Xi Zhu and Wujiang Xu and Wenyue Hua and Mingyu Jin and Zelong Li and Shuyuan Xu and Ruosong Ye and Yingqiang Ge and Yongfeng Zhang},
  year={2024},
  howpublished={arXiv:2403.16971},
  url={https://arxiv.org/abs/2403.16971}
}

@misc{agentscope2025,
  title={AgentScope 1.0: A Developer-Centric Framework for Building Agentic Applications},
  author={Dawei Gao and Zitao Li and Yuexiang Xie and Weirui Kuang and Liuyi Yao and Bingchen Qian and Zhijian Ma and Yue Cui and Haohao Luo and Shen Li and Lu Yi and Yi Yu and Shiqi He and Zhiling Luo and Wenmeng Zhou and Zhicheng Zhang and Xuguang He and Ziqian Chen and Weikai Liao and Farruh Isakulovich Kushnazarov and Yaliang Li and Bolin Ding and Jingren Zhou},
  year={2025},
  howpublished={arXiv:2508.16279},
  url={https://arxiv.org/abs/2508.16279}
}

@misc{langgraphPersistence,
  title={LangGraph documentation: Persistence},
  author={{LangChain}},
  year={2026},
  howpublished={Official documentation},
  url={https://docs.langchain.com/oss/python/langgraph/persistence}
}

@misc{mcp2025transport,
  title={Model Context Protocol: Transports}, author={{Model Context Protocol Contributors}},
  year={2025}, howpublished={Specification, revision 2025-06-18},
  url={https://modelcontextprotocol.io/specification/2025-06-18/basic/transports}}

@misc{mcp2025lifecycle,
  title={Model Context Protocol: Lifecycle}, author={{Model Context Protocol Contributors}},
  year={2025}, howpublished={Specification, revision 2025-06-18},
  url={https://modelcontextprotocol.io/specification/2025-06-18/basic/lifecycle}}

@misc{agui2026events,
  title={Agent User Interaction Protocol: Events}, author={{AG-UI Contributors}},
  year={2026}, howpublished={Official documentation, accessed September 23, 2026},
  url={https://docs.ag-ui.com/concepts/events}}
\endgroup
\appendix
\section{Deployment, Capacity, and Route Availability}
\label{app:serving-evidence}
The following experiments examine where computation occurs, how serving capacity changes with the workload, and how route state controls the availability of distributed services.

\subsection{Cloud capacity under a fixed completion deadline}
\label{sec:rightsizing-measured}
Data-local execution is useful for capacity planning when it allows the same offered workload to meet its deadline with a smaller Cloud CPU ceiling. We evaluate this relationship using two configurations that retain the same edge data source and Cloud receipt service. A byte-identical capability binary computes 32 SHA-256 rounds either beside the source or after fetching it into a Cloud container. The endpoint kernels and bridge builds differ, so the experiment compares these deployment configurations; edge capacity is available in addition to the constrained Cloud resources.

\paragraph{Workload and resource controls.}
The experiment crosses 1-KiB and 16-MiB inputs with Cloud CPU ceilings of 0.25, 0.5, 1, and 2 vCPU and a fixed 512-MiB memory limit. Each cell contains four paired batches with balanced execution order. After two warm-up requests and a three-second settling interval, each batch offers 20 tasks at one-second intervals through at most eight caller workers. Response time starts at planned arrival and therefore includes any wait for a caller worker. Tasks receive no application retry and remain in the denominator if they miss their deadline. A batch passes when all 20 tasks return independently reconstructed digests and accepted receipts within two seconds of planned arrival. Applied CPU and memory limits are checked before and after each batch; resource sampling excludes warm-up.

\begin{table}[htbp]\centering\small
\caption{Deadline-constrained placement. Every batch schedules 20 tasks; each
cell has four batches per mode. Passing requires all 20 exact, receipted tasks
within two seconds of scheduled arrival. Edge / Cloud refer to processing
location, while the CPU ceiling always applies to the Cloud workload.}
\label{tab:rightsizing}
\begin{tabular}{llrrrr}\toprule
Source & Cloud ceiling & Edge batches & Cloud batches & Edge timely & Cloud timely\\\midrule
1 KiB & 0.25 & 4/4 & 4/4 & 80/80 & 80/80 \\
1 KiB & 0.5 & 4/4 & 4/4 & 80/80 & 80/80 \\
1 KiB & 1 & 4/4 & 4/4 & 80/80 & 80/80 \\
1 KiB & 2 & 4/4 & 4/4 & 80/80 & 80/80 \\
16 MiB & 0.25 & 4/4 & 0/4 & 80/80 & 0/80 \\
16 MiB & 0.5 & 4/4 & 1/4 & 80/80 & 51/80 \\
16 MiB & 1 & 4/4 & 3/4 & 80/80 & 79/80 \\
16 MiB & 2 & 4/4 & 4/4 & 80/80 & 80/80 \\
\bottomrule\end{tabular}\end{table}
\begin{figure}[htbp]\centering
\includegraphics[width=\linewidth]{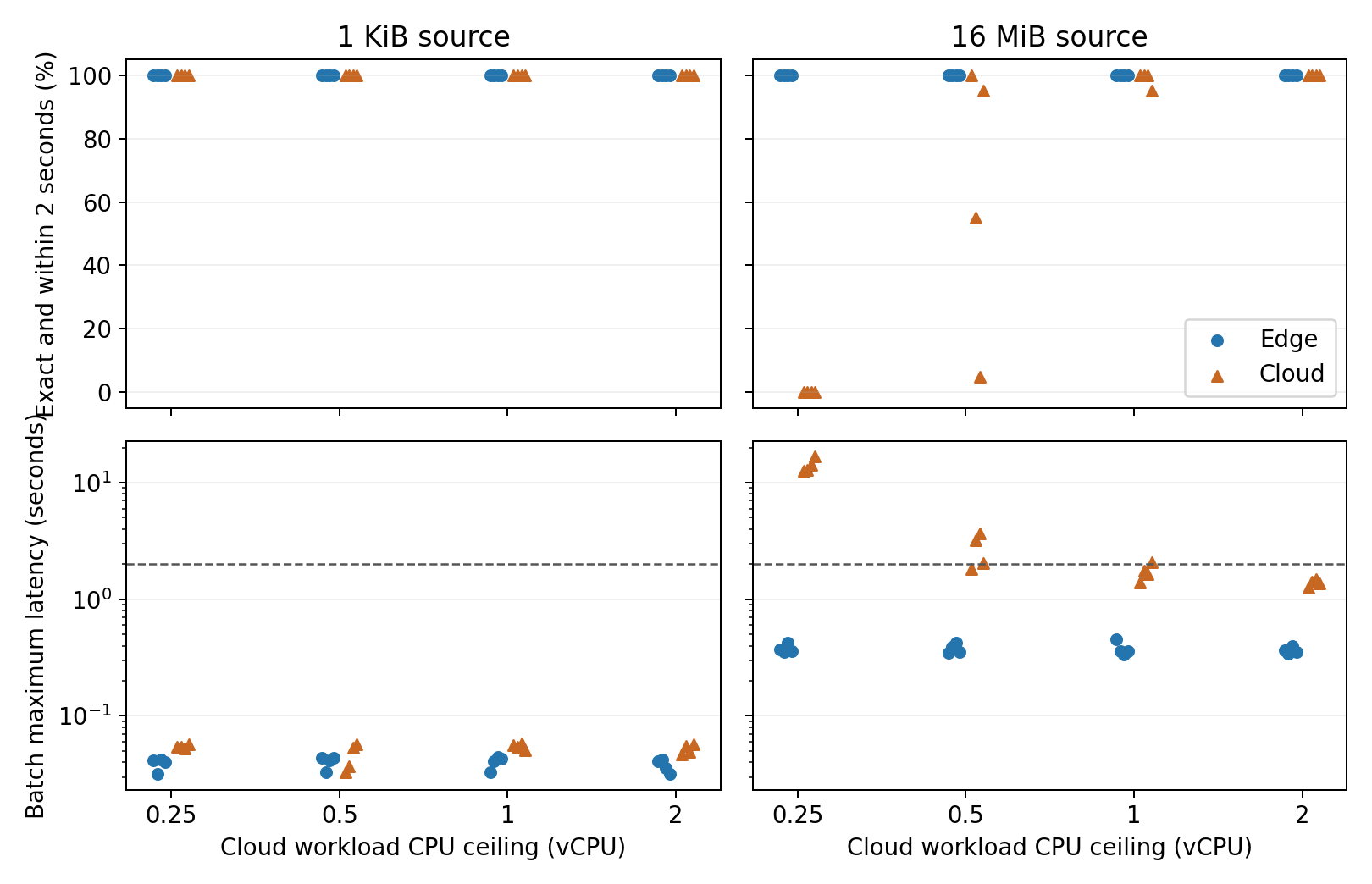}
\caption{Per-batch timely completion fraction (top) and maximum arrival-to-completion latency (bottom, logarithmic scale). The dashed line marks the two-second deadline; horizontal jitter separates repeated observations.}
\end{figure}

\paragraph{Capacity and workload dependence.}
All 1,280 tasks return exact outputs, and 1,170 meet the deadline. For 1-KiB inputs, both configurations pass at the lowest tested Cloud ceiling, 0.25 vCPU. For 16-MiB inputs, all edge-processing batches pass at 0.25 vCPU, whereas Cloud processing first passes all four batches at 2 vCPU. At the 0.25-vCPU large-input setting, median Cloud workload CPU time is 7.266 s per batch for Cloud processing and 0.557 s for edge processing; all four Cloud-processing batches experience CPU throttling. Together, these observations connect the reduced Cloud workload to deadline feasibility rather than only to lower utilization.

The counters include the dedicated endpoint, bridge, and observer; shared services, the Docker daemon, and VM overhead are excluded. The four-point grid identifies the lowest tested ceiling satisfying every batch at one arrival rate on the shared host. Finer resource thresholds and provisioned allocations are separate quantities. Section~\ref{sec:cost-evidence-bridge} relates the measured feasible ceilings to expenditure under explicit allocation assumptions.

\subsection{Repeatability of the large-input capacity result}
\label{sec:rightsizing-replication}
The large-input result in Section~\ref{sec:rightsizing-measured} motivates a follow-up over the complete 16-MiB stratum. Selected after inspection of the initial results, this repeat fixes all four CPU ceilings before execution, including the previously successful 2-vCPU setting. Its purpose is to test repeatability of the observed workload regime; initial and follow-up batches retain their separate identities.

Four balanced pairs per ceiling yield 32 batches and 640 scheduled tasks. The capability binary, endpoint bridge builds, 32 digest rounds, 512-MiB Cloud memory limit, one-second arrivals, eight caller workers, warm-up, and two-second deadline are unchanged. The input and result validators are identical to the initial study. Compilation and packaging occur outside the measurement window; other shared-host activity remains uncontrolled.

\begin{table}[htbp]\centering\small
\caption{Separate initial and repeat outcomes for the same 16 MiB stratum.
Each campaign has four batches and 80 tasks per mode/ceiling. A passing batch
requires all 20 exact outputs and receipts within two seconds.}
\label{tab:rightsizing-replication}
\begin{tabular}{llrrrr}\toprule
Cloud & Processing & \multicolumn{2}{c}{Passing batches} & \multicolumn{2}{c}{Timely tasks}\\
ceiling & location & Initial & Repeat & Initial & Repeat\\\midrule
0.25 & Edge & 4/4 & 4/4 & 80/80 & 80/80 \\
0.25 & Cloud & 0/4 & 0/4 & 0/80 & 0/80 \\
0.5 & Edge & 4/4 & 4/4 & 80/80 & 80/80 \\
0.5 & Cloud & 1/4 & 3/4 & 51/80 & 77/80 \\
1 & Edge & 4/4 & 4/4 & 80/80 & 80/80 \\
1 & Cloud & 3/4 & 3/4 & 79/80 & 78/80 \\
2 & Edge & 4/4 & 4/4 & 80/80 & 80/80 \\
2 & Cloud & 4/4 & 4/4 & 80/80 & 80/80 \\
\bottomrule\end{tabular}\end{table}

All 640 outputs are exact, and 555 arrive within the deadline. Both experiments identify 0.25 vCPU for edge processing and 2 vCPU for Cloud processing as the smallest tested ceilings passing every large-input batch. At intermediate Cloud-processing ceilings, timely-task counts vary: the 0.5-vCPU setting improves from 51/80 to 77/80, while the 1-vCPU setting changes from 79/80 to 78/80. Thus the endpoint capacity result repeats, but behavior near the deadline remains sensitive to execution conditions. Table~\ref{tab:rightsizing-replication} and Figure~\ref{fig:eval-capacity} show the initial and repeated observations separately.

\subsection{Computation and data movement across execution locations}
\label{sec:measured-placement}
To explain how deployment changes Cloud demand, we vary input size and computation intensity independently. Source data reside on the OpenWrt VM. Both configurations use the same static Go capability, request schema, deterministic digest, and summary receipt: edge execution reads the local file, whereas Cloud execution fetches that file before computing. Each endpoint receives requests through its fixed LAN bridge and a single eligible leased route. The Cloud container has two CPUs and 512 MiB of memory; the OpenWrt VM has two vCPUs and approximately 1 GiB. Different kernels and bridge builds make this a comparison of complete configurations.

The schedule crosses 1 KiB, 1 MiB, and 16 MiB with one or 32 digest rounds. Each of six cells contains ten order-balanced pairs of 20-task serial batches, preceded by two excluded warm-up requests per endpoint. Input bytes satisfy $b_i=(31i+7)\bmod 251$, and each round computes $h_j=\mathrm{SHA256}(h_{j-1}\Vert b)$ from 32 initial zero bytes. Expected digests are reconstructed independently. There are no application retries.

\begin{table}[htbp]
\centering\small
\caption{Placement batch medians. CPU is measured at each endpoint for 20 attempts; latency is the median of batch-level median attempt durations, including failures. C/E denotes Cloud/edge execution location. Ten paired batches per row.}
\label{tab:placement}
\begin{tabular}{@{}lrrrrrr@{}}
\toprule
Input / rounds & \multicolumn{2}{c}{Cloud CPU (s)} & \multicolumn{2}{c}{Edge CPU (s)} & \multicolumn{2}{c}{Latency (ms)} \\
 & C & E & C & E & C & E \\
\midrule
1 KiB / 1 & 0.080 & 0.040 & 0.030 & 0.030 & 16.4 & 26.4 \\
1 KiB / 32 & 0.073 & 0.041 & 0.026 & 0.031 & 13.1 & 25.3 \\
1 MiB / 1 & 0.483 & 0.039 & 0.086 & 0.054 & 60.3 & 24.4 \\
1 MiB / 32 & 0.844 & 0.056 & 0.091 & 0.437 & 77.9 & 43.1 \\
16 MiB / 1 & 5.806 & 0.052 & 0.851 & 0.323 & 659.5 & 43.1 \\
16 MiB / 32 & 11.314 & 0.222 & 0.925 & 6.403 & 984.2 & 334.8 \\
\bottomrule
\end{tabular}
\end{table}
\begin{figure}[htbp]
\centering
\includegraphics[width=\linewidth]{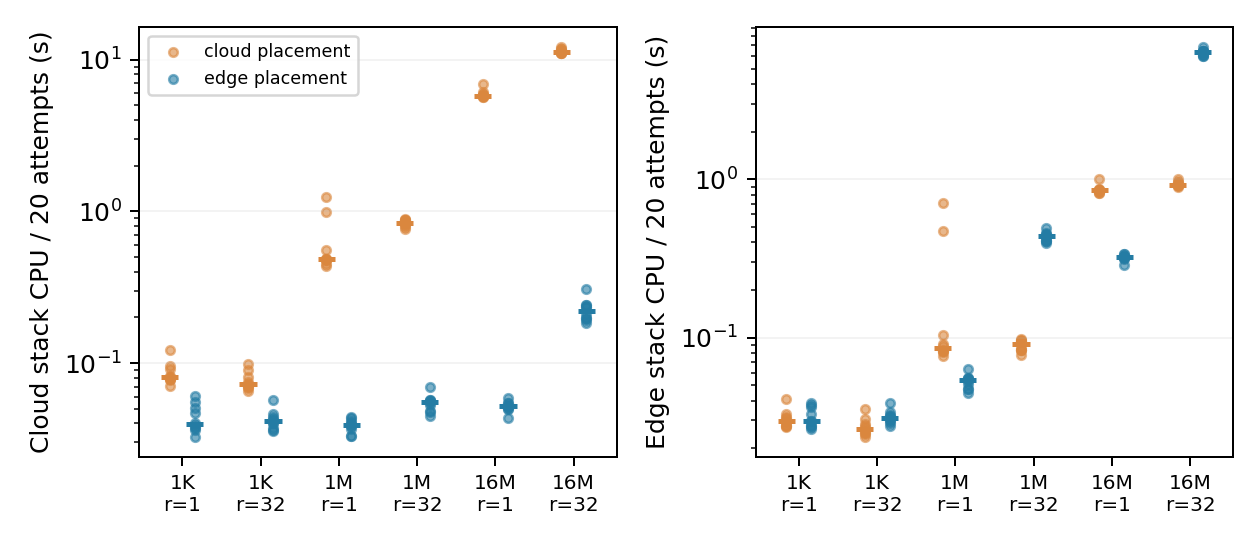}
\caption{Endpoint CPU under Cloud and edge placement. Each point represents twenty attempted tasks, including the batch containing a failed task; bars mark medians. Both axes are logarithmic. K and M denote KiB and MiB.}
\label{fig:placement}
\end{figure}

\paragraph{Work displacement.}
Of 2,400 attempts, 2,399 produce exact results: 1,200/1,200 at the edge and 1,199/1,200 in Cloud. One 1-MiB, one-round Cloud fetch terminates after 30.001 s at the configured fetch deadline; its underlying cause is unresolved. Its duration and resource use are included. For 16 MiB and 32 rounds, the paired median Cloud CPU change is $-11.102$ s per batch (95\% interval $[-11.368,-10.911]$), accompanied by a 5.470-s increase in edge CPU ($[5.134,5.565]$). For 1-KiB inputs, edge execution instead increases paired median attempt time by 12.025 and 11.135 ms for one and 32 rounds. Computation and data locality therefore interact with fixed dispatch costs.

\paragraph{Transfer and resource accounting.}
A complete Cloud-processing batch fetches 20 KiB, 20 MiB, or 320 MiB of source data. Edge-processing batches transfer no raw source bytes to Cloud during the measured interval; both configurations send 2,980--3,080 bytes of summary receipts. These are application-body counts. Both endpoints receive the source during excluded warm-up. The measured reduction is therefore in repeated transfers during steady execution; it leaves the initial data distribution unchanged.

Dedicated cgroups count the endpoint, bridge, child processes, and observer. Shared services, routing control, VM, and daemon costs are excluded. RSS is summed every 10 ms and may include repeated shared mappings or miss short peaks. Subsequent idle observations measure 0.0242--0.0268 Cloud CPU seconds and 0.0194--0.0212 edge CPU seconds per wall second; no baseline is subtracted. This sampling floor matters most for short tasks and the failed fetch. Cloud peak-RSS reduction for 16-MiB inputs is only about 2.1--2.2 MiB, showing that CPU displacement does not imply proportional resident-memory reduction. Confidence intervals use 5,000 resamples of the ten complete pairs in each cell.

\subsection{Workload-dependent deployment and expenditure}
\label{sec:priority-deployment}
\begin{figure}[htbp]\centering
\includegraphics[width=\linewidth]{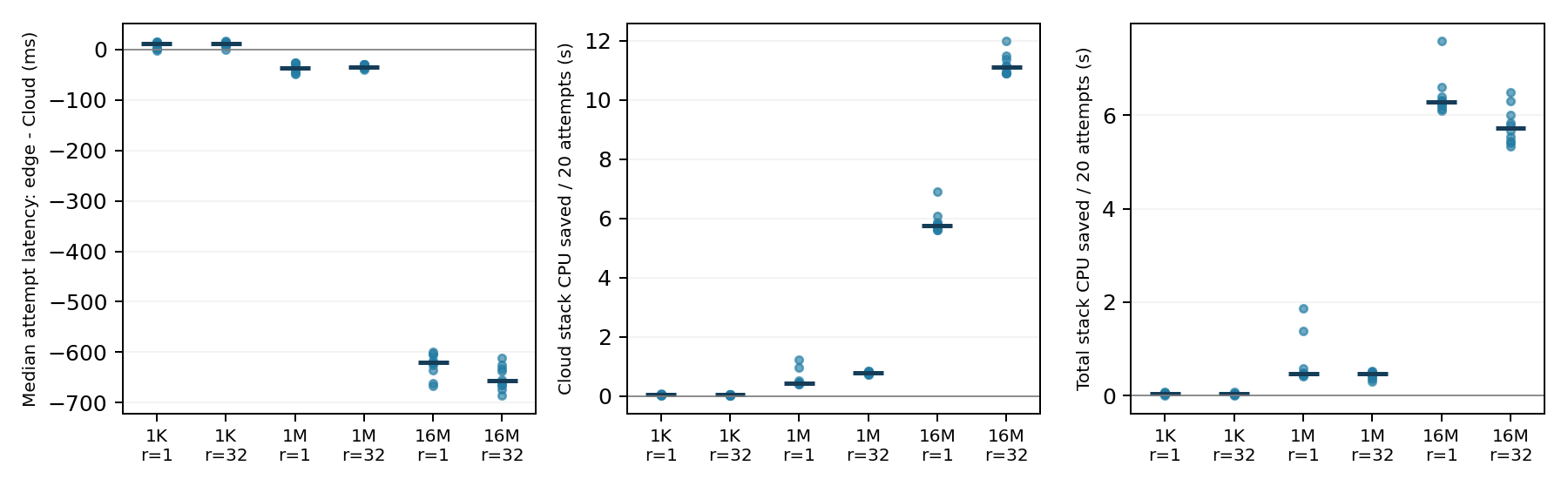}
\caption{Reanalysis of the deployment pairs. Each point is one batch
pair; horizontal marks are cell medians. The three panels separate latency,
Cloud CPU and total stack CPU. Lower latency differences favor the edge;
positive CPU savings indicate fewer measured CPU seconds.}
\label{fig:priority-deploy}\end{figure}

The placement measurements reveal two regimes. With small inputs, dispatch and coordination costs can outweigh local processing benefits. With larger inputs, avoiding source transfer and central computation reduces Cloud work and observed completion time. Table~\ref{tab:priority-deployment} and Figure~\ref{fig:priority-deploy} summarize the same 60 pairs as Section~\ref{sec:measured-placement}, including its failed Cloud fetch; they are alternative views of that experiment.

\begin{table}[htbp]\centering\small
\caption{Measured deployment tradeoff. Latency is ms; Cloud CPU saved is
seconds per 20-attempt paired batch. Ten pairs per row.}
\label{tab:priority-deployment}
\begin{tabular}{lrrrr}\toprule
Input & Digest rounds & Edge ms & Cloud ms & Cloud CPU saved s\\\midrule
1 KiB & 1 & 26.36 & 16.38 & 0.040 \\
1 KiB & 32 & 25.35 & 13.14 & 0.035 \\
1 MiB & 1 & 24.41 & 60.33 & 0.443 \\
1 MiB & 32 & 43.10 & 77.92 & 0.792 \\
16 MiB & 1 & 43.12 & 659.50 & 5.759 \\
16 MiB & 32 & 334.82 & 984.24 & 11.102 \\
\bottomrule\end{tabular}\end{table}

\paragraph{From resources to expenditure.}
Resource use affects expenditure through the billing boundary. For deployment $j$, a component model is
\begin{equation}
C_j=p_c A_{c,j}+p_m A_{m,j}+p_{in}B_{in,j}+p_{out}B_{out,j}
       +C_{model,j}+C_{edge,j}+C_{fixed,j}.
\end{equation}
Here $A_c$ and $A_m$ denote allocated CPU-time and memory-time, $B_{in}$ and $B_{out}$ denote traffic crossing priced boundaries, and the remaining terms cover model service, incremental edge operation, and fixed costs. Busy CPU and sampled RSS are not allocated resource time. In this workload, the avoided source transfer enters Cloud. Free ingress and an unchanged always-on allocation therefore yield no reduction in those Cloud bill components.

\begin{figure}[htbp]\centering
\includegraphics[width=.85\linewidth]{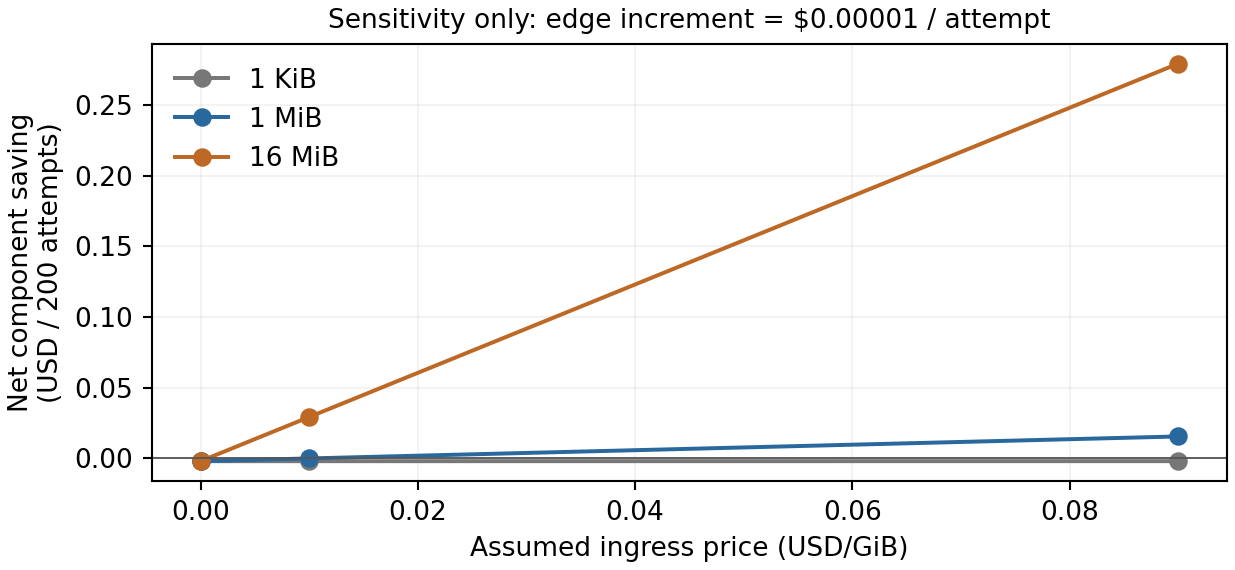}
\caption{Sensitivity of the source-transfer cost component to assumed ingress prices and additional edge cost, for one digest round. Negative values indicate that additional edge cost exceeds the avoided source-transfer charge.}
\label{fig:priority-cost}\end{figure}

A component sensitivity analysis applies
$\Delta C=p_{in}\Delta B_{in}-n c_{edge}$
to the measured source bytes, using assumed ingress prices of 0, 0.01, and 0.09 USD/GiB and incremental edge costs of 0, 0.000001, and 0.00001 USD per attempt. For 200 attempts on 16-MiB inputs, 3.125 GiB of avoided ingress gives 0.27925 USD under the assumed 0.09-USD/GiB price and a 0.002-USD edge increment. At zero ingress price the component difference is zero or negative, depending on edge cost. Figure~\ref{fig:priority-cost} illustrates this dependence; these prices are analytical assumptions rather than measured tariffs or invoices. Section~\ref{sec:cost-evidence-bridge} separately relates the deadline-feasible CPU ceilings to a possible allocation change.

\subsection{Placement of a complete model-and-tool workflow}
\label{sec:full-agent-placement}
A capability can be data-local while its surrounding Agent still depends on remote models and caller-owned devices. We therefore move the complete Agent control loop between OpenWrt and Cloud Docker. Both use identical Agent source, Python executable, SDK, device bindings, and one-CPU/512-MiB limits. Each execution independently issues two Qwen/Qwen3-8B requests, reads a Computer file, computes a digest, opens and observes Android Settings, receives a scripted confirmation, and writes and rereads an output artifact. Inference remains remote.

Two input regimes use 1 KiB with one digest round and 64 KiB with 256 rounds. Four pairs per regime balance execution order. Independent validation checks the final artifact, two Computer reads, one write, two Mobile commands, and the model-request records. The controller submits requests and supplies the scripted confirmation; reasoning remains in the evaluated Agent.

\begin{table}[htbp]\centering\small
\caption{Complete Agent deployment: four paired tasks per input stratum.
Values are means. CPU is the handler process's user+system time; peak RSS is
the process lifetime value reported at completion, not a per-task allocation
or whole-platform total.}
\label{tab:full-agent-placement}
\begin{tabular}{rlrrrrr}\toprule
KiB & Deployment & Passed & Total(s) & Model(s) & CPU(s) & RSS(MiB)\\\midrule
1 & OpenWrt/native & 4/4 & 21.58 & 6.54 & 0.121 & 33.5 \\
1 & Cloud/Docker & 4/4 & 24.11 & 9.63 & 0.135 & 79.7 \\
64 & OpenWrt/native & 4/4 & 23.56 & 9.28 & 0.123 & 33.5 \\
64 & Cloud/Docker & 4/4 & 21.47 & 7.15 & 0.146 & 79.7 \\
\bottomrule\end{tabular}\end{table}
\begin{figure}[htbp]\centering
\includegraphics[width=\linewidth]{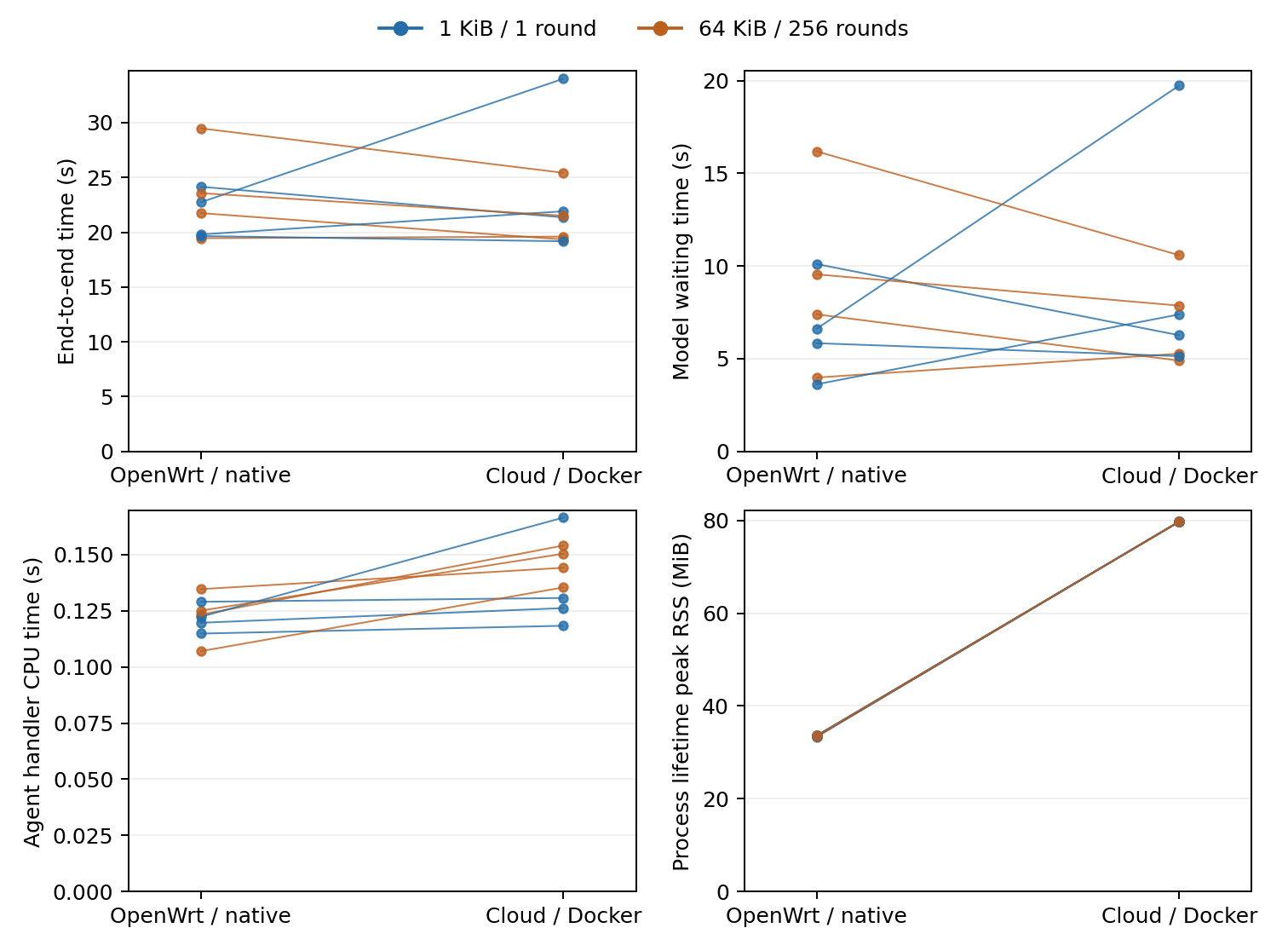}
\caption{Eight paired complete workflows. Lines connect the same task at the two execution locations. Model waiting and Agent CPU are reported separately; peak RSS is the process-lifetime maximum.}
\label{fig:full-agent-placement}\end{figure}

All 16 tasks produce correct outputs, and all 32 model requests match their Gateway records. Every task uses 319 tokens, totaling 5,104 tokens. The faster deployment changes with input regime, and model-request waiting changes in the same direction. Native execution has lower measured handler CPU and process peak RSS in this experiment, but the different kernels, hosting paths, and fresh upstream requests prevent assigning the difference to container isolation alone.

Process RSS, cgroup memory, and observed shared-service CPU retain separate accounting boundaries. In the initial Cloud process-tree series, changes in process membership make nine of sixteen intervals unavailable, leaving only one complete pair; these values remain missing. Section~\ref{sec:full-agent-resources} reports a separate repetition with child-process accounting. The deployment result demonstrates that the same model-and-tool workflow executes at either location; the configured token use is unchanged, and billable allocation changes are not measured.

Preparatory runs exposed device availability and result-framing incompatibilities before the formal workload. The formal comparison uses the same corrected SDK on both sides. Those preparatory outcomes remain separate from the sixteen reported executions.

\subsection{Shared-service resource use and workflow time decomposition}
\label{sec:full-agent-resources}
Short-lived workers complicate resource measurement because an endpoint snapshot can miss processes created and terminated between samples. We repeat the eight placement pairs in Section~\ref{sec:full-agent-placement} while grouping all 17 Cloud service roots and their descendants in a Windows Job object. Cumulative counters include exited children. Agent code, Python, SDK, task inputs, model route, devices, resource ceilings, and deadlines remain unchanged.

The group contains shared-service maintenance and console helpers; PostgreSQL, device runtimes, remote inference, Docker VM, and daemon work remain outside it. Agent cgroups are measured separately. All 16 workflows and 32 model requests complete correctly, and all 16 resource intervals are available. During the execution windows, 281 processes enter and 279 leave the service group. The initial experiment retains its nine missing intervals; the repetition supplies a separate, complete process-accounting series.

\begin{table}[htbp]\centering\small
\caption{Resource measurements for four tasks per row. Values are means of client time, Cloud accounting-window duration, process-group CPU, and the modeled idle-adjusted CPU estimate. CPU seconds sum across processes.}
\label{tab:full-agent-resources}
\begin{tabular}{rlrrrrr}\toprule
KiB & Placement & Passed & Task(s) & Window(s) & CPU(s) & Adjusted(s)\\\midrule
1 & Native & 4/4 & 18.70 & 19.04 & 22.99 & 15.57 \\
1 & Docker & 4/4 & 22.60 & 22.98 & 25.21 & 9.58 \\
64 & Native & 4/4 & 21.21 & 21.53 & 19.55 & 3.45 \\
64 & Docker & 4/4 & 23.79 & 24.11 & 27.77 & 4.61 \\
\bottomrule\end{tabular}\end{table}
\begin{figure}[htbp]\centering
\includegraphics[width=\linewidth]{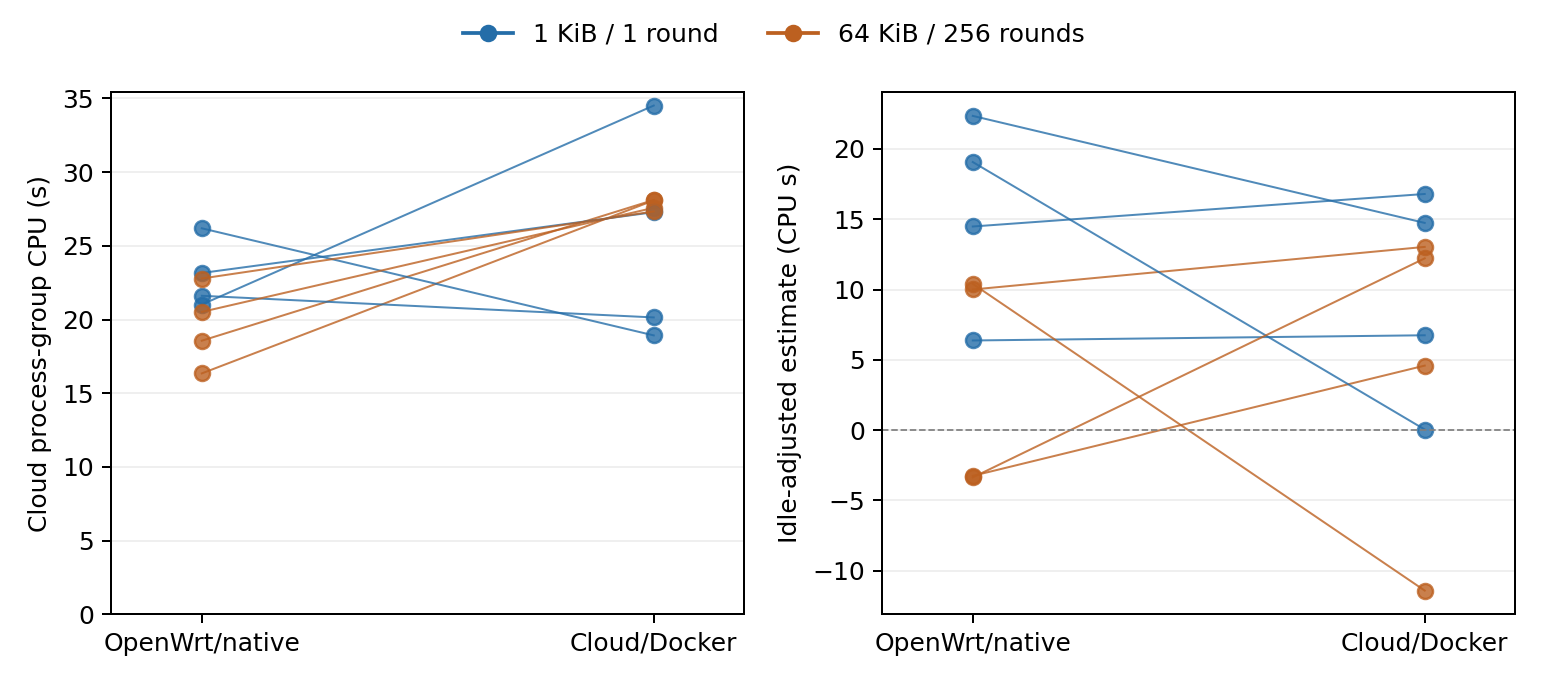}
\caption{Paired Cloud process-group CPU and idle-adjusted estimates. The latter retain negative values because sampled background work and model waiting contribute to the accounting interval.}\label{fig:full-agent-resources}\end{figure}

\paragraph{Background work.}
Let $C,T$ denote active-window CPU and duration, and $C_0,T_0$ the adjacent idle values. The background-adjusted estimate $C-(C_0/T_0)T$ assumes a stationary idle rate. Raw group CPU is lower for native placement in both input regimes, whereas the small-input ordering reverses after adjustment. Negative adjusted values are retained. The change in ordering shows that attribution of shared-service CPU depends on the assumed background rate.

\paragraph{Model waiting and local execution.}
Across both experiments, the 32 tasks form 16 matched pairs. We decompose each duration as
$T=M+(H-M)+(T-H)$,
where $M$ sums the two synchronous model-request intervals and $H$ is handler time. The residual terms cover tool operations, interaction, local computation, admission, dispatch, and completion observation. Durations use local monotonic clocks. Model-request intervals include Gateway, network, and upstream waiting, so $M$ measures the full observed model-request path within the unchanged workflow.

\begin{table}[t]\centering\small
\begin{tabular}{lrrrr}\toprule
Campaign & KiB & $\Delta T$ (s) & $\Delta M$ (s) & $\Delta(T-M)$ (s) \\
\midrule
Initial & 1 & -2.530 & -3.089 & +0.559 \\
Initial & 64 & +2.095 & +2.124 & -0.029 \\
Repeat & 1 & -3.896 & -3.401 & -0.495 \\
Repeat & 64 & -2.573 & -2.704 & +0.131 \\
\bottomrule\end{tabular}
\caption{Mean paired native-minus-Docker differences, four pairs per row.
Negative values mean less observed wall time on native. The final column is
an arithmetic residual, not a causal deployment effect.}
\label{tab:full-agent-timing-decomposition}\end{table}
\begin{figure}[t]\centering
\includegraphics[width=\linewidth]{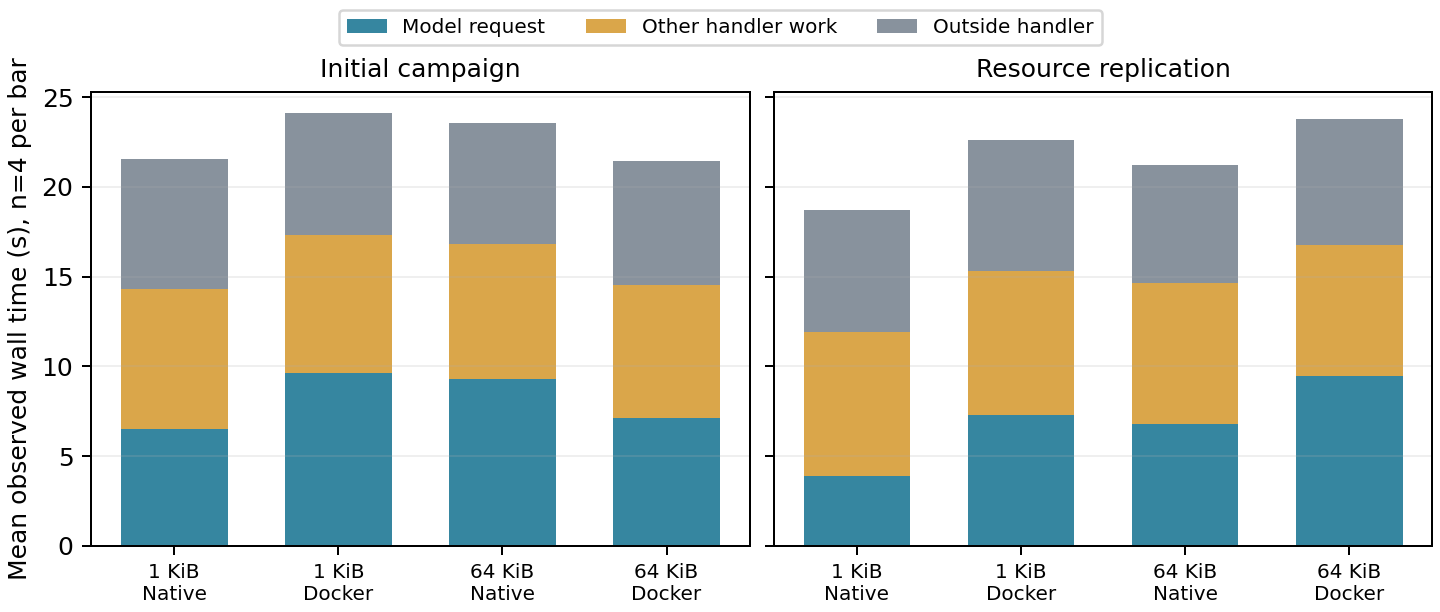}
\caption{Wall-time decomposition of the initial and repeated placement experiments. Each bar averages four executions and separates measured model-request time from the arithmetic residual.}
\label{fig:full-agent-timing-decomposition}\end{figure}

Model-request intervals account for much of the direction and magnitude of the observed end-to-end differences. Mean non-model differences change sign across the four experiment/input groups, whereas the process-memory comparison retains its own accounting boundary. Separating these components reveals which part of each complete workflow changes with deployment.

\subsection{Portability across multi-step task families}
\label{sec:diverse-agent}
We extend the fixed two-request workflow to twelve synthetic tasks: four expense summaries, four inventory reconciliations, and four Android accessibility checklists. The model selects tools and arguments at every step, while domain tools perform deterministic calculations. Expense and inventory outputs have independent literal reference results; checklist membership is reconstructed from actual device observations. Empty expense selections, matching inventories, and missing labels can be correct findings, provided the requested application is observed.

Both placements use the same Agent source, Python, SDK, Qwen route, devices, and one-CPU/512-MiB limits. Each task executes independently at both locations. Within each family, two pairs are native-first and two Docker-first; the complete schedule is shuffled. Each handler attempt allows twelve model requests and 540 seconds without application inference retry. Three tasks also appear in development runs, so this fixed workload evaluates execution portability rather than held-out reasoning accuracy.

\begin{table}[htbp]\centering\scriptsize
\begin{tabular}{llrrrrrr}\toprule
Family & Placement & Seen/planned & Task & Verified & GW & Time (s; n) & CPU (s; n) \\
\midrule
Expenses & Native & 4/4 & 4 & 4 & 20 & 24.06 (4) & 0.077 (4) \\
Expenses & Docker & 4/4 & 4 & 4 & 20 & 28.21 (4) & 0.132 (4) \\
Inventory & Native & 4/4 & 4 & 4 & 25 & 37.42 (4) & 0.090 (4) \\
Inventory & Docker & 4/4 & 4 & 4 & 24 & 31.45 (4) & 0.143 (4) \\
Screen & Native & 4/4 & 4 & 4 & 32 & 39.98 (4) & 0.138 (4) \\
Screen & Docker & 4/4 & 4 & 4 & 32 & 37.71 (4) & 0.194 (4) \\
\bottomrule\end{tabular}
\caption{Controlled multi-step tasks. Task counts successful workflows; Verified additionally requires complete execution records. GW counts associated gateway records, including repeated calls. Times cover observed terminal outcomes, with the available sample size shown.}
\label{tab:diverse-agent}\end{table}
\begin{figure}[htbp]\centering
\includegraphics[width=\linewidth]{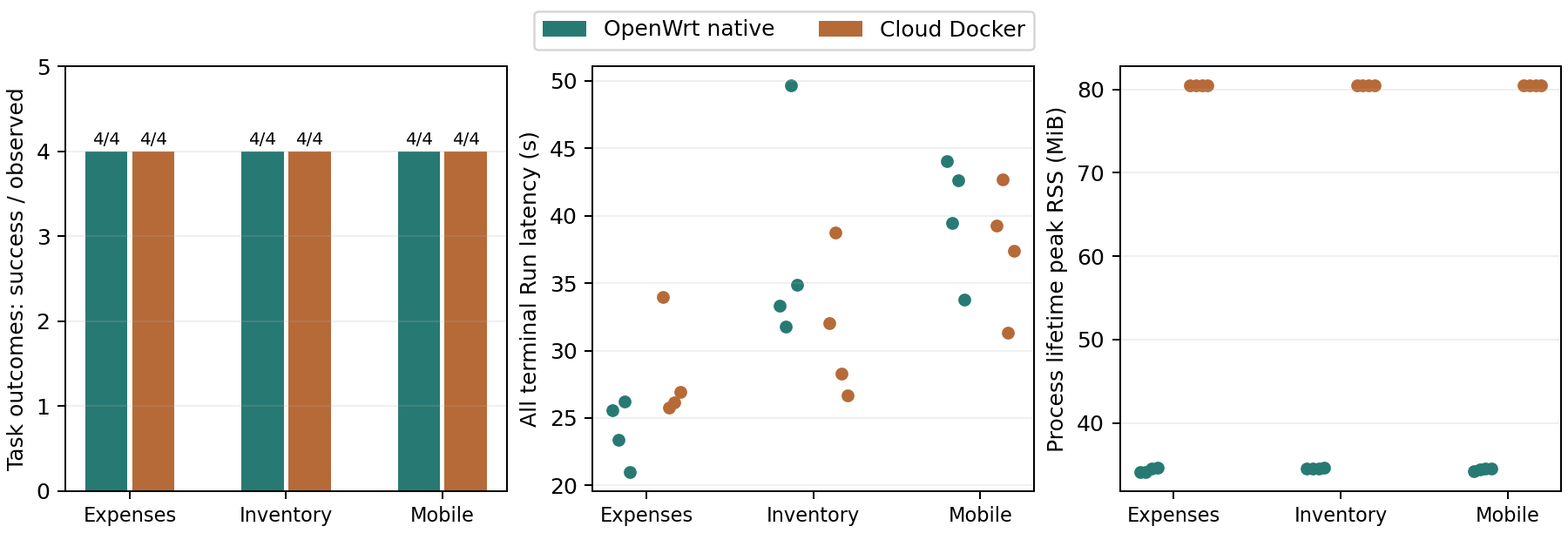}
\caption{Execution of the three model-driven task families. Circles indicate task success, crosses indicate unsuccessful or unresolved outcomes, and labels show successful/observed Runs. RSS is the process-lifetime peak.}
\label{fig:diverse-agent}\end{figure}

All 24 scheduled workflows satisfy the artifact, confirmation, device, and terminal-decision requirements. Their 153 Gateway records comprise 77 native and 76 Docker model requests. Native/Docker mean completion times are 24.06/28.21 s for expenses, 37.42/31.45 s for inventory, and 39.98/37.71 s for checklists. The faster placement changes by family because realized model trajectories and waiting differ. Native mean process peak RSS is 34.33--34.54 MiB across families, compared with 80.46 MiB for Docker. These process measurements exclude remote inference and shared-platform costs.

Separate preliminary runs expose an action-feedback interaction: the original checklist handler repeatedly rejected a content argument for a zero-argument writer, while an explicit argument error allowed the model to select the valid action. Four of six original tasks and all six repeated tasks succeeded. The accepted arguments, prompt, budget, and confirmation requirement were unchanged. These tasks informed handler development and are reported as preliminary task-specific observations. The paired results above use the common corrected handler in both placements.

\subsection{Execution configuration with a common kernel and userland}
\label{sec:native-isolation-measured}
The cross-location experiments vary both deployment and operating environment. To examine the execution configuration separately, we run the same LAN bridge, local digest backend, and driver in a native chroot and a Docker container created from the same exported image. Both share a kernel boot, Python and libc, binaries, and affinity to CPUs 0 and 1. The container has separate PID, network, and mount namespaces with no external network. The treatment is this combined namespace and launch configuration.

Three payload sizes (1, 64, and 1,024 KiB) and concurrency levels one and eight form six cells. Ten balanced pairs per cell execute 20 warm-up requests followed by 2,000 measured requests per arm. All 120 arms complete, and all 240,000 measured replies match the independently reconstructed byte count and digest.

\begin{table}[htbp]\centering\small
\caption{Native/Docker request latency. Each value is the median of ten arm
medians (ms). The last column is the median within-pair Docker-minus-native
difference and its exploratory 95\% percentile bootstrap interval.}
\label{tab:native-isolation-latency}
\begin{tabular}{lrrrr}\toprule
Payload & Concurrency & Native & Docker & Paired difference [interval]\\\midrule
1 KiB & 1 & 0.971 & 1.037 & +0.026 [-0.085, +0.175] \\
1 KiB & 8 & 6.546 & 7.595 & +0.639 [+0.274, +1.174] \\
64 KiB & 1 & 0.925 & 1.035 & +0.102 [+0.037, +0.168] \\
64 KiB & 8 & 7.376 & 8.224 & +0.846 [-0.001, +1.849] \\
1024 KiB & 1 & 1.672 & 1.858 & +0.160 [+0.083, +0.247] \\
1024 KiB & 8 & 11.860 & 12.855 & +1.187 [+0.338, +1.607] \\
\bottomrule\end{tabular}\end{table}
\begin{figure}[htbp]\centering
\includegraphics[width=\linewidth]{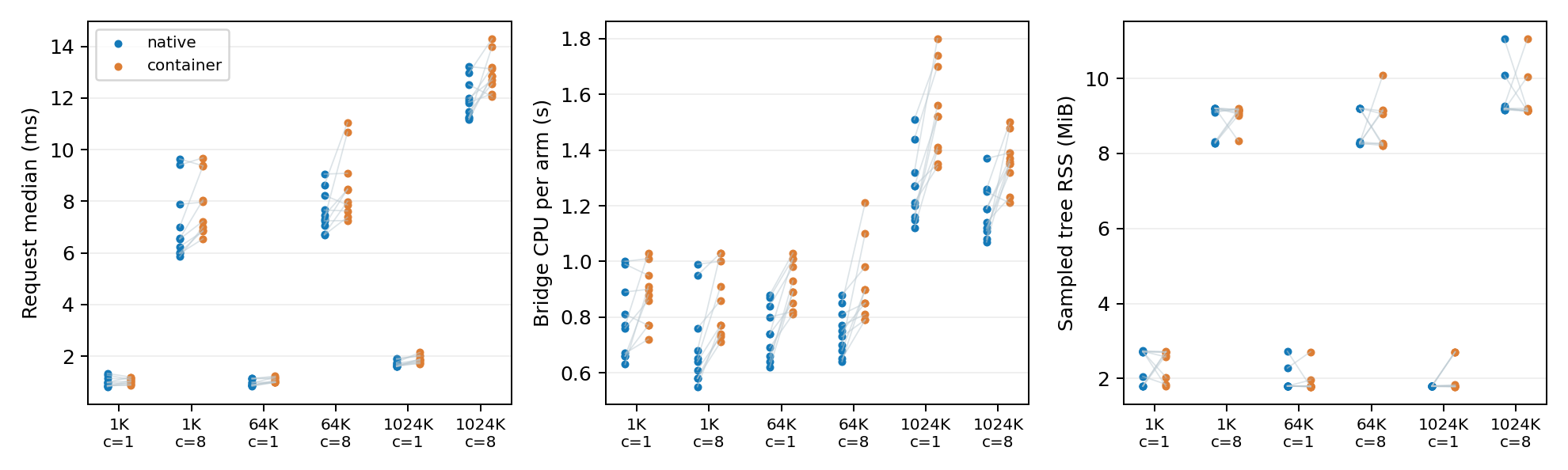}
\caption{All scheduled matched bridge observations. Lines join the two arms of
one scheduling pair; labels K denote KiB. CPU and RSS are bridge-process
measurements with the exclusions stated in the text.}
\label{fig:native-isolation}\end{figure}
\begin{table}[htbp]\centering\small
\caption{Median bridge CPU time per 2,000-request arm and median sampled peak process-tree RSS. Accounting excludes the daemon, driver, backend, and VM.}
\begin{tabular}{lrrrrr}\toprule
 & & \multicolumn{2}{c}{CPU seconds} & \multicolumn{2}{c}{Peak tree RSS (MiB)}\\
Payload & Concurrency & Native & Docker & Native & Docker\\\midrule
1 KiB & 1 & 0.765 & 0.890 & 2.715 & 2.666 \\
1 KiB & 8 & 0.645 & 0.815 & 9.170 & 9.150 \\
64 KiB & 1 & 0.740 & 0.910 & 1.801 & 1.793 \\
64 KiB & 8 & 0.740 & 0.875 & 8.299 & 8.668 \\
1024 KiB & 1 & 1.240 & 1.520 & 1.801 & 1.797 \\
1024 KiB & 8 & 1.165 & 1.355 & 9.207 & 9.148 \\
\bottomrule\end{tabular}\end{table}

Paired median Docker-minus-native latency differences range from 0.026 to 1.187 ms and are positive in all six cells. Four exploratory 95\% intervals exclude zero; two include it. Bridge CPU differences range from 0.075 to 0.255 s per 2,000 requests, while all peak-RSS intervals include zero. The results thus identify a small configuration-dependent latency/CPU cost without a clear process-memory difference.

CPU counters include the bridge parent and reaped children at 10-ms tick resolution. Memory is the maximum sampled process-tree RSS sum at five-millisecond intervals. The driver, backend, Docker daemon, and VM are outside these resource counters. Startup observations approach the five-millisecond polling granularity and describe a warm runtime. Intervals use 5,000 resamples of the ten whole scheduling pairs per cell, with no simultaneous-coverage adjustment. Shared-host activity remains a source of variation. The measured treatment is the combined namespace and launch configuration, with individual namespace costs left unseparated.

\subsection{Indexing cost under an unchanged route-selection policy}
\label{sec:matched-route}
A routing index is beneficial only if it accelerates the same selection policy. Both arms therefore share route objects, eligibility checks, scoring, and query order; a read-only view disables only index pointers in the scan arm. For every query, the selected route and score, or the absence of an eligible route, are compared outside the timed lookup interval.

Each platform runs eleven workload cells with ten balanced paired trials of 4,000 queries. Route counts vary from 100 to 50,000. At 10,000 routes, additional cells vary candidates per intent, missing queries, expired routes, and query concentration. The reference distribution has ten candidates per intent, uniform selection, and no missing or expired routes. Across both platforms, 220 pairs contain 880,000 query instances; all 2,640,000 repeated equivalence checks agree.

\begin{table}[htbp]
\centering\small
\caption{Within-platform index ablation. p95 entries are medians across ten paired trials (microseconds). Speedup is the median paired mean-time ratio with a trial-bootstrap 95\% interval.}
\label{tab:matched-route}
\begin{tabular}{@{}lrrrr@{}}
\toprule
Platform & Routes & Index p95 & Scan p95 & Mean-time speedup \\
\midrule
Windows & 100 & 0.300 & 0.600 & 2.5 [2.4, 2.8] \\
Windows & 1,000 & 0.400 & 6.250 & 20.4 [17.8, 22.0] \\
Windows & 10,000 & 0.950 & 193.600 & 250.4 [209.4, 274.9] \\
Windows & 50,000 & 2.150 & 3397.700 & 1773.3 [1336.6, 1964.3] \\
OpenWrt VM & 100 & 0.371 & 1.051 & 3.1 [2.8, 3.3] \\
OpenWrt VM & 1,000 & 0.544 & 9.619 & 18.3 [15.6, 20.1] \\
OpenWrt VM & 10,000 & 0.942 & 133.618 & 158.4 [117.9, 187.0] \\
OpenWrt VM & 50,000 & 1.401 & 1957.289 & 1132.1 [775.4, 1355.8] \\
\bottomrule
\end{tabular}
\end{table}

For trial $r$, the speedup is
\begin{equation}
S_r=\overline{t}_{\mathrm{scan},r}/\overline{t}_{\mathrm{index},r}.
\end{equation}
We report the median trial ratio and percentile intervals from 5,000 resamples of the ten whole pairs. Figure~\ref{fig:route-matched} shows the within-OpenWrt effect. Increasing candidates per intent reduces the benefit because the surviving candidates still require scoring. Missing-query and fully expired workloads retain their no-result semantics. At 10,000 routes, the indexes require 262,144 additional bytes, exposing the lookup-time/storage tradeoff.

The shortest indexed intervals approach the timer floor, particularly for Windows misses. Each platform is analyzed separately: absolute host/VM differences include operating-environment effects, while the controlled comparison within a platform isolates index availability.

\subsection{Lease renewal and stale-route withdrawal}
\label{sec:measured-leases}
Leased registration trades continuous renewal traffic for bounded stale-route eligibility. We examine that tradeoff on the installed OpenWrt daemon using ten balanced pairs. Each route has a five-second lease and is observed every 0.5 s for twelve seconds. The treatment renews the same lease every two seconds; the control stops renewal. The endpoint and lookup semantics remain fixed.

Renewed routes remain eligible in all 240 observations. Each unrenewed route is eligible for the first ten observations and absent for the remaining fourteen, giving 100/240 eligible lookups. The first absent lookup occurs 5.004--5.289 s after registration, an observation interval that includes management-path timing. Twenty wrong-tenant lookups return no route. These are control-plane observations of administrative registrations; the endpoint is not invoked.

The result shows how renewal preserves route eligibility while lease expiry removes stale entries. It leaves a distinct recovery question: once a daemon has lost the registration itself, renewal must be followed by re-registration. Section~\ref{sec:router-recovery} evaluates that transition using actual business calls.

\subsection{Route reconstruction and atomic configuration updates}
\label{sec:router-recovery}
A router must handle both the loss of volatile registrations and changes to persistent configuration. We evaluate these mechanisms separately on the same OpenWrt VM. A local gateway uses the daemon's IPC interface, and the SDK reaches it through a fixed SSH management path. Requests contain 1,024 bytes and a pair-specific nonce; response digests and handler receipts independently establish execution. The local administrative gateway omits production JWT authorization, keeping the experiment focused on route lifecycle behavior.

\paragraph{Re-registration after restart.}
Six balanced pairs differ only in SDK re-registration after a missing-route response. Both use five-second leases, a 0.6 renewal fraction, and the same retry policy. After ten successful baseline calls, the routing daemon restarts while the SDK, gateway, handler, and tunnel remain live. Serial probes continue at a nominal 200-ms cadence for thirty seconds after the restart action returns. No other registration-refresh path can recreate the route.

\begin{table}[htbp]
\centering\small
\caption{Six matched OpenWrt restart pairs. Fresh probes start after the service action returns; baseline probes are excluded from these counts. Unrecovered controls are censored at the observation bound.}
\label{tab:router-restart}
\begin{tabular}{@{}lrrr@{}}
\toprule
SDK re-registration & Recovered arms & Successful fresh probes & Censored \\
\midrule
Disabled & 0/6 & 0/900 & 6 \\
Enabled & 6/6 & 890/900 & 0 \\
\bottomrule
\end{tabular}
\end{table}
\begin{figure}[htbp]
\centering
\includegraphics[width=\linewidth]{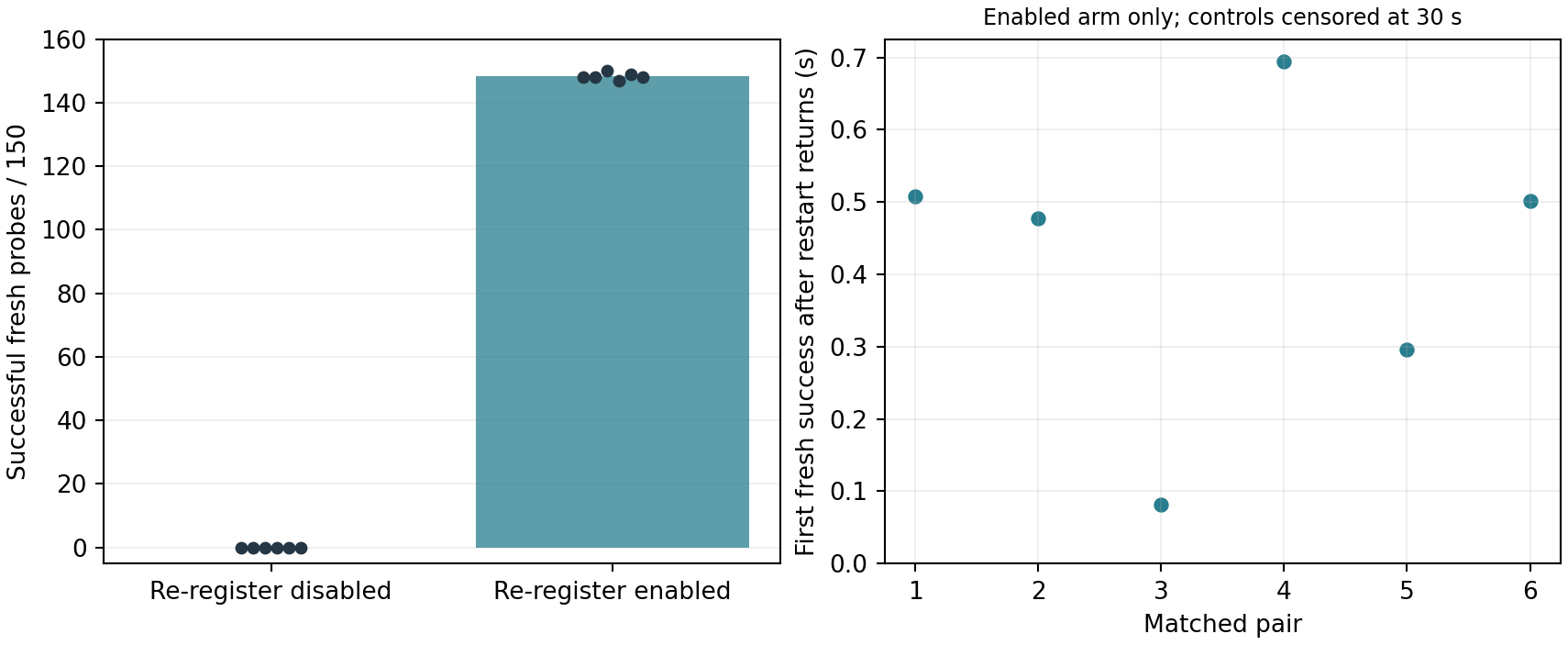}
\caption{Re-registration after router restart. Each point is one arm in six pairs; bars mark means. Recovery time begins when the restart action returns. Recovery times are shown only for arms that recover within the observation window.}
\end{figure}

Re-registration restores service in 6/6 treatment arms, with 890/900 subsequent probes succeeding; controls recover in 0/6 arms and complete 0/900 probes. The complete schedule has 1,932 probes and 1,022 matching handler receipts. First successful post-restart responses occur 0.082--0.695 s after the restart action returns. This timing excludes the action's duration; controls remain unavailable through the thirty-second observation window. New route identities and successful business responses jointly establish reconstruction of the lost registration.

\paragraph{Configuration replacement.}
A separate six-pair experiment redirects a static route from handler A to handler B while preserving a dynamic SDK route. Valid and invalid candidates differ only in whether the route version satisfies the positive-version requirement. Ninety alternating static/dynamic calls span preparation, reload, subsequent use, and restoration in each arm.

\begin{table}[htbp]
\centering\small
\caption{Static-route update scenarios: six attempts per candidate condition. The same dynamic route is retained across hot reload. Successful probes include static and dynamic calls before, during, after and following restoration.}
\label{tab:router-reload}
\begin{tabular}{@{}lrrrr@{}}
\toprule
Candidate & Accepted & Successful probes & Degraded & Diagnostic restored \\
\midrule
Invalid & 0/6 & 540/540 & 6/6 & 6/6 \\
Valid & 6/6 & 540/540 & 0/6 & 6/6 \\
\bottomrule
\end{tabular}
\end{table}

All 1,080 calls produce matching handler receipts. All six valid candidates are accepted and select B; all six invalid candidates are rejected and retain A. Dynamic-route and daemon identities remain unchanged. Invalid updates enter a degraded configuration state until restoration, without disrupting the sampled business calls. This supports atomic replacement for the tested validation error. The six pairs resolve availability at the serial probing cadence; shorter interruptions can fall between observations.

\subsection{Route reflection across an interrupted intermediary}
\label{sec:peer-propagation-measured}
Local registration is insufficient when an invocation must traverse an intermediary router. We form an A--B--C peer chain with three copies of the installed daemon on one OpenWrt VM. Each has a private process root and route table; A and C have no direct adjacency. Peer sessions use TLS 1.3 and a shared experimental CA over isolated loopback. Within each pair, only B's route-reflection option changes.

The schedule contains four balanced pairs for each of 4 and 32 source routes. A publishes the route set and withdraws half; B is then stopped. During the interruption, A replaces all remaining old routes. Restarting B uses the same configuration and a new boot epoch, without republishing the replacements at A. Publication, withdrawal, interruption, and offline update each have a three-second observation window, followed by six seconds for reconnection. Heartbeats are one second, the miss limit is three, graceful retention is ten seconds, and source leases are 120 seconds. Validation compares exact route identities, attributes, and paths, including tenant and withdrawal behavior.

\begin{table}[htbp]\centering\small
\caption{Route state at the destination peer. Each row contains four arms. Expected-state counts include absent routes in the reflection-disabled control; publication and recovery use all source routes as denominators. Recovery time is the median observed bound for reflection-enabled arms.}
\label{tab:peer-propagation}
\begin{tabular}{rlrrrr}\toprule
Routes & Reflector & Oracle arms & Published at C & Recovered at C & Bound (s)\\\midrule
4 & Off & 4/4 & 0/16 & 0/16 & --- \\
4 & On & 4/4 & 16/16 & 16/16 & 1.295 \\
32 & Off & 4/4 & 0/128 & 0/128 & --- \\
32 & On & 4/4 & 128/128 & 128/128 & 1.576 \\
\bottomrule\end{tabular}\end{table}
\begin{figure}[htbp]\centering
\includegraphics[width=\linewidth]{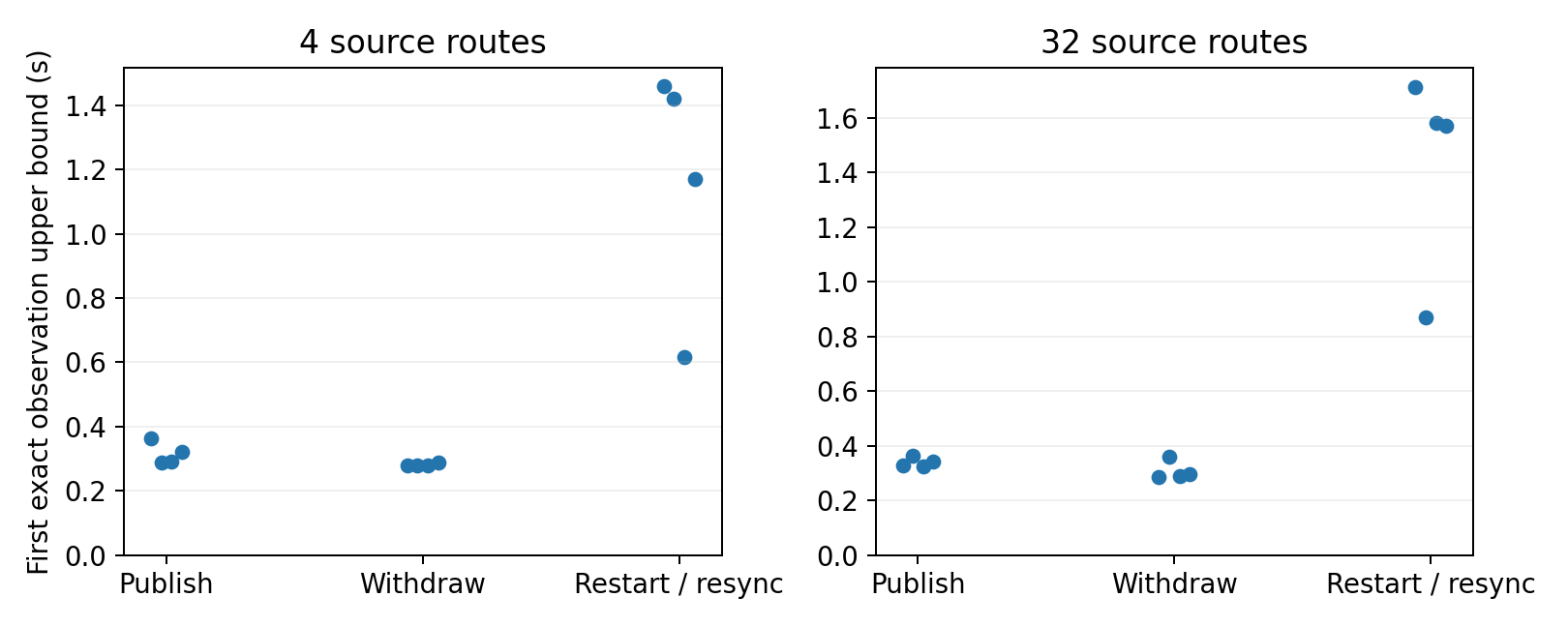}
\caption{Time from the reflection action to the first complete destination route-set observation. Measurements include management operations and polling. Reflection-disabled arms have no available destination route and therefore no recovery-time point.}
\end{figure}

All sixteen arms match their expected route states. B learns A's routes in both configurations; C receives them with path $[A,B]$ only when reflection is enabled. C rejects old-capability lookups while B is absent and converges to the exact replacement set after reconnection. Withdrawn routes do not reappear. Reflection-enabled arms record eight updates and two withdrawals for four source routes, and 64 updates and sixteen withdrawals for 32 routes across B's two process lifetimes.

An earlier attempt encountered eight management-transport failures before route publication; it remains separate from the complete sixteen-arm experiment. The result concerns actual protocol sessions between logical routers sharing one VM. With reflection enabled, the downstream router acquires and reconstructs the intended route set. Continuous churn and physical-node timing are not varied here, and backend execution is examined next.

\subsection{Discovery, trust, and route eligibility}
\label{sec:discovery-measured}
Discovering an endpoint and authorizing its advertised capability are separate steps. We test their relationship with two installed OpenWrt daemons and DNS-SD announcements parsed by the installed discovery service. Each arm uses a fresh externally disconnected network namespace, private-CA mutual TLS, and no open-mesh admission. A controlled publisher sends twelve identical-within-pair announcements at one-second intervals; route candidates are obtained through packet parsing rather than inserted directly.

Four balanced pairs in each same-domain and foreign-domain stratum compare automatic admission disabled with the same-domain admission policy. The consumer has no static peer entry for the publisher, while the return adjacency is fixed. Capabilities have 120-second leases, candidate lifetimes are five seconds, and peer grace is ten seconds. Twenty-five snapshots separated by one-second sleeps track candidate detection, authenticated adjacency, and route eligibility through announcement expiry.

\begin{table}[htbp]\centering\small
\begin{tabular}{llrrrr}\toprule
Candidate domain & Admission & Discovered & Peer & Route & Verified\\\midrule
Same & Off & 4/4 & 0/4 & 0/4 & 4/4 \\
Same & Same-domain & 4/4 & 4/4 & 4/4 & 4/4 \\
Other & Off & 4/4 & 0/4 & 0/4 & 4/4 \\
Other & Same-domain & 4/4 & 0/4 & 0/4 & 4/4 \\
\bottomrule\end{tabular}
\caption{Four discovery arms per cell. Verified includes policy-consistent rejection; usable routes are counted separately. Route validation matches identity, tenant, peer origin, and path vector.}
\end{table}
\begin{figure}[htbp]\centering
\includegraphics[width=\linewidth]{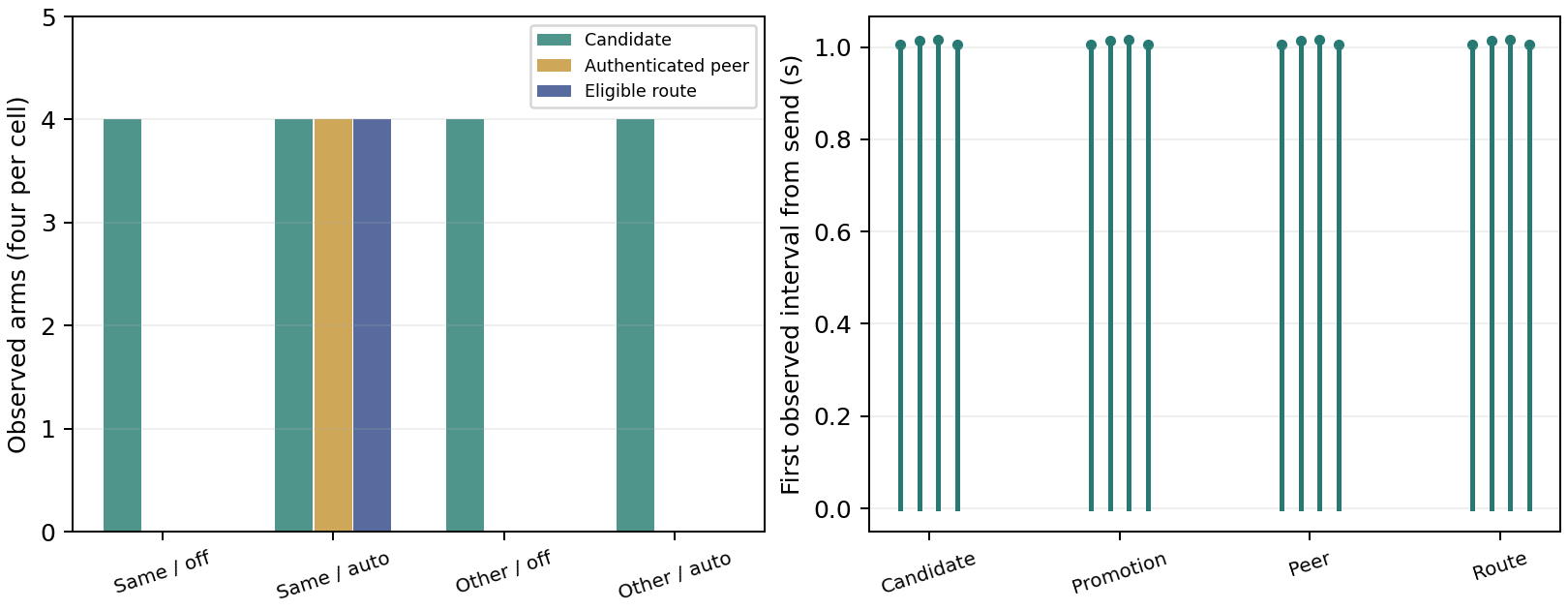}
\caption{Discovery and admission outcomes. The right panel shows observed stage intervals for the four eligible same-domain automatic-admission arms.}
\end{figure}

All sixteen arms satisfy their expected outcomes. Receiving an announcement produces a candidate in every arm, but only policy-eligible candidates become authenticated, selectable routes. After publication stops, candidate, promoted adjacency, and route state disappear within the observed sequence. The experiment therefore connects discovery to trust-controlled availability while distinguishing rejection from successful admission.

The measurements cover protocol execution on one VM and include polling uncertainty. Publication is supplied by the controlled DNS-SD sender, and the announced backend is not called in this experiment. Section~\ref{sec:peer-invoke-measured} extends the same route lifecycle to bidirectional backend execution.

\subsection{Invocation through discovered peer routes}
\label{sec:peer-invoke-measured}
To connect route eligibility with useful execution, we invoke a deterministic echo capability through the learned peer route in both directions. The path consists of client IPC lookup, source daemon, authenticated peer tunnel, destination daemon, and target-local HTTP backend. The client uses the returned route and target identities without a direct-HTTP fallback. Each backend records its received body; an independent decoder matches the nonce, exact output, and opposite-node receipt.

Four balanced pairs vary automatic admission while retaining peer identities, certificates, backend bytes, and observation phases. Fresh daemons run in an isolated namespace on the same VM. The peer tick is 50 ms, discovery polling one second, candidate lifetime five seconds, peer grace ten seconds, and route lease 120 seconds. Twelve announcements create the discovery window. Both directions are probed before publication, during availability, after expiry, and after a second publication followed by four seconds of waiting. Each call has a five-second timeout and no retry.

\begin{table}[!ht]\centering\small
\begin{tabular}{llrrr}\toprule
Admission & Phase & Probes & Route found & Backend verified\\\midrule
Off & Before & 8 & 0 & 0 \\
Off & Available & 8 & 0 & 0 \\
Off & Expired & 8 & 0 & 0 \\
Off & Recovered & 8 & 0 & 0 \\
Same-Domain & Before & 8 & 0 & 0 \\
Same-Domain & Available & 8 & 8 & 8 \\
Same-Domain & Expired & 8 & 0 & 0 \\
Same-Domain & Recovered & 8 & 8 & 8 \\
\bottomrule\end{tabular}
\caption{Each cell includes four arms and both directions. No-route observations
are policy/lifecycle outcomes and are not counted as successful executions.}
\end{table}
\begin{figure}[!ht]\centering
\includegraphics[width=\linewidth]{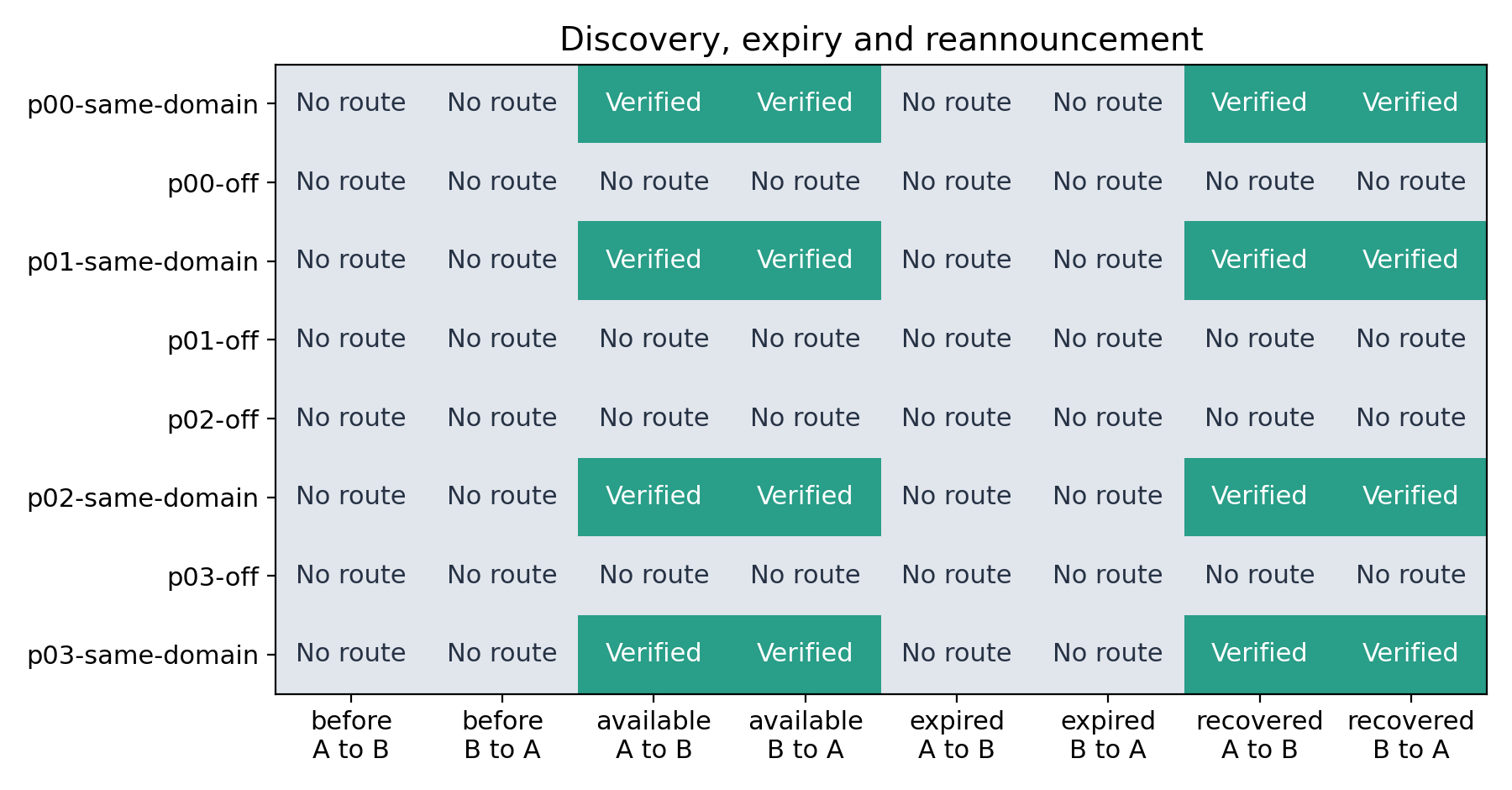}
\caption{Directional invocation outcomes. Successful execution requires both HTTP completion and a matching backend receipt.}
\end{figure}

The eight arms produce 64 directional probes. Sixteen expect a usable route and each reaches exactly one opposite-node backend with the correct body. The other 48 expect no route and are rejected as specified. Expiry removes callability despite live backends; reannouncement restores it. Thus the experiment links admission policy and route lifetime to actual execution, rather than inferring execution from a populated table.

The topology contains two logical routers on one OpenWrt VM, and the workload is sequential HTTP echo. Calls issued in each phase show that expiry withdraws access and reannouncement restores it. No business operation spans the interruption, so the observed recovery concerns new-call availability. Development calls used to establish valid invocation framing are separate from the fixed schedule.

\subsection{Model-driven delegation across logical edge nodes}
\label{sec:cross-edge-agent}
We next replace the echo backend with a worker Agent operating on its own local file. A caller Agent delegates the task and finishes after receiving an artifact receipt. The worker uses four real model decisions to read, calculate, write, and finish; the caller makes two. Calculation uses exact integer tools, allowing independent SQL reconstruction of the expected report.

Two SDK processes and their routing daemons share one OpenWrt VM. The peer path uses a learned route and mutual-TLS tunnel; a direct-HTTP reference reaches the same worker without peer authorization. Both use the same model-network adapter. Eight balanced pairs cover expense and inventory tasks, three- and nine-row inputs, and both directions. Source, SDK, input bytes, Qwen/Qwen3-8B configuration, temperature zero, disabled thinking, and 128-token response limits are common. Each arm has fresh identities, outputs, and model calls. Three-row tasks were used in development; nine-row inputs were not.

\begin{table}[!ht]
\centering\small
\caption{Real Agent peer/direct comparison. End-to-end times include six model decisions per arm; resource samples cover both logical nodes.}
\label{tab:cross-edge-agent}
\begin{tabular}{@{}lrrrr@{}}\toprule
Path & Success & Calls & Median (s) & Median CPU (s) \\
\midrule
Peer tunnel & 8/8 & 48 & 10.023 & 0.0853 \\
Direct HTTP & 8/8 & 48 & 9.523 & 0.0808 \\
\bottomrule\end{tabular}
\end{table}
\begin{figure}[!ht]
\centering\includegraphics[width=\linewidth]{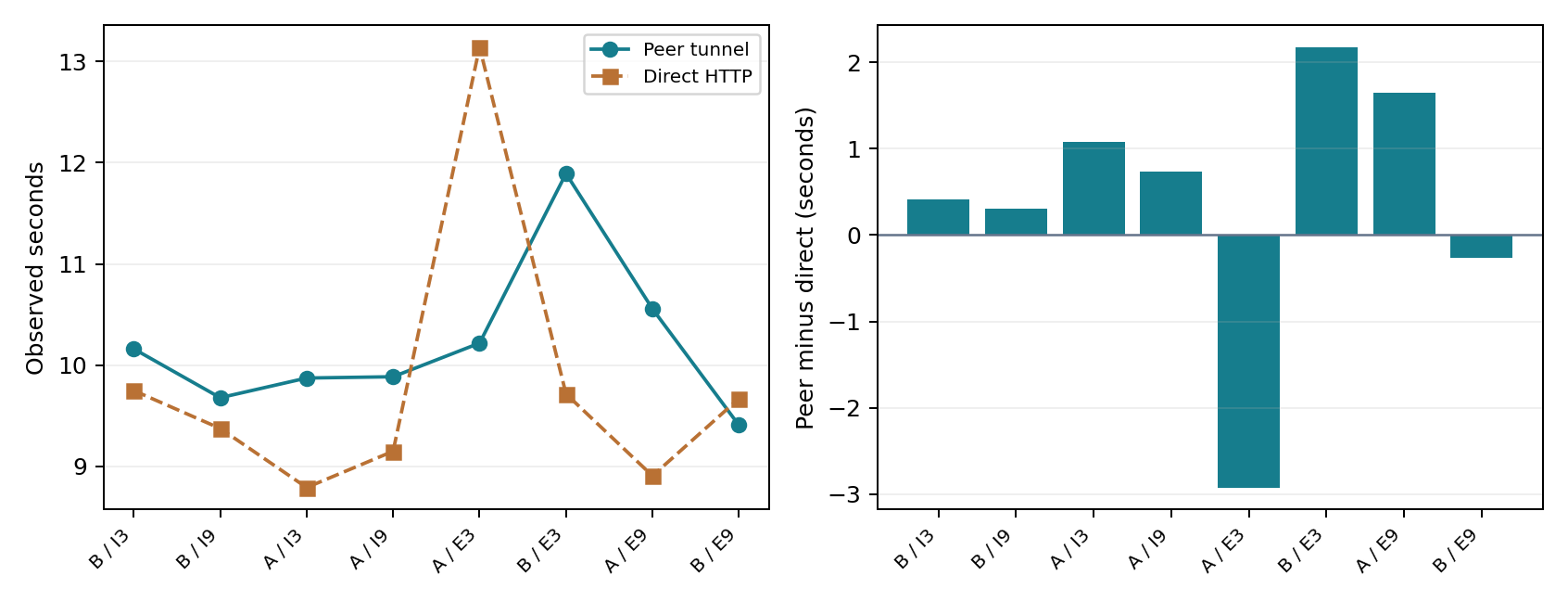}
\caption{Eight peer/direct pairs with fresh execution and equal model-call counts. Labels identify direction, task family, and input size.}
\label{fig:cross-edge-agent}
\end{figure}

All sixteen workflows complete correctly, with all 96 model requests linked to Gateway records. Peer invocation counters increase once at each endpoint only for peer arms. Across eight pairs, peer-minus-direct completion time has median 0.574 s, mean 0.396 s, and range $[-2.923,2.176]$ s. These descriptive differences include independently realized model trajectories and waiting, as well as the added peer service. Endpoint process-RSS sums have medians of 86.49 MiB for peer and 86.94 MiB for direct execution; they exclude remote inference and shared platform resources.

\paragraph{Availability after peer loss.}
A separate four-arm experiment stops the destination daemon while its worker remains live. Both outage calls fail after route loss, with no worker model requests or journal records. After restarting the daemon and observing stable route availability, both newly submitted tasks complete correctly. The fourteen model requests across these four arms match the expected caller/worker identities. This restores access for new work, while the two failed original calls remain failed. The synchronous path has a thirty-second timeout; worker cancellation and interrupted-task continuation require separate lifecycle mechanisms.

\subsection{Application delivery through IPv6 and Relay paths}
\label{sec:priority-transport}
We examine whether a registered service is callable through the two supported transport configurations. Both enter through Cloud and reach the same SDK TLS backend. The IPv6 path uses a global prefix supplied through upstream relay and neighbor-discovery proxying. The Relay path additionally uses an SSH reverse-TCP laboratory bridge, with TLS verification retained at the original endpoints. Timing includes this asymmetric laboratory path, while delivery outcomes are checked against the common backend.

Twelve balanced pairs cover 1-, 16-, and 32-KiB bodies. Each mode restart is followed by at most 300 seconds of registration-readiness observation with a thirty-second synchronization interval. Application timing begins afterward, so it describes the first request after readiness. Concurrency is one, retries are disabled, and paired calls share body bytes and digest validation. Independent handler receipts distinguish execution from registration alone.

\begin{table}[htbp]\centering\small
\caption{Production transport requests. Durations are median seconds;
``All'' includes failed attempts, ``Exact'' includes only independently verified
outputs, and ``Ready'' is untimed registration waiting. A dash means no exact output.}
\label{tab:priority-transport}
\begin{tabular}{rlrrrr}\toprule
KiB & Path & Exact/attempts & All s & Exact s & Ready s\\\midrule
1 & IPv6 & 3/4 & 0.726 & 0.776 & 34.043 \\
1 & Relay & 4/4 & 0.831 & 0.831 & 37.290 \\
16 & IPv6 & 4/4 & 0.677 & 0.677 & 33.873 \\
16 & Relay & 0/4 & 1.116 & -- & 37.152 \\
32 & IPv6 & 4/4 & 0.684 & 0.684 & 33.693 \\
32 & Relay & 0/4 & 0.792 & -- & 35.021 \\
\bottomrule\end{tabular}\end{table}
\begin{figure}[htbp]\centering
\includegraphics[width=\linewidth]{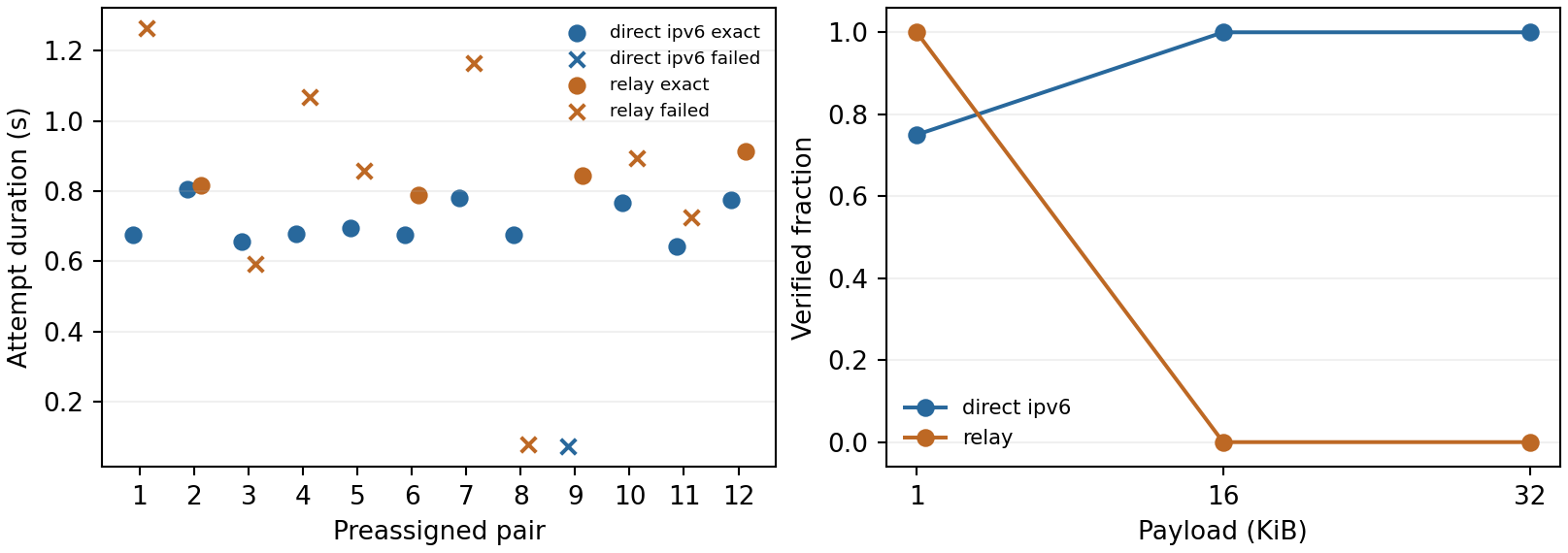}
\caption{Application delivery by body size. Crosses identify failed attempts; the second panel reports exact-output fractions.}
\label{fig:priority-transport}\end{figure}

Large-body Relay calls encounter node-disconnection resets, and some requests report an unavailable deployment after readiness. The failed calls in Table~\ref{tab:priority-transport} show that registration availability and payload delivery can diverge. Section~\ref{sec:relay-followup} examines request-size and flow-control handling to explain these outcomes. Both paths carry payload through Cloud; the byte counters measure their application-body transfers, excluding lower-layer framing.

\subsection{Stream-local rejection and bounded response flow control}
\label{sec:relay-followup}
Oversized requests should fail without disrupting unrelated streams. The synchronous request limit is 16,384 bytes including its envelope, so a 16-KiB business body already exceeds it. We compare two build-matched implementations that differ only in RESET handling: immediate stream-state release and bounded retention of sequence/credit state to drain previously authorized in-flight frames. The latter retains at most 64 compact records and accepts neither out-of-order nor over-credit frames.

Four triplets per arm send a 1-KiB request, an oversized 16- or 32-KiB request, and another 1-KiB request. Both arms use the same toolchain, TLS dependencies, backend, and authorization. All eight normal calls per arm complete with unique receipts, and all four oversized calls are rejected before backend execution. Immediate release produces four protocol errors and eight new TCP/TLS handshakes. Bounded draining produces stream-local rejection with no new errors or handshakes. Normal calls also succeed in the control after reconnection. Bounded draining preserves the existing session and avoids those additional handshakes while maintaining the same normal-call completion count.

\paragraph{Response-credit boundary.}
A separate preliminary experiment completes 6/12 calls: responses beyond the initial 4-KiB send credit fail. A credit-aware sender retains the bounded response, transmits available credit, and resumes on credit updates under the original deadline. With this implementation, a fixed balanced comparison uses 1-, 4-, and 7-KiB bodies and four pairs per size after a three-second settling interval.

\begin{table}[htbp]\centering\small
\caption{Repaired common-bound deployment. Median seconds include only exact
outputs; all attempts remain in the success denominator. $\Delta$ is the
median Relay-minus-IPv6 difference over complete exact-output pairs, not the
difference of the two marginal medians.}
\label{tab:relay-common-bound}
\begin{tabular}{rlrrr}\toprule
KiB & Path & Exact/attempts & Exact median s & $\Delta$ s\\\midrule
1 & IPv6 & 4/4 & 0.647 & -- \\
1 & Relay & 4/4 & 0.786 & +0.139 \\
4 & IPv6 & 4/4 & 0.651 & -- \\
4 & Relay & 4/4 & 0.810 & +0.163 \\
7 & IPv6 & 4/4 & 0.710 & -- \\
7 & Relay & 4/4 & 0.739 & +0.029 \\
\bottomrule\end{tabular}\end{table}
\begin{figure}[htbp]\centering
\includegraphics[width=\linewidth]{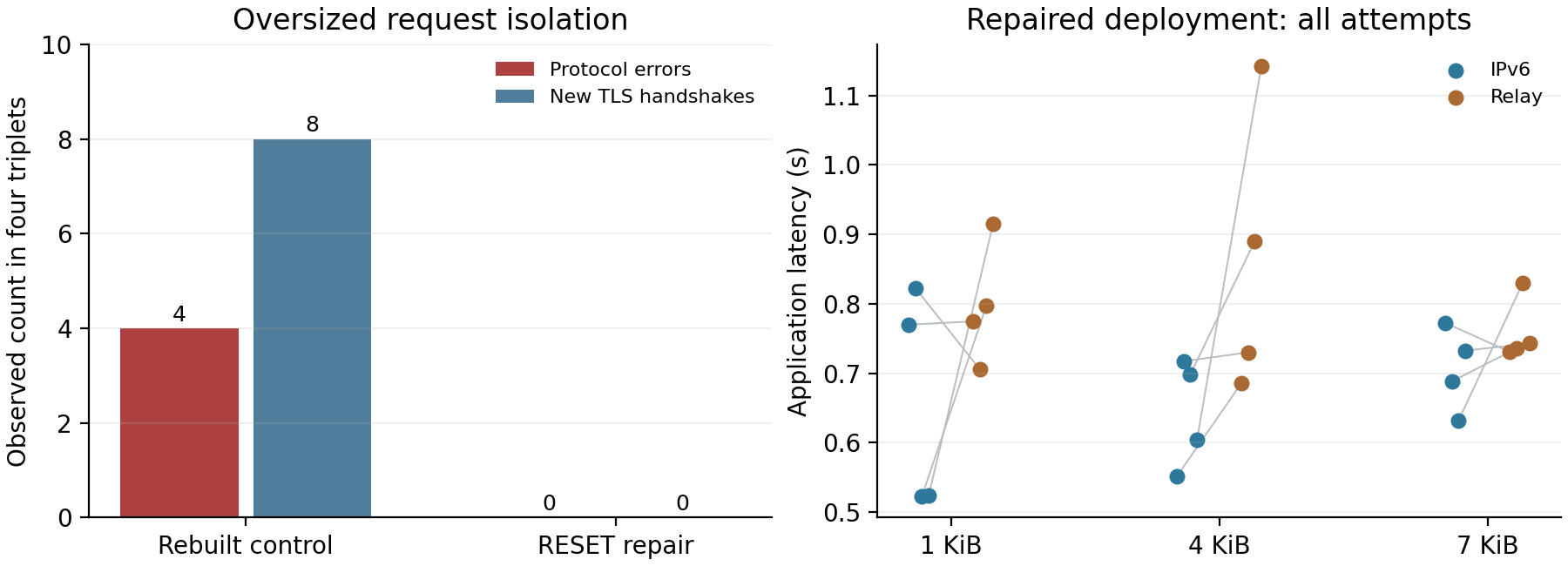}
\caption{Stream-local rejection and bounded transport delivery. Left: scenario outcomes. Right: individual attempt durations, with crosses identifying failures.}
\label{fig:relay-followup}\end{figure}

The table reports this final payload range separately from the earlier oversized-request study. Build-matched RESET controls identify the isolation mechanism, whereas the IPv6/Relay timing observations still include the Relay-only SSH bridge and shared Cloud processing. Together, the experiments show why request rejection and flow control must preserve connection state at different boundaries.

\subsection{Repeated invocation under bounded concurrency}
\label{sec:transport-concurrency-measured}
Repeated calls exercise admission, dispatch, execution, and settlement together. We run the same echo-and-digest SDK tool through the same Cloud service, caller, and TLS backend with 1- and 7-KiB bodies at concurrency one and four. Each of four cells contains four balanced pairs. After fresh registration, a three-second settling interval, and three warm-up calls, each arm issues 32 measured requests without application retry. The caller authenticates before each arm, outside timing. Both paths pass through Cloud; Relay also retains the laboratory SSH bridge.

An independent validator checks exact content, nonce, unique handler receipt, target, and selected path. Goodput counts verified completions per batch wall time, including unsuccessful calls and worker/session setup. Latency summarizes successful requests. Paired differences use the four batch-median differences per cell, rather than treating calls within a batch as independent repetitions.

\begin{table}[htbp]\centering\small
\caption{Finite application batches. Each cell schedules 128 calls per path.
Latency is the median of four batch medians; goodput is the median of four
batch values. Paired columns show IPv6 / Relay.}
\label{tab:transport-concurrency}
\begin{tabular}{llrrrr}\toprule
Payload & Conc. & IPv6 valid & Relay valid & Latency (ms) & Goodput (/s)\\\midrule
1 KiB & 1 & 128/128 & 128/128 & 577.5 / 770.6 & 1.73 / 1.29 \\
1 KiB & 4 & 128/128 & 128/128 & 1501.1 / 2460.4 & 2.62 / 1.60 \\
7 KiB & 1 & 128/128 & 128/128 & 647.5 / 835.5 & 1.54 / 1.16 \\
7 KiB & 4 & 128/128 & 128/128 & 982.7 / 1495.9 & 3.96 / 2.67 \\
\bottomrule\end{tabular}\end{table}
\begin{figure}[htbp]\centering
\includegraphics[width=\linewidth]{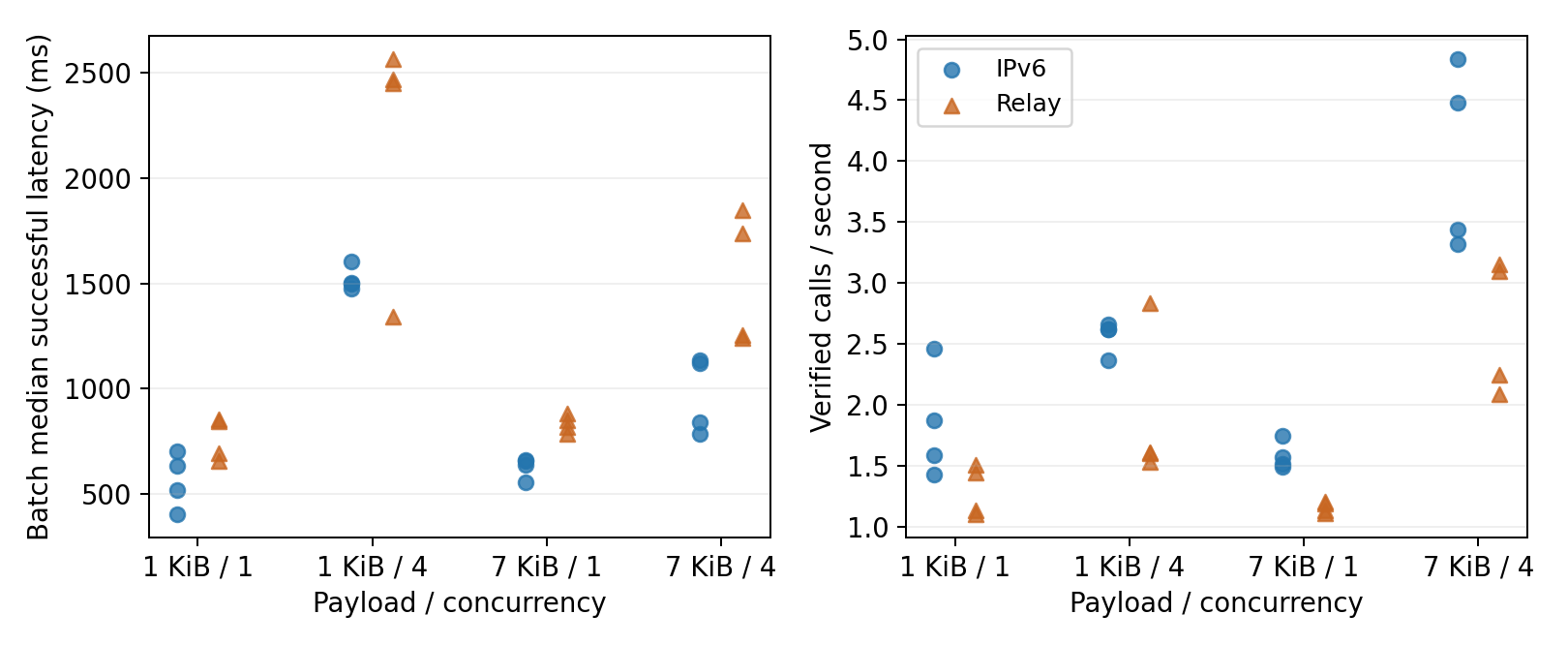}
\caption{Per-batch completion and timing summaries, with four batches per path and workload cell.}
\end{figure}

The final configuration completes all 1,024 measured calls across 32 arms. Median paired Relay-minus-IPv6 batch latencies are +183.5, +953.5, +210.6, and +535.2 ms in table order. These include the whole service path and the asymmetric laboratory bridge, so they characterize the deployed configurations rather than propagation delay alone.

The earlier complete configuration returns 1,022/1,024 successful calls; two admission-transaction deadlocks occur before SDK execution. An earlier pilot also contains two settlement failures after execution. Revised admission and settlement lock scopes preserve invocation serialization while avoiding the conflicting shared-row locks. The two complete configurations remain separate, and the final observations describe bounded concurrency after that change. An incomplete, credential-limited collection is retained separately from both complete schedules. Transport timing is compared within each complete configuration; the version change is evaluated through its admission and settlement outcomes.

\par\bigskip
\section{Workflow and Service Comparisons}
\label{app:external-evidence}
The workflow and service experiments share business operations with their respective baselines. Detailed settings identify the common execution components and the additional service mechanisms included in each path.

\subsection{Service invocation latency and memory feasibility}
\label{sec:external-service-measured}
We compare Nexus and Dapr on their shared service-invocation function. Two instances invoke the same byte-echo backends through Nexus peer IPC/tunnels or Dapr's local HTTP API. Dapr uses official version 1.18.4 runtime and Sentry binaries. Both product configurations enable mutual TLS and actual discovery on the same two-vCPU, 1-GiB OpenWrt VM; plaintext direct HTTP provides a transport reference. Fresh, externally disconnected loopback namespaces isolate each arm. Entry interfaces, discovery, certificate issuance, and supporting processes remain product-specific.

\paragraph{Invocation experiment.}
Each experiment uses all six orders of the three paths. After a ten-second startup interval and one warm-up per direction, each arm issues three requests per direction at 128, 1,024, and 8,192 bytes. Two complete experiments contain 648 measured requests and 72 separate warm-ups. The second was scheduled after a discovery failure in the first, and both are reported. Validation requires exact bytes and one opposite-backend receipt.

Nexus uses a 100-ms peer tick and same-domain discovery admission. Dapr uses its mDNS resolver, two sidecars, and Sentry, with one Go scheduler processor per process. Timing includes each local entry path, input loading, and instrumentation. Product-process RSS includes two routing daemons, two message-bus processes, and discovery for Nexus, and two sidecars plus Sentry for Dapr; backends are recorded separately.

\begin{table}[!ht]
\centering\small
\caption{Two fixed external-baseline campaigns. Latency is conditional on a
verified successful request; failed calls remain in the success denominator.
RSS sums include all product processes but can double-count shared mappings.}
\label{tab:external-service}
\begin{tabular}{@{}lrrrr@{}}\toprule
Path & Verified/scheduled & p50 (ms) & p95 (ms) & RSS sum (MiB) \\
\midrule
Direct HTTP & 216/216 & 0.144 & 0.346 & --- \\
Dapr & 207/216 & 1.239 & 2.655 & 292.26 \\
Nexus & 207/216 & 99.169 & 148.770 & 9.73 \\
\bottomrule\end{tabular}
\end{table}
\begin{figure}[!ht]
\centering\includegraphics[width=\linewidth]{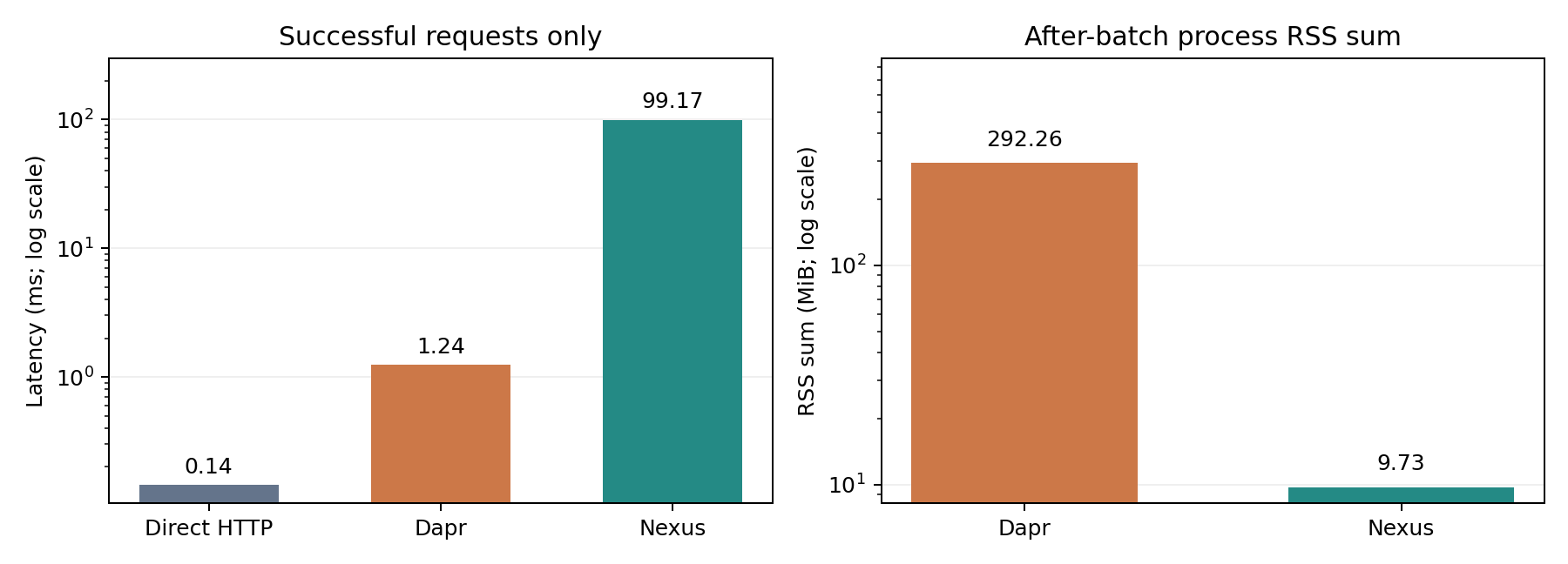}
\caption{Successful-call latency and sampled process RSS in the two service experiments, on logarithmic axes. The direct-HTTP reference has no intermediary product process; its caller and backend lie outside this RSS series.}
\label{fig:external-tradeoff}
\end{figure}

Dapr has lower successful-call latency, while Nexus has the smaller sampled process-RSS sum. Each product completes 207/216 measured calls. Nine Dapr calls fail during interface discovery; nine Nexus calls fail when the target backend cannot bind its experimental address. The underlying intermittent namespace/interface behavior remains undetermined. Timing is computed over successful calls, with the two incidents retained in their respective completion counts. RSS sums sample the listed product processes and can count shared mappings repeatedly; peak and unique memory are different quantities.

\paragraph{Equal memory budgets.}
A separate grid applies aggregate cgroup limits of one CPU and 16, 32, 64, 128, or 256 MiB, with swap disabled and a 180-second arm bound. Two opposite-order pairs per ceiling schedule eighteen measured calls and two warm-ups per arm. The resource group includes runtimes, issuer where required, backends, and measurement processes. Kernel counters identify out-of-memory termination; calls prevented from starting remain in the scheduled denominator.

\begin{table}[!ht]
\centering\small
\caption{Equal enforced memory budgets. N/D denote Nexus/Dapr; a complete arm
finishes both warmups and all 18 measured calls. Startup failures count as
uncompleted scheduled requests. Kernel OOM counts refer to arms.}
\label{tab:external-memory}
\begin{tabular}{@{}rrrrrr@{}}\toprule
MiB & N complete & D complete & N calls & D calls & OOM N/D \\
\midrule
16 & 2/2 & 0/2 & 36/36 & 0/36 & 0/2 \\
32 & 2/2 & 0/2 & 36/36 & 0/36 & 0/2 \\
64 & 2/2 & 2/2 & 36/36 & 36/36 & 0/0 \\
128 & 2/2 & 2/2 & 36/36 & 36/36 & 0/0 \\
256 & 2/2 & 2/2 & 36/36 & 36/36 & 0/0 \\
\bottomrule\end{tabular}
\end{table}
\begin{figure}[!ht]
\centering\includegraphics[width=.90\linewidth]{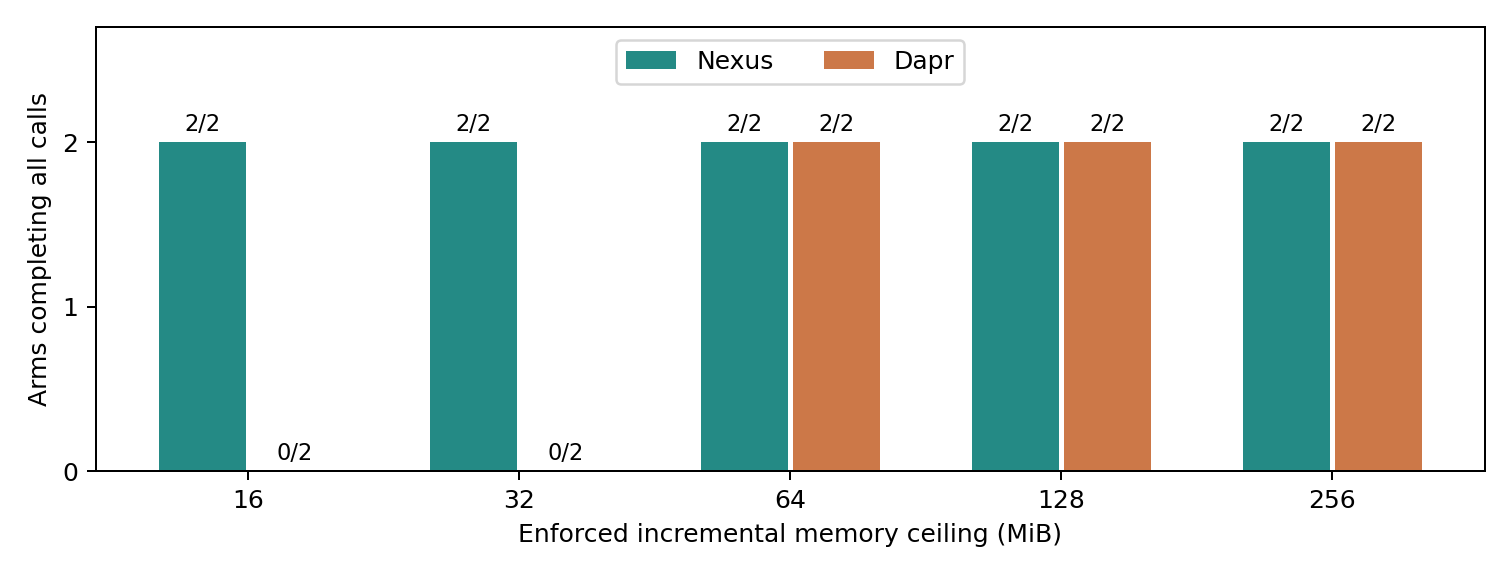}
\caption{Completed workloads under equal incremental-runtime memory ceilings. Twenty arms cover five ceilings and two opposite-order pairs per ceiling.}
\label{fig:external-memory}
\end{figure}

Nexus completes all ten arms, including both at 16 MiB, with incremental peaks of 10.31--10.77 MiB. Dapr terminates through kernel OOM at 16 and 32 MiB and completes both arms at each higher ceiling, with successful-arm peaks of 61.94--64.98 MiB. The grid therefore identifies 16 and 64 MiB as the smallest tested feasible ceilings. Executables are pre-staged in host memory, allowing some shared pages to be charged outside the new groups. These are warm incremental-runtime measurements. Taken with the separate unconstrained invocation experiment, they describe service latency and memory feasibility at different operating points; an equal-latency resource comparison would require both measurements under the same service requirement.

\subsection{Fixed-policy Agent workflows with a shared device backend}
\label{sec:dify-agent-measured}
A common model-and-tool configuration allows us to compare workflow execution while controlling the task interface. Nexus runs a direct SDK loop; Dify 1.17.1 uses native loop, LLM, code, and HTTP nodes with provider version 0.0.68. Both call the same Qwen/Qwen3-8B source through the Nexus gateway and the same authenticated Computer and Android tools. The shared adapter executes individual actions; all model decisions remain in the respective workflow.

Twelve synthetic tasks comprise four expense summaries, four inventory reconciliations, and four Android checklists. Two opposite-order repetitions per task yield 24 pairs and 48 serial arms. Inputs, two-message policy, temperature zero, 256-token response limit, disabled thinking, and twelve-decision budget are matched. Fresh paths and a reset Android Home state prevent cross-arm reuse. Invalid actions consume the budget, and neither inference nor transport retries are added. Three tasks were used for prompt calibration; the nine remaining tasks are reported separately. Success requires an independently validated report, write/readback hash, device records, and matching confirmation.

\begin{table}[!ht]
\centering\small
\caption{Nexus and Dify workflow outcomes. Non-calibration counts cover the nine tasks not used to revise the prompt. Attempt duration includes successful and failed termination; model calls and invalid actions include failed decisions.}
\label{tab:dify-agent}
\begin{tabular}{@{}lrrrrr@{}}\toprule
Implementation & Success & Non-calib. & Attempt p50 (s) & Calls & Invalid \\
\midrule
Nexus SDK & 24/24 & 18/18 & 25.06 & 159 & 9 \\
Dify + shared tools & 22/24 & 16/18 & 31.10 & 168 & 15 \\
\bottomrule\end{tabular}
\end{table}

Nexus completes 24/24 tasks and Dify 22/24. Among the 22 jointly successful pairs, the median Dify-minus-Nexus completion-time difference is 3.88 s and the mean is 4.80 s. The median is 3.67 s among sixteen successful non-calibration pairs and 3.42 s in eighteen equal-call-count pairs. Figure~\ref{fig:dify-agent-paired} presents observed completion times and model decisions. These differences include independently realized model outputs and upstream waiting, even when call counts match.

The two unsuccessful Dify executions stop at different stages. One reaches the shared confirmation tool, which becomes unavailable for an undetermined transient reason. The other writes a report but exhausts the model-decision budget on malformed or invalid subsequent actions. Neither incident is localized to Dify's core. The nominal 540-second budget is cooperative: both implementations use a 300-second model-read cap but differ in remaining-budget handling and cancellation. The completion-time differences consequently include these workflow policies as well as the independently realized model requests.

Nexus usage links 115,235 tokens to Gateway request identities; Dify nodes report 124,367 tokens without the same direct request-ID linkage. Both use a zero configured model tariff. The resource series also has a specific scope: 47 complete Dify cgroup snapshots have median 2,768.4 MiB, while 28 individual snapshot errors leave some group totals unavailable. The snapshots include the selected fifteen-service profile and charged cache. Token counts and resource observations are therefore reported by their source and accounting boundary, separately from the paired task outcomes; they do not measure a complete-product memory or billing difference.

\par\bigskip
\section{Environment Abstraction and Action Semantics}
\label{app:environment-evidence}
These experiments follow an action from invocation-scoped authority through device execution and result delivery. Resource adapters, live grant changes, retained results, browser observations, and terminal state expose different aspects of that interaction.

\subsection{Application-side resource adaptation}
\label{sec:sdk-resource}
A shared SDK can remove protocol-handling code from a business workflow while retaining its device semantics. We compare SDK helpers with handwritten HTTP adapters for Computer reads, Android open/observe actions, asynchronous confirmation, and artifact write/readback. Both retain the same hosted handler, context injection, credentials, TLS support, conditional recovery interface, and Display publisher. Only the application-side resource adapter changes; SDK hosting and the shared service infrastructure remain in both paths.

Eight pairs use four 1-KiB and four 64-KiB inputs with identical within-pair bytes, devices, grants, and distinct output paths. Workspace timeout is 35 seconds, Mobile deadline thirty seconds, command polling 250 ms, and question polling 500 ms; each Run has a 120-second observation bound. The HTTP adapter retains bounded retry and stable command identities. These deterministic tasks make no model requests.

\begin{table}[htbp]
\centering\small
\caption{Resource integration. Times are seconds; medians cover four
Runs per row. Request counts cover the instrumented handler through readback,
before final output publication, rather than all platform traffic.}
\label{tab:sdk-resource}
\begin{tabular}{@{}llrrrr@{}}
\toprule
Input & Adapter & Exact output & End-to-end & Handler & Requests \\
\midrule
1 KiB & SDK & 4/4 & 12.797 & 6.335 & 21--25 \\
1 KiB & HTTP & 4/4 & 13.088 & 6.664 & 23--25 \\
64 KiB & SDK & 4/4 & 13.051 & 6.553 & 23--25 \\
64 KiB & HTTP & 4/4 & 13.047 & 6.192 & 21--28 \\
\bottomrule
\end{tabular}
\end{table}
\begin{figure}[htbp]
\centering
\includegraphics[width=\linewidth]{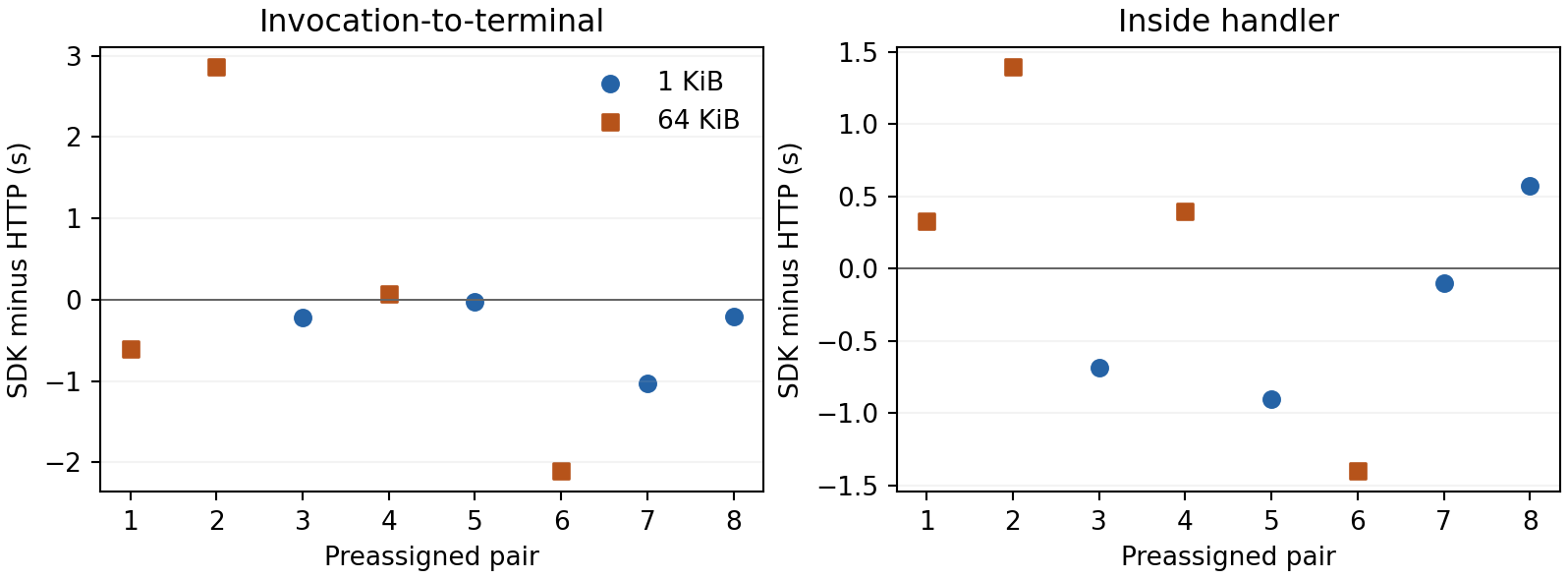}
\caption{Paired SDK-minus-HTTP duration differences. Positive values indicate longer SDK execution.}
\label{fig:sdk-resource-duration}
\end{figure}

All sixteen workflows produce exact artifacts, with 48 successful Computer commands, 32 Mobile commands, and sixteen answered confirmations. Excluding comments, docstrings, imports, and shared infrastructure, the authored resource adapters contain fourteen nonblank physical lines with SDK helpers and 99 with HTTP. Both use the same 62-line orchestration and measurement harness. The count measures protocol code authored for this adapter, excluding shared dependencies; developer time is not measured.

In this initial deployment the operation journal is inactive in both arms, and no operation records are created. Moreover, randomized order places the SDK first in seven of eight pairs. The observed durations therefore describe this integration configuration without isolating adapter latency or journal cost. Section~\ref{sec:sdk-active} reports a separate, order-balanced configuration with active journaling and device-compatible operation identities.

\subsection{Operation-identity consistency across resource adapters}
\label{sec:sdk-active}
Heterogeneous interfaces may encode request identities differently, but recovery requires each device request to remain linked to the same logical operation. We evaluate this linkage with an active operation journal. Both SDK and HTTP adapters deterministically derive device-compatible identifiers from journal operation keys, using the same mapping, handler, caller, permissions, and devices. This configuration differs from the earlier inactive-journal deployment, so its timing is analyzed within the new comparison.

Eight balanced pairs span 1-KiB and 64-KiB inputs, with two pairs in each execution order per size. Every workflow reads Computer data, opens and observes Android, obtains a confirmation, and writes and verifies its final artifact. All sixteen Runs complete correctly, producing 48 successful Computer commands, 32 Mobile commands, and sixteen confirmations.

\begin{table}[htbp]
\centering\small
\caption{Active-journal resource integration. Each row contains four Runs; times are median seconds. Request counts use the pre-publication boundary defined in Table~\ref{tab:sdk-resource}.}
\label{tab:sdk-active}
\begin{tabular}{@{}llrrrr@{}}
\toprule
Input & Adapter & Exact output & End-to-end & Handler & Requests \\
\midrule
1 KiB & SDK & 4/4 & 15.474 & 8.382 & 39--41 \\
1 KiB & HTTP & 4/4 & 13.924 & 7.222 & 37--40 \\
64 KiB & SDK & 4/4 & 14.466 & 7.685 & 34--41 \\
64 KiB & HTTP & 4/4 & 14.723 & 6.852 & 35--39 \\
\bottomrule
\end{tabular}
\end{table}

The database contains 128 successful journal entries: eight ordered operations per Run covering presentation, Mobile access, interaction events, artifact write, and final output. Every Computer write uses its associated journal key, and all 32 Mobile request identifiers match the deterministic mapping from their journal entries. Matching the stored operation keys to actual device requests establishes the linkage from recorded recovery state to device execution.

A preliminary active-context run rejects Mobile identifiers that do not satisfy the device request format. The paired experiment uses the same compatible mapping in both adapters and records the device--operation linkage on healthy paths. Fault-induced reuse and effect uniqueness are examined in Sections~\ref{sec:worker-recovery}, \ref{sec:ambiguous-effect}, and~\ref{sec:mobile-result-recovery}. The authored adapters contain fourteen and 99 lines under the common counting rule. Their whole-workflow median ordering changes by input size; journal-overhead estimates are confined to the journal on/off experiment because the earlier inactive-journal deployment also differs in SDK and service context.

\subsection{Model-directed composition of shared resource adapters}
\label{sec:priority-real-composition}
To include actual model interaction, eight balanced SDK/HTTP pairs execute a two-decision Qwen workflow. Four pairs use 1-KiB Computer inputs and four use 64 KiB. The first model decision selects a composite inspection-and-recording tool, which reads Computer data, opens and observes Android Settings, obtains a scripted confirmation, and writes and rereads an artifact. The second decision summarizes the observed application and artifact equality.

Both arms use the same image, Qwen/Qwen3-8B source, prompt, temperature zero, disabled thinking, 128-token output limit, device permissions, and active journal. Each model call has a ninety-second timeout and no automatic retry. The HTTP adapter shares SDK hosting, trust, and recovery infrastructure, and both arms execute fresh inference. The task exercises JSON selection of one composite tool, followed by a result-summary decision.

\begin{table}[htbp]\centering\small
\caption{Real Qwen workflow comparison. Four independent arm executions per
row. Times are medians over attempted workflows, in seconds. ``Model'' sums
two inference calls; ``Handler'' measures inside the hosted composite tool.}
\label{tab:priority-composition}
\begin{tabular}{rlrrrr}\toprule
KiB & Adapter & Exact/attempts & Workflow s & Model s & Handler s\\\midrule
1 & SDK & 4/4 & 30.825 & 8.675 & 9.412 \\
1 & HTTP & 4/4 & 27.374 & 5.586 & 9.106 \\
64 & SDK & 4/4 & 30.924 & 9.572 & 8.228 \\
64 & HTTP & 4/4 & 30.221 & 9.574 & 9.187 \\
\bottomrule\end{tabular}\end{table}
\begin{figure}[htbp]\centering
\includegraphics[width=\linewidth]{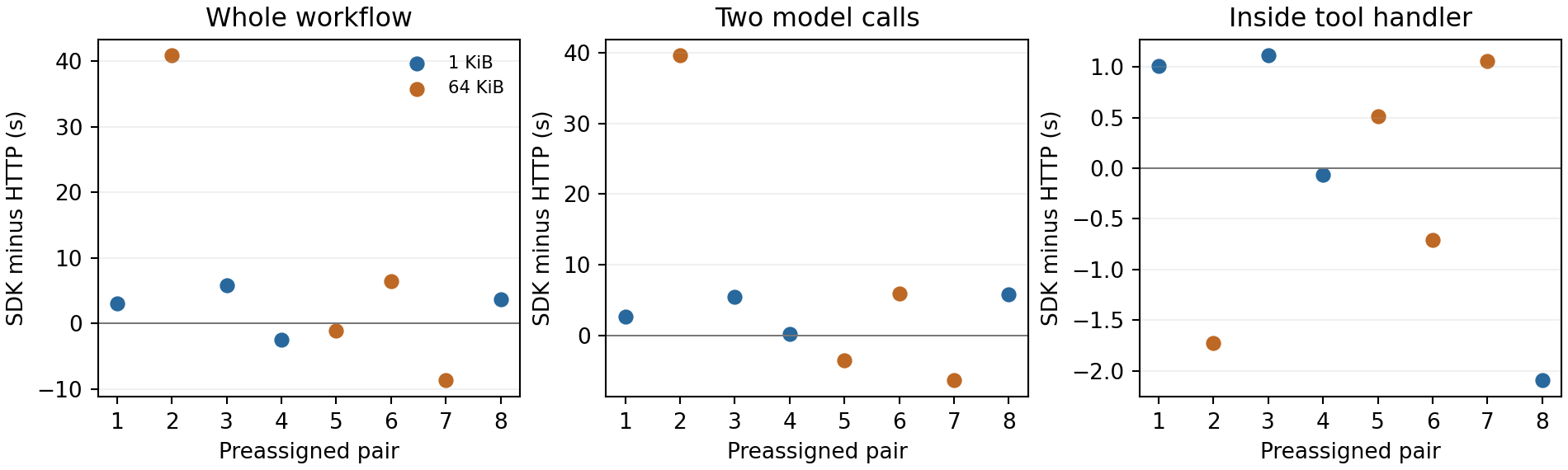}
\caption{Paired SDK-minus-HTTP differences in end-to-end, model-request, and composite-tool time. Each point represents one independently executed workflow pair.}
\label{fig:priority-composition}\end{figure}

All sixteen workflows pass independent artifact and device validation, yielding 48 Computer commands, 32 Mobile commands, and 128 journal operations. All 32 model requests match Gateway records and total 5,256 tokens. Each adapter uses 327 tokens for a 1-KiB workflow and 330 for a 64-KiB workflow. The timing decomposition separates model-request intervals from handler work; the reported whole-workflow differences combine adapter execution with upstream variation.

The model controller and hosted tool use distinct tenant scopes, linked here by the experimental request trace rather than automatic model-charge attribution to the tool Run. Request-body instrumentation ends before output publication and excludes response and wire bytes. These conditions define a model-and-device integration result with unchanged observed token use and a zero configured model tariff.

\subsection{Delegation cost during healthy and interrupted connectivity}
\label{sec:matched-mobile}
Delegation adds service scheduling and authorization to device execution. We measure that cost by invoking the same Android Settings open-and-observe task through the authenticated Cloud device API and through a hosted Agent. The latter additionally performs worker scheduling, MCP invocation, run-scoped delegation, and reporting. Both use the same emulator, executor, package/heading validator, and foreground reset.

Ten balanced pairs run under healthy connectivity and ten under a nominal ten-second outage. The outage starts before submission and drops only emulator traffic to the Cloud reverse connection. Device permissions remain unchanged. Command waits are bounded to thirty seconds, hosted completion observation to ninety seconds, and caller polling to 250 ms. Time begins before submission and includes queueing and observation.

\begin{table}[htbp]
\centering\small
\caption{Same-device workflow outcomes. Completion requires both successful termination and the independent Settings oracle. Time is the median among successful attempts.}
\label{tab:matched-mobile}
\begin{tabular}{@{}llrrr@{}}
\toprule
Condition & Path & Successful & Failed & Time (s) \\
\midrule
Healthy & Device API & 10/10 & 0 & 4.872 \\
Healthy & Agent delegate & 10/10 & 0 & 15.236 \\
10-s loss & Device API & 10/10 & 0 & 14.799 \\
10-s loss & Agent delegate & 10/10 & 0 & 15.441 \\
\bottomrule
\end{tabular}
\end{table}

Both paths complete all ten healthy and all ten interrupted workflows. Healthy paired added time is 10.503 s (95\% interval $[9.534,11.705]$ s); under loss it is 0.797 s ($[0.271,1.361]$ s). All twenty outage arms record dropped packets, with actual intervals of 10.165--10.332 s. Because the outage begins before submission, worker queueing can overlap device disconnection and account for part of the smaller added time. Queueing and reconnection are not varied independently. The intervals resample the ten complete pairs and describe the experienced cost of the additional admission, delegation, and reporting path.

\paragraph{Availability after inactivity.}
Three separate scenarios submit fresh authorized commands after the inactivity intervals shown below. Device heartbeats and polling continue, while no device command occurs during the interval.

\begin{table}[htbp]
\centering\small
\caption{One observed run per idle interval; scenario outcomes without statistical intervals.}
\label{tab:idle}
\begin{tabular}{@{}rrrrl@{}}
\toprule
Target idle (s) & Actual idle (s) & Completion (s) & Commands & Oracle \\
\midrule
60 & 62.607 & 16.636 & 2 & Pass \\
600 & 601.205 & 10.020 & 2 & Pass \\
3600 & 3601.081 & 11.714 & 2 & Pass \\
\bottomrule
\end{tabular}
\end{table}

The post-idle observations submit new calls while heartbeat and polling services remain active. The one-hour observation uses a later Cloud context implementation and is reported separately. In preparation, a force-stop removes the Accessibility binding and prevents both paths from executing. That outcome identifies device authorization as a prerequisite for reconnection to support useful work.

\subsection{Computer execution through two admission interfaces}
\label{sec:measured-computer}
The same Computer workflow is reachable through chat-run and tool-invocation admission. A hosted handler reads caller-selected input, computes SHA-256, writes a JSON artifact, reads it back, and publishes the digest. An independent device-side reader checks byte count, digest, nonce, exact JSON, and agreement with the displayed result. Both paths preserve the same non-ASCII workspace, capabilities, handler, and input bytes.

Ten balanced pairs per input size (1 KiB and 64 KiB) use 250-ms polling and a ninety-second observation bound. Durations include admission, worker scheduling, device operations, and completion observation. The measured quantity is completion through each admission and observation path; browser painting and SDK-only execution are not timed separately.

\begin{table}[htbp]
\centering\small
\caption{Same-Computer API comparison. All 40 attempts completed and passed
the independent artifact oracle. Paired differences are chat minus tool.}
\label{tab:computer-extended}
\begin{tabular}{@{}lrrrr@{}}
\toprule
Input & Pairs & Chat median (s) & Tool median (s) & Paired median (s) \\
\midrule
1 KiB & 10 & 10.634 & 9.239 & 0.702 \\
64 KiB & 10 & 9.320 & 8.420 & 0.955 \\
\bottomrule
\end{tabular}
\end{table}

The paired median-difference intervals are $[0.239,1.368]$ s for 1 KiB and $[0.658,1.756]$ s for 64 KiB, obtained from 5,000 resamples of whole pairs. The two input sizes retain separate estimates because their observed differences also include shared-host scheduling.

The initial forty-attempt schedule completed eight requests and rejected 32 at the tenant's rolling admission quota. A separate repeat uses a dedicated 5,000-run quota instead of 100; its successful observations are not pooled with the quota-limited experiment. Non-ASCII context transport was corrected before the reported repeat, with both entry paths using the same implementation. This separates task execution behavior from admission capacity and context compatibility.

\subsection{Continuation after a Computer Runtime interruption}
\label{sec:computer-restart}
A pending device operation may outlive the Runtime process that initially served its connection. Ten balanced pairs test this case with the same 1-KiB read/write/read workflow. The control keeps the Runtime alive; the treatment stops it immediately before submission and starts a replacement ten seconds later with the same paired identity. Cloud, worker, Agent, permissions, input, and validator remain unchanged. Completion is observed for ninety seconds.

All twenty Runs complete with exact artifacts and no handler retry. Healthy and interrupted median durations are 7.375 and 14.473 s. The paired median additional time is 7.148 s, with a 95\% pair-bootstrap interval of $[6.830,7.360]$ s. Actual restart gaps are 10.001--10.012 s. Because interruption begins before admission, worker queueing overlaps part of the device absence.

\begin{figure}[htbp]
\centering
\includegraphics[width=\linewidth]{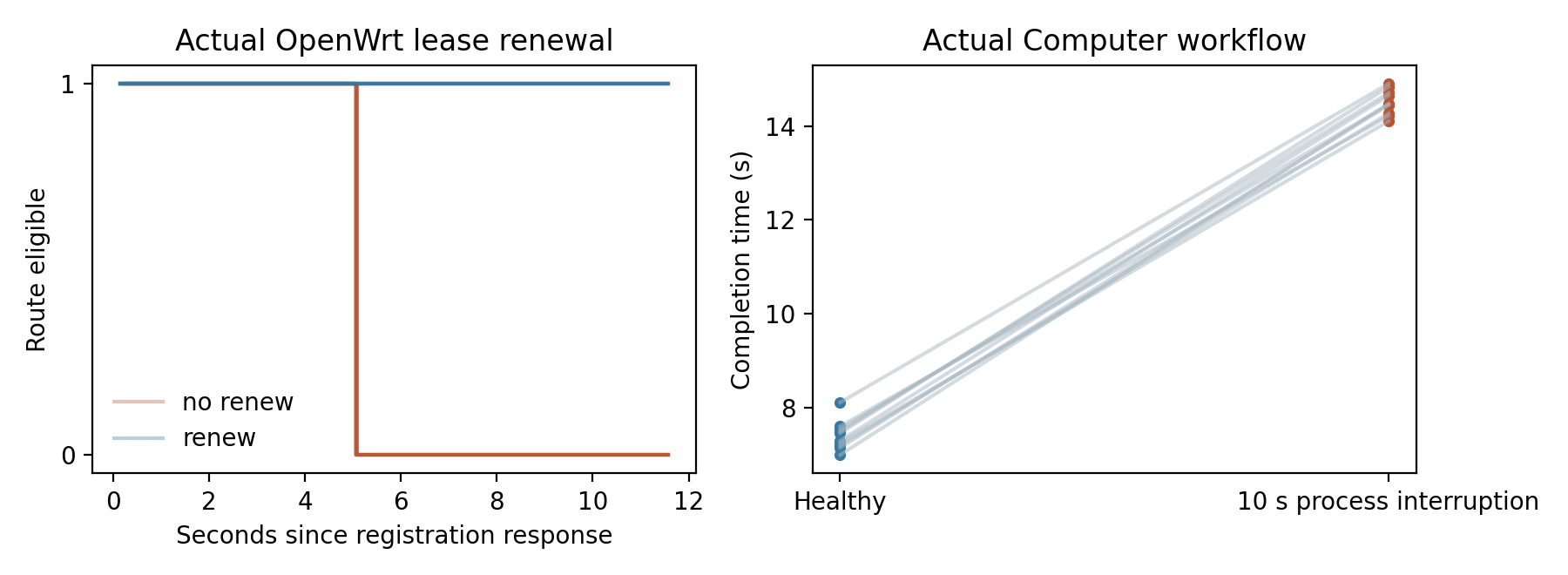}
\caption{Left: route eligibility with and without lease renewal, with ten sampled traces per arm. Right: paired Computer workflows with and without a ten-second Runtime interruption.}
\label{fig:recovery-extensions}
\end{figure}

The reported SDK uses a 35-second workspace timeout, covering the Cloud command wait and result observation. A preceding pilot using a three-second timeout failed its interrupted arm before the Runtime returned. The resulting experiment demonstrates completion of a waiting operation after reconnection within its deadline; the Agent handler itself survives. Across this study and the separate Computer-interface experiment, 180 command records correspond to two reads and one write per successful Run. Result-loss recovery and repeated file replacement require additional observations, provided in Section~\ref{sec:computer-ack}.

\subsection{Cross-environment composition within one invocation}
\label{sec:measured-composition}
A single reporting task combines five operations: reading the caller's Computer file, opening Android Settings, observing its package and heading, writing a Computer artifact, and reading that artifact back. The artifact records the input digest, byte count, nonce, and observed device state. Run-scoped grants bind the same Agent to the selected Computer and Mobile environments.

A shuffled ten-attempt schedule uses five 1-KiB and five 64-KiB inputs with a ninety-second observation bound and the workspace-timeout configuration of Section~\ref{sec:computer-restart}. All ten attempts complete with exact artifacts and no handler retry. Independent command records show thirty Computer operations and twenty Mobile operations with the intended device identities. Median end-to-end durations are 11.366 and 11.462 s for the two sizes.

The deterministic handler removes model variability from this composition experiment. One invocation connects the Computer-derived digest, Android observation, and final artifact under the same resource bindings. The times characterize this workflow on two environments sharing the experimental host; there is no alternative implementation arm or independent-device network in this experiment.

\subsection{A model--tool--observation loop on the selected device}
\label{sec:real-agent}
A local reasoning loop calls Qwen/Qwen3-8B to choose a Settings inspection action, executes it through the selected adapter, and returns the actual bounded observation to the model. The second response is checked for the observed package and heading. Both arms use temperature zero, disabled thinking, a 128-token output limit, and fresh inference; the adapter is absent from the prompt.

Five paraphrases of this one semantic task are repeated twice per arm, producing twenty scheduled attempts with balanced shuffled order. Each attempt permits two model requests, each with a ninety-second deadline and no model substitution.

\begin{table}[htbp]
\centering\small
\caption{Real Qwen-driven micro-workflow. Ten attempts per adapter (five task paraphrases, two repeats).}
\label{tab:matched-live}
\begin{tabular}{@{}lrrr@{}}
\toprule
Path & Successful & Failed & Median time (s) \\
\midrule
Device API & 10/10 & 0 & 11.590 \\
Agent delegate & 10/10 & 0 & 20.665 \\
\bottomrule
\end{tabular}
\end{table}
\begin{figure}[htbp]
\centering
\includegraphics[width=\linewidth]{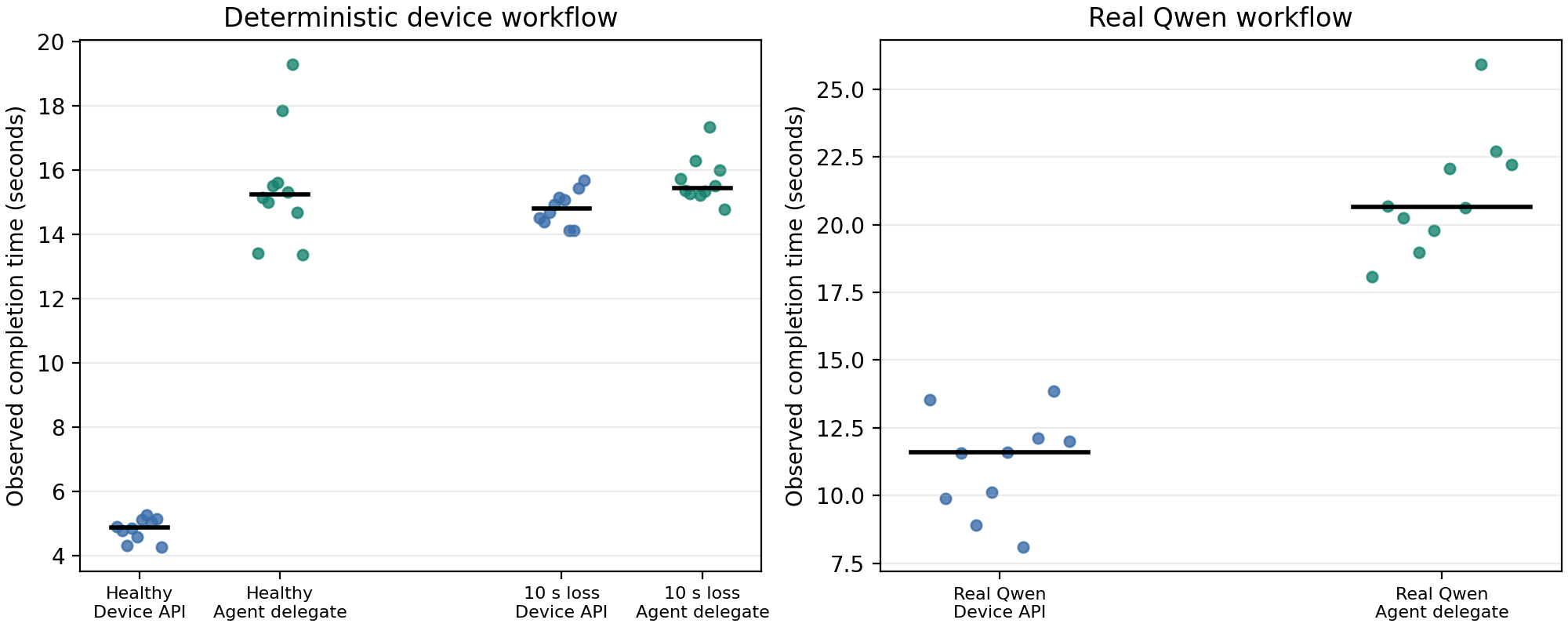}
\caption{Successful workflow completion times, with one point per execution and median bars. Left: the deterministic device task under healthy and interrupted connectivity. Right: fresh Qwen requests around the same device task. Tables retain all attempt denominators.}
\label{fig:workflow-latency}
\end{figure}

All forty model requests return successfully with usage records, totaling 4,588 input and 360 output tokens. The table reports independently validated workflow outcomes separately from successful model responses. Five paraphrases provide repeated linguistic forms of one semantic task, and timing includes both model waiting and the selected submission path. Usage is provider-reported under a configured free route; upstream invoices are not measured.

\subsection{Authorization at invocation and action-dispatch boundaries}
\label{sec:measured-authority}
Authorization must remain meaningful after a Run has been admitted. We use the Computer file workflow to distinguish initial grants, display access, and revalidation before later operations. Inputs, deployed handler, device, and output validator remain fixed while caller grants change through the service interface.

\paragraph{Initial grants and result access.}
Ten scenarios have three repetitions each. Full and restored grants complete all six corresponding workflows. Missing-write and revoked grants prevent artifacts in all six attempts; missing-write Runs execute one read, whereas fully revoked Runs create no device commands. These checks include SDK enforcement. Display scenarios separately require the correct Run, tenant context, and authenticated caller as well as a valid proof.

\begin{table}[htbp]
\centering\small
\caption{Authorization scenarios with three repetitions per row. Outcomes distinguish action execution from permission-consistent refusal.}
\label{tab:authority-scenarios}
\begin{tabularx}{\linewidth}{@{}>{\raggedright\arraybackslash}X>{\raggedright\arraybackslash}Xr@{}}
\toprule
Condition & Observed outcome & Count \\
\midrule
Full grant / restored grant & Run completes, exact artifact & 3/3 each \\
Missing write permission & Run fails, read occurs, no write & 3/3 \\
Grant revoked before submission & Run fails, no device command & 3/3 \\
Valid Display proof & HTTP 200, intended Run visible & 3/3 \\
Missing / wrong-Run / tampered proof & HTTP 404, no Run payload & 3/3 each \\
Wrong tenant with valid proof & HTTP 404, no Run payload & 3/3 \\
Unauthenticated with valid proof & HTTP 401, no Run payload & 3/3 \\
\bottomrule
\end{tabularx}
\end{table}

A same-tenant API-key principal is then tested against a human caller's completed Run. It cannot read with the human's proof, read without a proof, or obtain a new proof; each condition returns three denials. Interleaved owner reads succeed, giving twelve expected observations. Another three-observation scenario tests natural proof expiry: initial access succeeds, the old proof fails after 3,605.624 s, and a newly issued proof restores access. Caller login is refreshed before the expiry probe, isolating the display proof's lifetime.

\paragraph{Live grant changes.}
Six balanced pairs start with full workspace authority and an interrupted Computer Runtime. After the first authorized read is queued, the treatment removes workspace grants and the control reaffirms them. Both restart the same Runtime, while the handler retains its initially captured context. All six controls complete. All six revoked-grant Runs finish the queued read but fail before a write command is created; no artifact appears. Recorded times place revocation between read creation and completion.

The shared Runtime interruption lets the grant change determine the difference in subsequent dispatch. The queued read completes under its earlier authorization, while current grants govern creation of the later write. Operation records locate the stopping point at that dispatch decision. The separate same-tenant experiment distinguishes a human principal from an API-key principal; it does not vary authority between two human accounts.

\subsection{Credential invalidation and authorized reconnection}
\label{sec:mobile-credential}
Network recovery cannot restore an invalid credential. The Android client therefore backs off on transient transport errors but stops synchronization after authentication rejection. Six balanced pairs compare a retained device token with a token rotated at the server while the running client still holds the old credential. Both enqueue the same Settings action with a 120-second lifetime after confirming that the device can execute an observation.

For ten seconds, the experiment observes command and client state without restarting the application or changing network connectivity. It then probes the old token. Rotated-token arms explicitly re-pair through the normal device interface and continue observing the same queued command; a fresh observation verifies the resulting Settings state.

\begin{table}[htbp]
\centering\small
\caption{Android credential intervention, six arms per condition. Completion after re-pairing concerns the same queued command; it is explicit repair, not automatic retry.}
\label{tab:mobile-credential}
\begin{tabular}{@{}lrrrr@{}}
\toprule
Condition & Completed at window end & Old token rejected & Auth failure & Final success \\
\midrule
Retain & 6/6 & 0/6 & 0/6 & 6/6 \\
Rotate & 0/6 & 6/6 & 6/6 & 6/6 \\
\bottomrule
\end{tabular}
\end{table}

Retained credentials complete 6/6 commands during the 10.125--11.155-s observation windows. Rotated credentials complete 0/6; all old-token probes are rejected and all clients enter authentication-failed state. After explicit re-pairing, all six original commands complete and all six observations identify Settings. Completion follows explicit restoration of authority: the six rotated-token clients resume their original queued commands only after re-pairing.

A separate unpaired-token scenario waits naturally beyond its ten-minute expiry. After 604.976 s, the old token is rejected; a fresh token for the same record permits pairing and actual open/observe execution. The ten-minute lifetime belongs to the unpaired credential; established device sessions use their own credentials. Earlier preparatory runs contained Accessibility failures associated with observation during re-pairing; the reported comparison performs UI inspection before confirmation and uses noninterfering state observation afterward. The six pairs and one expiry scenario characterize these explicit credential transitions on the emulator.

\subsection{Action completion without durable result retention}
\label{sec:mobile-ack}
A device can execute an action while Cloud still lacks its result. We examine this distinction using the baseline Android client without a durable result outbox. Six balanced blocks each contain healthy transport, result loss, and acknowledgment loss. The same Accessibility-driven Apply action writes an independent on-device click record and visible marker; the record observes effects but does not suppress duplicates.

A command-specific proxy either forwards traffic, discards the result before Cloud receives it, or forwards the result and discards Cloud's acknowledgment. Each arm is observed for at least thirty seconds after the first result submission. A fresh observation is submitted after five seconds. After the fault window, result-loss arms receive an explicit new Apply command, while the original command identity remains unchanged.

\begin{table}[htbp]
\centering\small
\caption{Android effect/result boundary, six attempts per condition. Original command state is distinguished from successful new work. Only the result-lost condition subsequently issues an explicit new Apply command.}
\label{tab:mobile-ack}
\begin{tabular}{@{}lrrrr@{}}
\toprule
Condition & Original complete & Unresolved & Fresh work & Clicks: initial to final \\
\midrule
Healthy & 6/6 & 0/6 & 6/6 & 6 to 6 \\
Result lost & 0/6 & 6/6 & 6/6 & 6 to 12 \\
Acknowledgment lost & 6/6 & 0/6 & 6/6 & 6 to 6 \\
\bottomrule
\end{tabular}
\end{table}

Healthy and acknowledgment-loss commands complete in 6/6 cases each. With the result lost, all six original commands remain running despite six recorded clicks. All eighteen fresh observations succeed. The six explicit new-command retries complete and increase the result-loss click count from six to twelve, leaving the originals unresolved. Actual windows span 30.214--31.698 s, and each condition has six wire result attempts.

The experiment separates connection liveness, device execution, and recorded completion. Acknowledgment loss leaves Cloud's result intact. Result loss instead leaves the original command pending; submitting a new Apply command repeats the device effect without completing that original record. This historical client/server configuration motivates the same-server outbox comparison in Section~\ref{sec:mobile-result-recovery}, but its outcomes are analyzed separately. The debug action provides a directly countable effect; other applications and power-loss durability are not tested.

\subsection{Durable receipt replay and subsequent-command delay}
\label{sec:mobile-result-recovery}
A durable receipt can reconcile an executed action without issuing a second action. We compare no-outbox and encrypted-outbox Android clients against the same idempotent receipt server. Both use the same Accessibility action and independent, non-deduplicating click record. The outbox client persists one pairing-bound completion receipt before reporting, retries it before polling another command, and clears it after acknowledgment.

Two opposite-order pairs cover each of four conditions: healthy transport, result loss, acknowledgment loss, and result loss followed by process replacement. Faults select the original command and use a twelve-second gate. A fresh observation is submitted after five seconds. Process replacement occurs after the first result submission, when the receipt is already durable. Each arm continues for at least 45 seconds after release, without creating a replacement Apply command.

\begin{table}[!ht]
\centering\small
\caption{Android receipt-recovery outcomes, with two pairs per condition. Completion refers to the original command; subsequent-observation time is median created-to-completed duration in seconds.}
\label{tab:mobile-result-recovery}
\begin{tabular}{@{}lrrrr@{}}\toprule
 & \multicolumn{2}{c}{Original completed} & \multicolumn{2}{c}{Fresh observe (s)} \\
Scenario & Baseline & Outbox & Baseline & Outbox \\
\midrule
Healthy & 2/2 & 2/2 & 0.82 & 1.19 \\
Result loss & 0/2 & 2/2 & 1.72 & 6.86 \\
ACK loss & 2/2 & 2/2 & 0.66 & 7.43 \\
Result loss + process exit & 0/2 & 2/2 & 4.27 & 5.74 \\
\bottomrule\end{tabular}
\end{table}
\begin{figure}[!ht]
\centering\includegraphics[width=\linewidth]{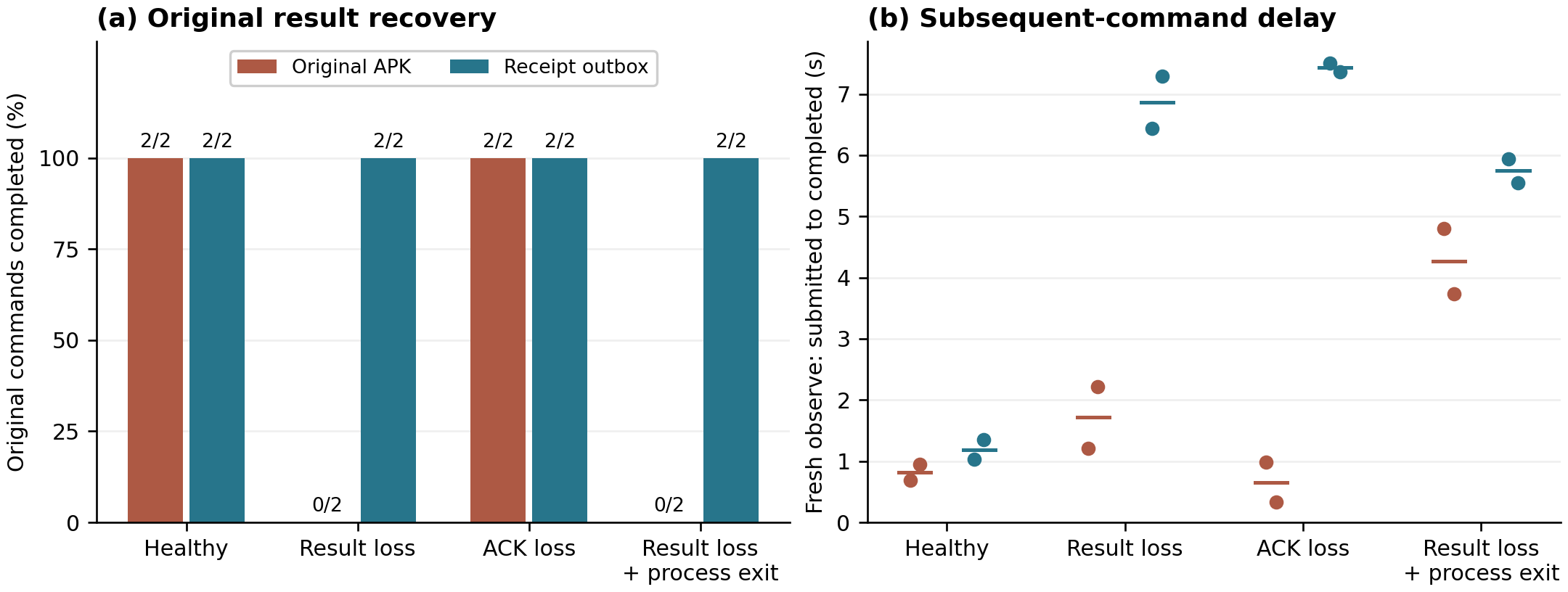}
\caption{Original-command recovery and the delay of subsequent observations. Points are individual durations and horizontal marks are medians of two observations.}
\end{figure}

The outbox client completes all eight original commands and clears its pending receipt in every arm. Across both clients, all sixteen actions produce exactly one click, all subsequent observations complete, and all markers match. Each of six outbox fault arms receives an accepted post-release receipt. In contrast, the no-outbox client leaves both result-loss originals and both result-loss/process-replacement originals unresolved. Wire-body identity, command records, process identities, and the click log independently connect replay with the same action.

Waiting for receipt acknowledgment delays later commands. Under acknowledgment loss, Cloud has already completed the command, so replay reconciles the client's pending receipt while preserving that completion. Under result loss, the retained receipt supplies Cloud with the missing result of the original action. All injected failures follow durable receipt publication and preserve pairing authority. The action-to-persistence interval, changes in pairing authority, and permanent receipt rejection are not exercised here. These eight pairs share one server and are analyzed separately from the earlier client/server configuration.

\subsection{Retained Computer results across connection and process loss}
\label{sec:computer-ack}
Computer recovery combines Cloud redelivery of an unfinished command with Runtime retention of its completed result. Eight balanced blocks test healthy execution, result loss, result loss plus Runtime replacement, and acknowledgment loss. All use the same paired device, Agent, grants, 1-KiB read/write/read workflow, and ninety-second observation bound.

A proxy selects the artifact-write command. Result-loss conditions discard its responses and close the connection for twelve seconds after the first result frame. The restart condition additionally replaces the Runtime while preserving pairing and its encrypted journal. Acknowledgment loss instead forwards the result, observes Cloud acceptance, and drops the reply. The Agent Run remains live, and recovery uses the original command identity.

\begin{table}[htbp]
\centering\small
\caption{Computer file workflow, eight attempts per condition. Deliveries count repeated transmission of the same write command. Stable object means the post-write file identity and timestamps match every later result and the final artifact.}
\label{tab:computer-ack}
\begin{tabular}{@{}lrrrr@{}}
\toprule
Condition & Completed & Deliveries & Stable object & Median (s) \\
\midrule
Healthy & 8/8 & 8 & 8/8 & 7.854 \\
Result lost & 8/8 & 32 & 8/8 & 24.001 \\
Result lost + process restart & 8/8 & 32 & 8/8 & 23.698 \\
Acknowledgment lost & 8/8 & 8 & 8/8 & 9.256 \\
\bottomrule
\end{tabular}
\end{table}
\begin{figure}[htbp]
\centering
\includegraphics[width=.92\linewidth]{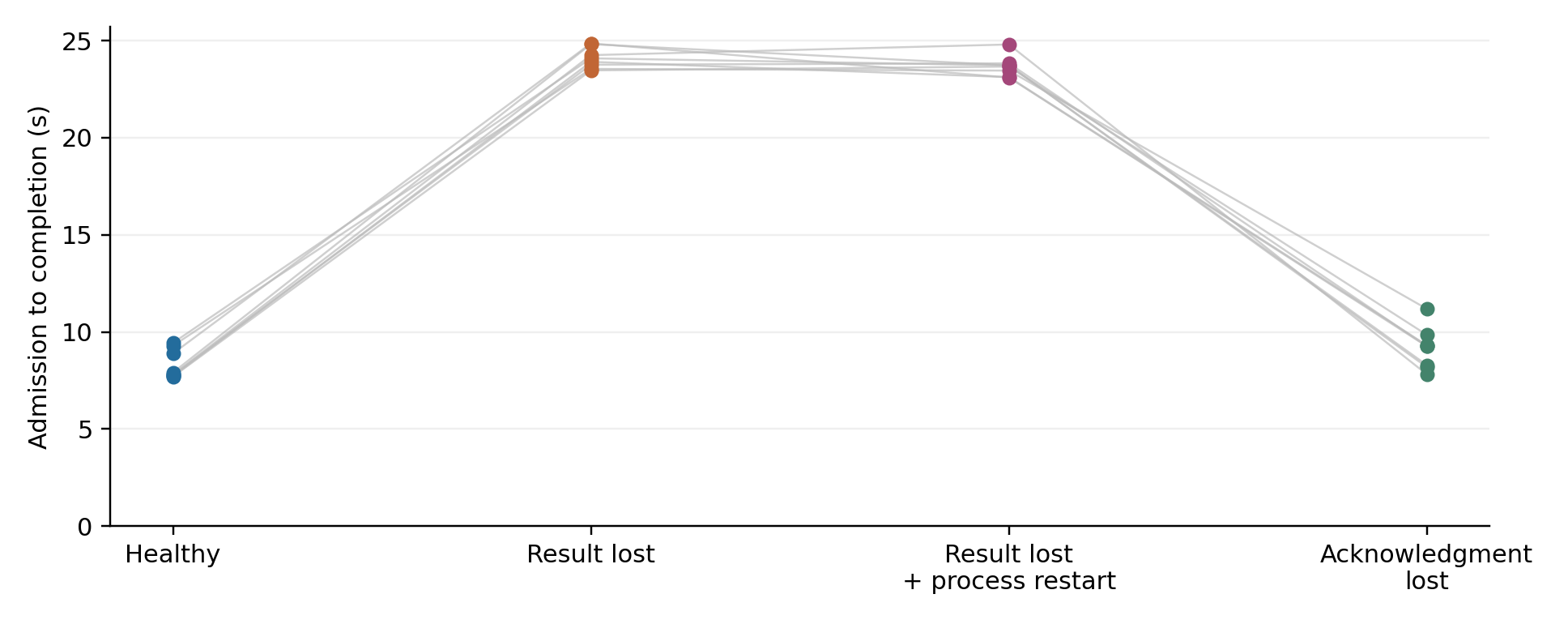}
\caption{Computer workflow completion times under the specified result-loss and restart conditions. Lines connect the same experimental block, not successive time points. The result-loss gate lasts twelve seconds.}
\label{fig:computer-ack}
\end{figure}

All 32 Runs complete with exact artifacts. Their 96 command records contain two reads and one write per Run. To detect repeated identical writes, an independent observer records file and volume identities, timestamps, size, and digest at result frames and after completion. All 32 target files retain their original post-write identity and timestamps. The Runtime's atomic file-replacement implementation makes this a stronger observation than content equality alone.

Median completion times are 7.854, 24.001, 23.698, and 9.256 s in condition order. The two result-loss conditions deliver the eight distinct write commands 32 times each, yet retain the original file objects. Median lost-result-to-accepted-result intervals are 16.314 and 16.022 s, including the twelve-second fault and retry cadence. Acknowledgment loss is shorter because Cloud has already recorded completion.

The measured recovery uses a result already created and retained in the encrypted journal. Cloud redelivery and Runtime replay recover the original write result even after Runtime replacement, while the file's identity and timestamps remain unchanged. The experiment keeps journaling enabled in every arm and varies loss and restart, so its timing quantifies those interruptions. Failures before result persistence, journal eviction or corruption, and arbitrary shell effects are not tested.

\subsection{Observation freshness and device-bound browser actions}
\label{sec:attached-browser-measured}
A browser action can be inconsistent with the Agent's observation even when the browser session is live. We test this boundary with a page initially selecting A. After observation, an external controller changes the DOM selection to B and records that change independently. The Agent's retained intention, Apply(A), can then be distinguished from the effect, Apply(B). A hosted Agent controls Chrome on a caller-bound Computer; an unselected Computer remains online as a placement control.

Two opposite-order blocks compare four conditions. The current-observation arm updates its intention and acts with revision R2. The stale arm obtains R2 but supplies the earlier R1 and retains its original intention. The unguarded arm omits the expected revision. The unobserved-change arm takes no second snapshot, leaving R1 as the latest recorded observation despite the changed DOM. A recovery gate measures effects after rejection and before refreshed authorization. Two additional blocks cover a live selected Runtime, its loss before browser opening, and a new Run after restoration, for fourteen serial arms.

\begin{table}[!ht]
\centering\small
\caption{Initial page-action effects in the scheduled revision comparison,
two attempts per condition. Stale-intention effects apply B although the
retained intention is A. Recovery effects are reported separately.}
\label{tab:attached-browser-effects}
\begin{tabular}{@{}lrrr@{}}\toprule
Condition & Current-plan effect & Rejected & Stale-plan effect \\
\midrule
Fresh observation and plan & 2/2 & 0/2 & 0/2 \\
Old R1 after R2 & 0/2 & 2/2 & 0/2 \\
Expected revision omitted & 0/2 & 0/2 & 2/2 \\
DOM changed, no new snapshot & 0/2 & 0/2 & 2/2 \\
\bottomrule\end{tabular}
\end{table}
\begin{figure}[!ht]
\centering\includegraphics[width=\linewidth]{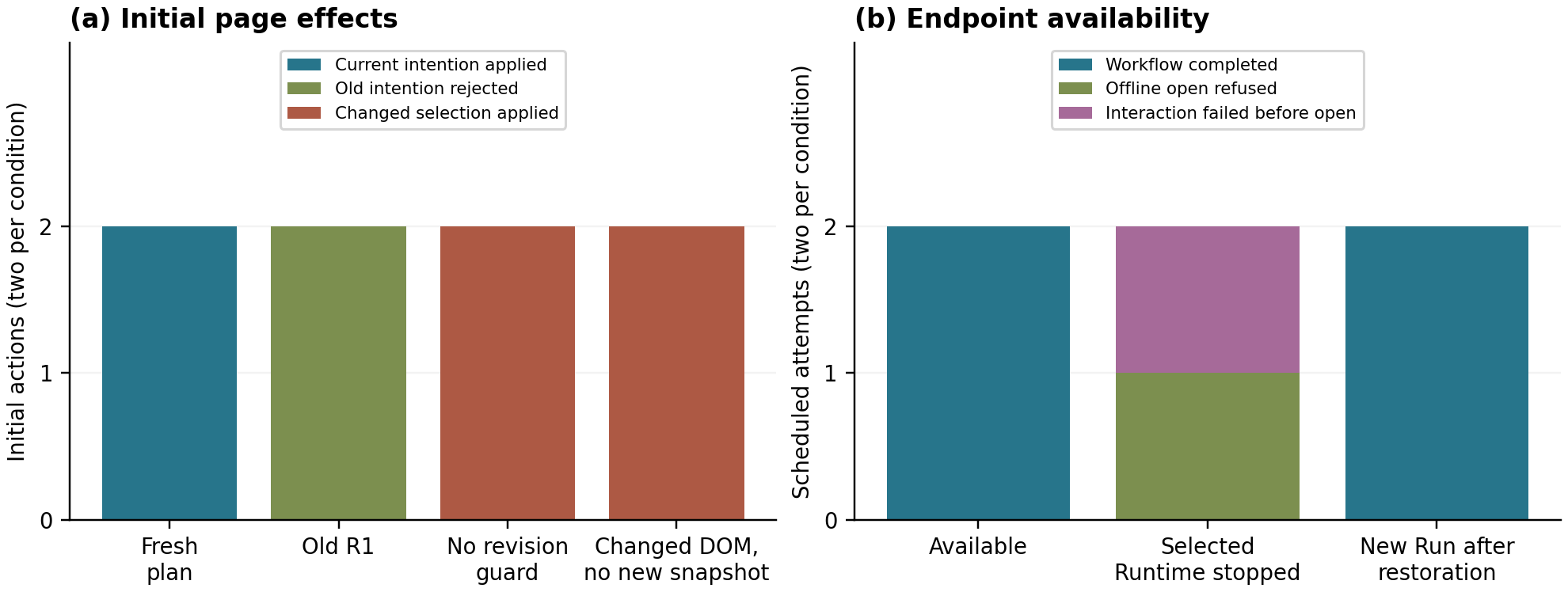}
\caption{Initial page effects and endpoint-availability outcomes. The left panel precedes authorized recovery; the right panel reports completed browser operations and explicit offline refusals.}
\label{fig:attached-browser-effects}
\end{figure}
\begin{figure}[!ht]
\centering\includegraphics[width=\linewidth]{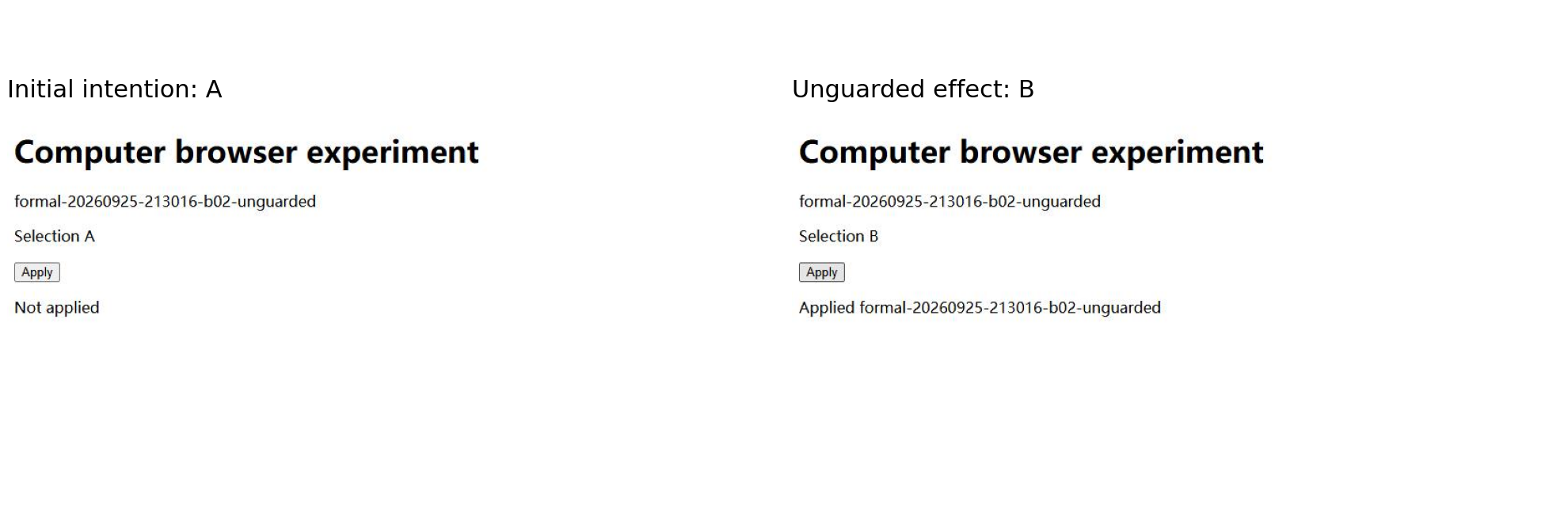}
\caption{Viewport observations before selection change and after the unguarded action. The Agent retains intention A while the independent page-effect record identifies B. The displayed regions preserve the captured pixels.}
\label{fig:attached-browser-screenshots}
\end{figure}

Both explicitly superseded revisions are rejected before any page effect. One of two subsequent stale recoveries succeeds; the other fails at its interaction gate. Unguarded and unobserved-change arms each produce both stale-intention effects. The revision check rejects an explicitly superseded observation, whereas a DOM change without a new snapshot leaves the earlier revision selectable. Available and restored-new-Run conditions each complete 2/2 workflows. Of two stopped-endpoint arms, one reaches browser open and receives the explicit offline refusal; the other fails at an earlier interaction gate. Action outcomes are determined from the independent page-effect records, alongside each Run's terminal state.

Command records, device/session identities, screenshot hashes, and an independent page-effect journal connect the results to the selected Computer. Both registrations share the same host, and observations return through Cloud. The tested software uses Chrome 153.0.8010.52 and Playwright 1.62.0.

\paragraph{Repeated conditions.}
The complete fourteen-condition schedule is repeated with the same SDK, browser, order, and device identity, adding observation-only callback instrumentation without retry. Both stale recoveries then produce one authorized effect, and both stopped endpoints explicitly reject browser open. The four effects in the unguarded and unobserved-change controls persist.

\begin{table}[!ht]
\centering\small
\caption{Browser outcomes in the initial experiment and the instrumented repetition. Each experiment has fourteen conditions, including two opportunities for authorized recovery and two for offline refusal.}
\label{tab:browser-diagnostic-repeat}
\begin{tabular}{@{}lrr@{}}\toprule
Outcome & Initial experiment & Instrumented repeat \\
\midrule
Scheduled Runs observed & 14/14 & 14/14 \\
Old-snapshot actions rejected & 2/2 & 2/2 \\
Authorized recovery effects & 1/2 & 2/2 \\
Explicit offline-open refusals & 1/2 & 2/2 \\
Available workflows completed & 2/2 & 2/2 \\
New Runs after restoration completed & 2/2 & 2/2 \\
Interaction failures & 2 & 0 \\
Stale-intention effects, no revision guard & 2/2 & 2/2 \\
Stale-intention effects, no new snapshot & 2/2 & 2/2 \\
\bottomrule\end{tabular}
\end{table}

The repetition records fifty browser commands, 36 viewport assets, and 646 callbacks; six callback rejections correspond to expected stale or offline conditions. One interrupted observation resumes the original pending Run and interaction without resubmitting the action, extending that gate's duration. The initial experiment contains two interaction failures despite accepted replies in the database; neither recurs in the instrumented repetition, and their causes remain undetermined. Across both experiments, revision checks reject superseded snapshots, while unguarded actions and unobserved DOM changes produce the same stale-intention effects. The repeated mechanism-level outcome identifies protection of recorded-observation consistency, distinct from continuous page-state freshness.

\subsection{Interactive terminal semantics and session continuity}
\label{sec:posix-terminal-measured}
Interactive programs need terminal state and foreground signal delivery in addition to byte streams. We compare a remote pseudoterminal with redirected pipes using the same child program, Bash, and Linux guest userland. The child independently persists its TTY status, dimensions, session and process-group identifiers, foreground group, parent shell identity, and inherited nonce before printing them. Nexus carries terminal input through its authenticated WebSocket to the selected Computer Runtime; the pipe reference uses a direct guest connection and omits terminal allocation and signal translation.

The PTY begins at 80 columns by 24 rows, then receives a resize to 132 by 41. A sixty-second foreground wait receives Ctrl-C; interruption must be observed within three seconds before another command is issued in the same shell. Other conditions detach only the caller or close the terminal, then reconnect or create a new terminal after three seconds and query the nonce. Reversing condition order gives eight observations. Shell startup-file handling and transport paths differ. The comparison therefore measures the child-visible terminal effects under each configuration; startup and transport time are not separate outcomes.

\begin{table}[!ht]
\centering\small
\caption{Independent effects in the corrected eight-condition repetition.
Each column has two scheduled opportunities. A dash denotes an outcome not
requested by that workload, rather than a failure.}
\label{tab:posix-terminal-effects}
\begin{tabular}{@{}lrr@{}}\toprule
Interactive effect & Nexus PTY & Pipe control \\
\midrule
Child receives TTY and foreground group & 2/2 & 0/2 \\
Requested dimensions observed in child & 2/2 & 0/2 \\
Ctrl-C interrupts the waiting child & 2/2 & 0/2 \\
Subsequent command in original shell & 2/2 & --- \\
\midrule
Retention effect & Reattach original & Close and recreate \\
\midrule
Same shell PID/session retained & 2/2 & 0/2 \\
In-memory exported nonce retained & 2/2 & 0/2 \\
\bottomrule\end{tabular}
\end{table}

In both PTY arms, the waiting child's process group matches the terminal foreground group, the interrupt reaches that child, and subsequent commands retain the original shell. Pipes report no TTY or dimensions and no interruption within the window. Reattachment preserves the shell and nonce; terminal recreation loses them. Eight distinct sessions and 24 independent child receipts establish the final repetition's effects.

The terminal-semantic results use eight distinct sessions. Three earlier experiment groups share sessions across conditions and are excluded from independent condition contrasts; five earlier WebSocket attachment timeouts have undetermined causes. The isolated-session repetition changes observation and session isolation, without a product-code change. The original shell remains live during the short caller detachment, preserving its session and exported state. Closing and recreating the terminal loses that state; recovery after Runtime loss is not tested.

\par\bigskip
\section{Invocation Continuity and Settlement}
\label{app:lifecycle-evidence}
The lifecycle experiments separate admitted work, external effects, recorded completion, delivered output, and settled usage. Failures are introduced at explicit boundaries to identify which state supports continuation.

\subsection{Committed-operation reuse after worker loss}
\label{sec:worker-recovery}
Worker recovery can reproduce a correct output either by repeating prior work or by reusing recorded results. We distinguish these paths through independent counts of external appends. Six blocks cross journal enabled/disabled with healthy/lost workers, yielding 24 Runs. Each handler executes six 200-ms-paced stages that append nonce, stage, and digest records to a file. The journal wraps the same callbacks and declares unknown append outcomes unsafe to reconcile. The callback itself performs no duplicate suppression.

The fault kills the Cloud worker supervisor and execution child after acknowledgment of stage three; in the journal arm this follows the operation commit. The Agent, Cloud service, and PostgreSQL remain live, so the original Agent may continue appending after worker loss. A replacement worker uses the same ninety-second lease and fifteen-second maintenance interval. Inputs, pacing, recovery capability, and fault boundary are common, with balanced order. The isolated database uses synthetic opening balances and the actual admission, execution, journal, and settlement paths.

\begin{table}[htbp]
\centering\small
\caption{Real worker-process loss and journal ablation: six attempts per cell. Times and duplicate counts are means; no model is called. Every completed task is charged once at 0.050000 synthetic USD.}
\label{tab:worker-recovery}
\begin{tabular}{@{}llrrrr@{}}
\toprule
Worker & Journal & Outputs & Appends & Duplicates & Completion (s) \\
\midrule
Healthy & Off & 6/6 & 6.0 & 0.0 & 7.260 \\
Healthy & On & 6/6 & 6.0 & 0.0 & 8.193 \\
Lost & Off & 6/6 & 12.0 & 6.0 & 108.414 \\
Lost & On & 6/6 & 6.0 & 0.0 & 108.618 \\
\bottomrule
\end{tabular}
\end{table}
\begin{figure}[htbp]
\centering
\includegraphics[width=\linewidth]{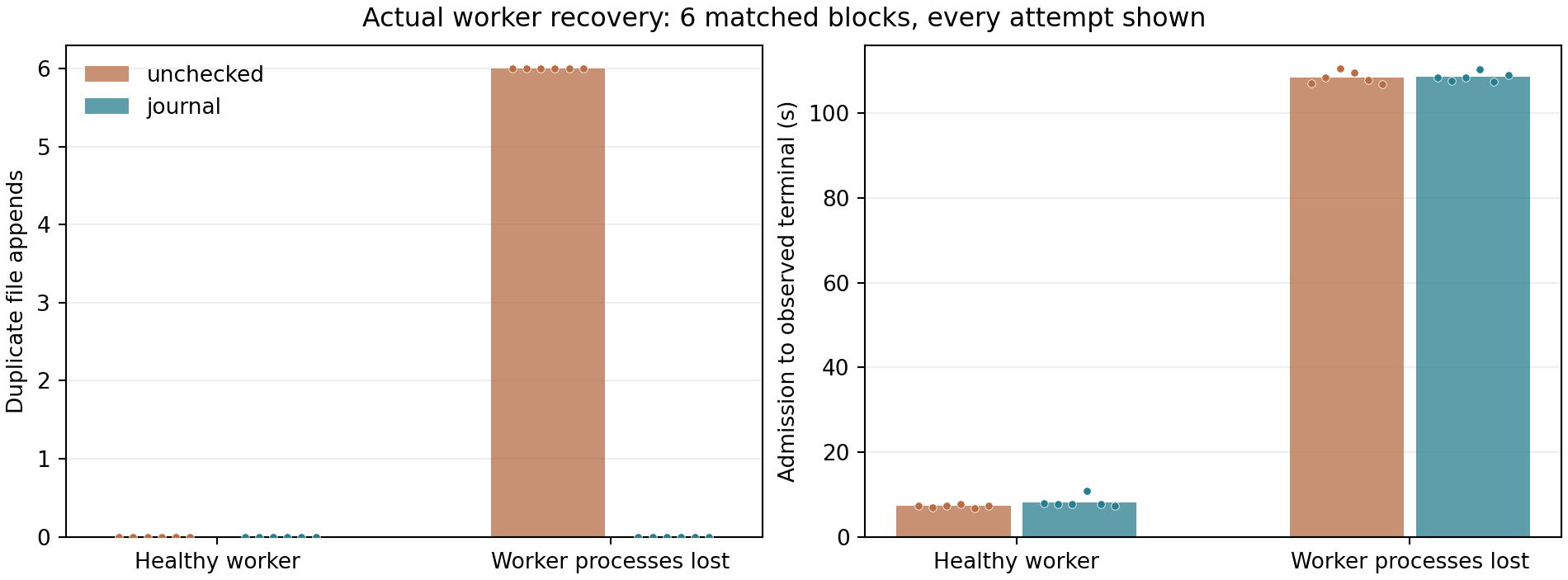}
\caption{Repeated appends and total completion time under worker loss. Points show individual Runs and bars show means.}
\end{figure}

All 24 outputs are exact, across 180 actual appends. Lost-worker Runs average six duplicate appends without journaling and zero with it. The paired mean journal overhead under healthy workers is 0.933 s (exploratory 95\% six-block bootstrap interval $[0.305,1.874]$ s); under worker loss, the mean completion-time difference is 0.204 s ($[-0.937,1.252]$ s). Fault-to-terminal intervals are 101.096--102.832 s without the journal and 101.649--103.223 s with it. Lease detection dominates recovery time, while the journal's primary benefit is preserved work.

Independent append counts, committed operation values, and Run/task/turn identities distinguish reuse from repeated execution. Each successful Run has one usage record and one debit/credit pair, all terminal reservations are released, and no replacement attempt creates a second charge. The synthetic consumer/provider balances end at 8.800000/0.960000 USD with 0.240000 USD platform share. This experiment covers acknowledged operation commits while the Agent survives. Section~\ref{sec:ambiguous-effect} moves the fault earlier, into the gap between an external effect and its completion record.

\subsection{Recovery between an external effect and its completion record}
\label{sec:ambiguous-effect}
A journal cannot infer whether an unrecorded external operation already took effect. We place an explicit crash barrier after an external service durably commits stage three and before the Agent returns from that callback. The external service survives and stores nonce/stage/digest rows in a transactional database. Optional stable-key deduplication returns an existing receipt instead of inserting another row.

Six balanced blocks compare four recovery configurations with the same six-stage handler, 200-ms pacing, external service, and crash point. Without a journal, prior stages execute again. A guarded journal pauses on the unknown result; after at least thirty seconds, an explicit caller retry accepts that uncertainty. A reconciliation declaration alone permits automatic replay while the resource ignores keys. Finally, the same declaration is combined with transactional resource deduplication. Cloud, worker, and database remain live while the Agent process is replaced. Each invocation uses a fixed 0.050000 synthetic-USD tariff.

\begin{table}[htbp]
\centering\small
\caption{Agent-process loss after the stage-three external commit. Six attempts per arm; effect and duplicate counts are totals across that arm. The guarded arm completes only after an explicit caller retry.}
\label{tab:ambiguous-effect}
\begin{tabular}{@{}lrrrr@{}}
\toprule
Arm & Final outputs & Pauses & Effect rows & Duplicates \\
\midrule
No journal & 6/6 & 0 & 54 & 18 \\
Guard + caller retry & 6/6 & 6 & 42 & 6 \\
Reconciliation flag alone & 6/6 & 0 & 42 & 6 \\
Journal + resource idempotency & 6/6 & 0 & 36 & 0 \\
\bottomrule
\end{tabular}
\end{table}
\begin{figure}[htbp]
\centering
\includegraphics[width=\linewidth]{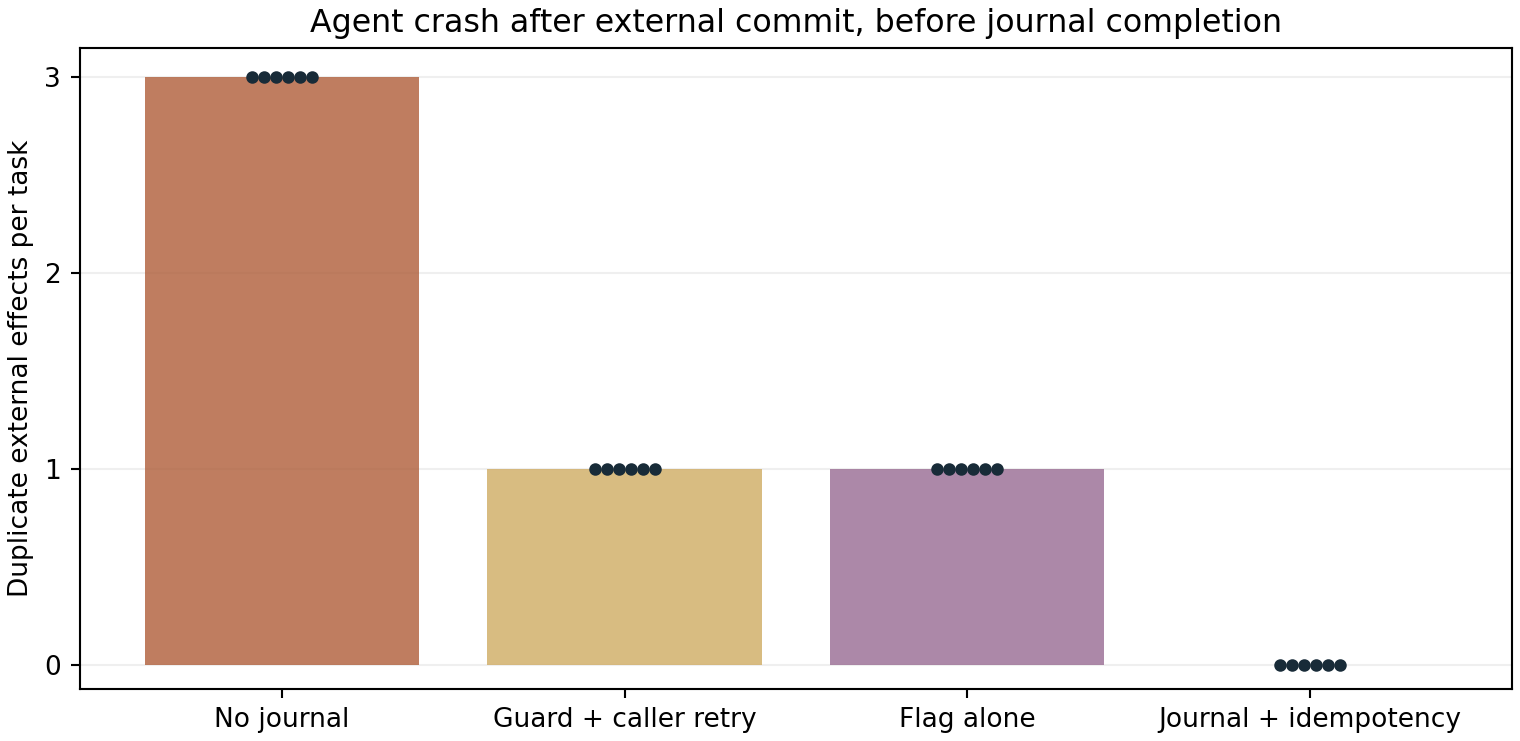}
\caption{Repeated external effects after Agent failure at the effect/recording boundary. Points show individual Runs and bars show means. The guarded arm includes a subsequent caller retry.}
\end{figure}

All 24 Runs eventually produce the exact logical output, across 174 external effect rows. No journal yields eighteen duplicate stages over six Runs. Guarded replay followed by caller retry and the unsupported reconciliation declaration each yield six duplicates. Journal plus resource deduplication yields zero. During every guarded observation window, exactly three rows remain and no additional effect occurs before caller intervention. The guard therefore suspends unsafe automatic work, while duplicate-free continuation depends on the external resource.

Independent resource rows, journal keys, request digests, process identities, and task records connect the injected crash to each observed effect. Every Run settles once, releases terminal reservations, and produces the expected synthetic balance transfers. The guarded configuration includes a deliberate wait and caller decision; the other configurations recover autonomously. Comparing their effect counts separates three contributions: reuse of committed stages, suspension of uncertain replay, and resource-side recognition of a repeated operation. The resource commits its effect independently of Cloud's completion record, which is precisely the interval exercised by this experiment.

\subsection{Cooperative cancellation and residual work}
\label{sec:measured-cancellation}
Task cancellation has an execution effect as well as a persisted state and a settlement outcome. We isolate the execution mechanism with a twelve-stage handler that waits 200 ms and appends a nonce/stage/digest record at each stage. The cooperative arm refreshes cancellation state before every append; the unchecked arm executes the same stages without that check. Cancellation is requested after three observed appends, matching completed work rather than elapsed time.

Six balanced pairs under each of fixed and reported pricing give 24 Runs. Both tariffs reserve 0.050000 synthetic USD, and reported-price handlers submit 0.032100 USD before their first effect. Actual admission, callbacks, cancellation, and settlement run against an isolated PostgreSQL instance. A dedicated execution thread exercises the handler path while leaving worker-supervisor recovery outside this experiment.

For append set $E_i$, cancellation-response time $t_i^c$, and terminal time $t_i^t$, we measure
\begin{equation}
R_i=\sum_{e\in E_i}\mathrm{I}[t_e>t_i^c],\qquad
S_i=12-|E_i|,\qquad T_i=t_i^t-t_i^c.
\end{equation}
Here $R_i$ counts residual effects, $S_i$ avoided stages, and $T_i$ terminal-state delay. Independent digest and timestamp checks validate 180 appends across all Runs.

\begin{table}[htbp]
\centering\small
\caption{Active cancellation, six pairs per tariff. U: unchecked handler; C: cooperative handler. Stage counts are scenario outcomes. Timing is the paired mean difference C$-$U, with an exploratory 95\% pair-bootstrap interval.}
\label{tab:cancellation}
\begin{tabular}{@{}lrrrr@{}}
\toprule
Tariff & Total stages U/C & Post-return U/C & $\Delta$ terminal (ms) & 95\% interval \\
\midrule
Fixed & 12/3 & 8.17/0 & -1206.8 & [-1279.6, -1139.6] \\
Reported & 12/3 & 8.67/0 & -1309.3 & [-1338.6, -1280.7] \\
\bottomrule
\end{tabular}
\end{table}
\begin{figure}[htbp]
\centering
\includegraphics[width=\linewidth]{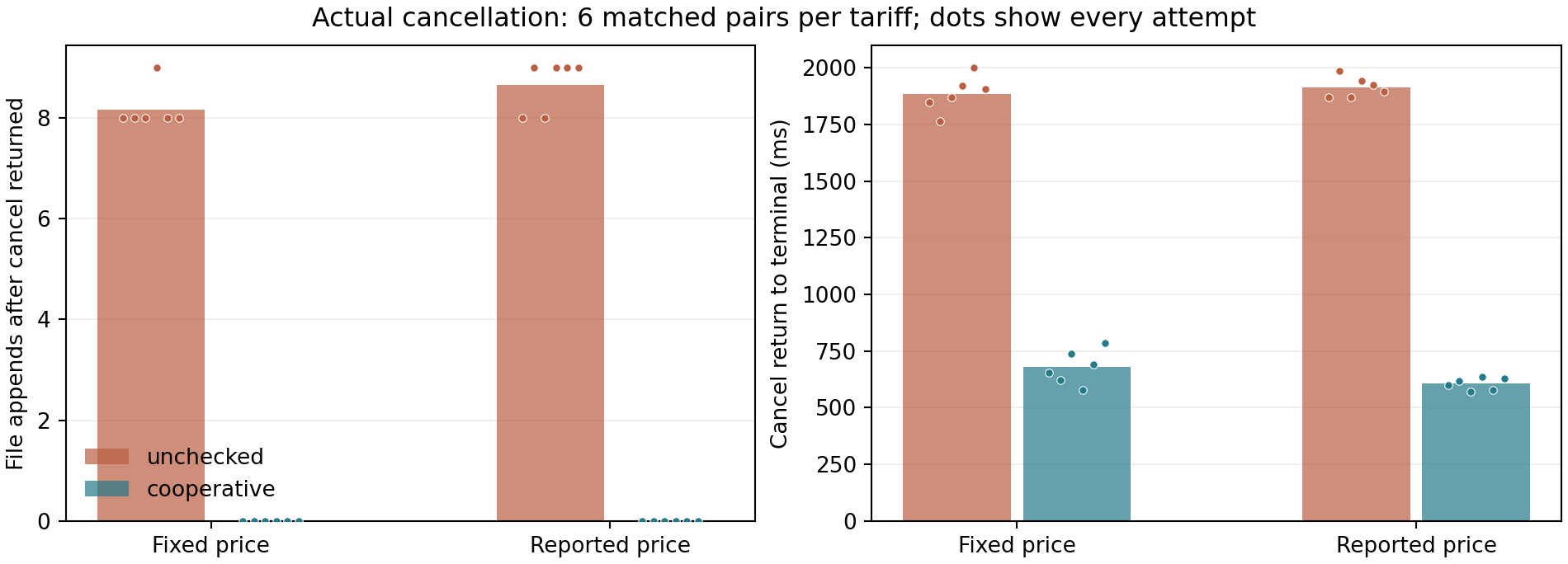}
\caption{Residual local effects and termination after cancellation. Points show individual attempts and bars show means.}
\label{fig:cancellation}
\end{figure}

Every cooperative handler stops after stage three, avoiding nine appends and producing no effect after the cancellation response. Unchecked handlers complete all twelve stages, including eight or nine appends after that response. Mean response-to-terminal time changes from 1,884.7 to 677.9 ms under fixed pricing and from 1,914.0 to 604.7 ms under reported pricing. Exploratory intervals resample six pairs per tariff.

All tasks reach the persisted cancelled state, including unchecked handlers that finish all twelve stages. Reservations are released with zero debit and credit, while the twelve accepted cost reports remain recorded separately. The append traces therefore distinguish a cancelled invocation from a handler that actually stops. In the cooperative arm, the pre-append check prevents every later stage in the tested schedule; the unchecked arm continues. This mechanism relies on the handler reaching a cancellation check and does not forcibly interrupt arbitrary code or an upstream model request. Healthy-path polling overhead is not measured by these cancellation-triggered runs.

\subsection{Incremental event delivery and finite history}
\label{sec:measured-stream}
Stream continuity depends on retaining the original producer; a replay cursor determines how much already-consumed output is sent again. Both clients in this comparison reconnect to the same retained task. The control requests history from event zero and deduplicates identifiers; the treatment requests only events after its last consumed identifier.

The SDK handler emits 32 digest-checked events and appends one independent stage record per event, with 20-ms pacing. Eight cells cross 64- and 4,096-byte payloads, interruption after event four or sixteen, and direct or LAN-bridge access. Ten balanced pairs per cell give 160 tasks. The client closes its connection, waits 100 ms, and reconnects. Both arms share retention settings of 512 tasks, 64 events and 1 MiB per task, and 300 seconds. All components run on real loopback sockets in one two-CPU, 512-MiB Linux container.

\begin{table}[htbp]
\centering\small
\caption{Consumed SSE bodies across initial and reconnect responses. Every direct and bridge repetition has the shown count; timing is analyzed separately.}
\label{tab:stream-bytes}
\begin{tabular}{@{}rrrrrr@{}}
\toprule
Text (B) & Cut event & Replay (B) & Cursor (B) & Reduction & Duplicates R/C \\
\midrule
64 & 4 & 7,631 & 6,791 & 11.0\% & 4/0 \\
64 & 16 & 10,165 & 6,791 & 33.2\% & 16/0 \\
4,096 & 4 & 152,783 & 135,815 & 11.1\% & 4/0 \\
4,096 & 16 & 203,701 & 135,815 & 33.3\% & 16/0 \\
\bottomrule
\end{tabular}
\end{table}
\begin{figure}[htbp]
\centering
\includegraphics[width=\linewidth]{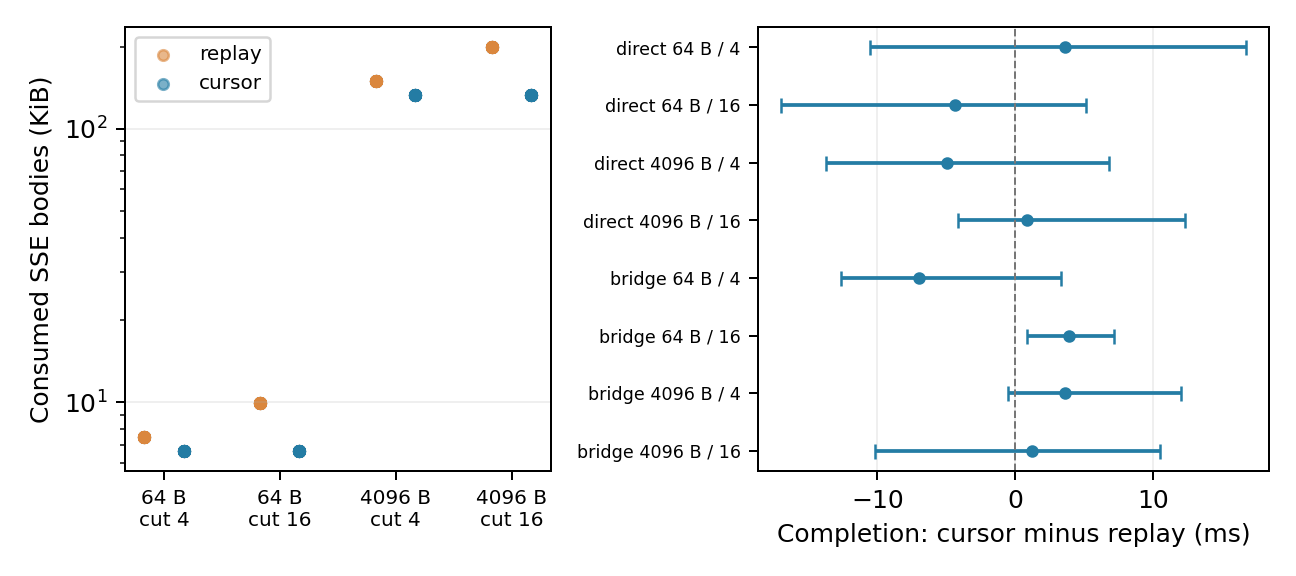}
\caption{Incremental stream delivery. Left: consumed SSE bodies, with overlapping points for identical direct and bridge counts. Right: median paired completion-time difference, cursor minus full replay, and 95\% intervals from 5,000 pair-bootstrap resamples. Each timing cell has ten pairs.}
\label{fig:stream-recovery}
\end{figure}

All 160 tasks reconstruct the exact output, with 5,120 ordered stage records and one producer execution per task. Every disconnect precedes final production. Full replay redelivers four or sixteen consumed events; cursor continuation redelivers none. At 4,096 bytes and event sixteen, consumed SSE body decreases from 203,701 to 135,815 bytes, a 33.3\% reduction including the initial prefix and reconnect response. The measure excludes wire overhead and unread buffered bytes.

Total completion time has no consistent reduction: seven of eight exploratory intervals include zero, while one small-payload bridge cell increases by 3.893 ms. First-new-event delay improves in some cells, but the paced producer remains the main completion dependency. The repeatable benefit is lower redundant delivery with unchanged output semantics.

\paragraph{History loss.}
Fifteen separate observations test retained history, changed payload, a four-event history window, natural retention expiry, and actual server replacement. Retained history supplies the remaining eight of twelve events; the other conditions reject incompatible or unavailable continuation. A further twelve balanced pairs compare replay policies after expiry or process loss.

\begin{table}[htbp]
\centering\small
\caption{Recovery after history loss, with six attempts per row. Counts include the original execution and any replacement. Duplicate appends count repeated business stages rather than repeated event delivery.}
\label{tab:stream-loss}
\begin{tabularx}{\linewidth}{@{}>{\raggedright\arraybackslash}Xlrrr@{}}
\toprule
Fault & Reconnect policy & Outputs & Producer starts & Duplicate appends \\
\midrule
Process replacement & Full replay & 6/6 & 12 & 72 \\
Process replacement & Cursor & 0/6 & 6 & 0 \\
Retention expiry & Full replay & 6/6 & 12 & 72 \\
Retention expiry & Cursor & 0/6 & 6 & 0 \\
\bottomrule
\end{tabularx}
\end{table}

After history loss, all twelve cursor-zero attempts reconstruct the output by starting a new producer and repeating every append. All twelve suffix-continuation attempts reject the missing task, produce no remaining output, and start no replacement producer. The policies consequently preserve different properties: restarting recovers the output at the cost of repeated effects, whereas suffix continuation preserves producer uniqueness but leaves the task incomplete. Retaining the original history allows both properties to coexist, as shown by the preceding 160-task experiment.

\subsection{Incremental delivery through protocol adapters}
\label{sec:native-adapter}
We extend the retained-history experiment to the installed OpenWrt adapter's MCP and A2A ingress paths. The adapter maps requests to a real SDK tool or A2A executor, using a shared tunnel and protocol implementation within each comparison. The evaluated MCP streaming response uses Nexus progress/result events, while A2A carries its stream-response objects. Standard-client framing is examined separately in Section~\ref{sec:mcp-client}.

Each fresh execution appends sixteen independent stage records with 25-ms pacing. MCP emits sixteen progress events and a final result; A2A emits task/status events, sixteen artifacts, and completion. Eight cells cross protocol, 64- or 4,096-byte text, and disconnection after four or eight events. Ten balanced pairs per cell compare full-history replay with suffix continuation after a 100-ms pause, yielding 160 tasks. SDK and adapter versions, input, path, and retention are common within pairs. The path uses FastMCP 3.4.7, MCP 1.25.0, A2A SDK 1.1.2, and protobuf 6.33.6.

\begin{table}[htbp]
\centering\small
\caption{Native-ingress consumed SSE bodies. Every repetition in a cell has these byte counts. Both arms reconstruct the same business output; duplicate delivered events are cut/zero for replay/cursor.}
\label{tab:native-adapter}
\begin{tabular}{@{}lrrrrr@{}}
\toprule
Protocol & Text (B) & Cut event & Replay (B) & Cursor (B) & Reduction \\
\midrule
MCP & 64 & 4 & 5,514 & 4,450 & 19.3\% \\
MCP & 64 & 8 & 6,578 & 4,450 & 32.4\% \\
MCP & 4,096 & 4 & 86,154 & 68,962 & 20.0\% \\
MCP & 4,096 & 8 & 103,346 & 68,962 & 33.3\% \\
A2A & 64 & 4 & 9,624 & 8,036 & 16.5\% \\
A2A & 64 & 8 & 11,388 & 8,036 & 29.4\% \\
A2A & 4,096 & 4 & 82,204 & 72,550 & 11.7\% \\
A2A & 4,096 & 8 & 100,096 & 72,550 & 27.5\% \\
\bottomrule
\end{tabular}
\end{table}
\begin{figure}[htbp]
\centering
\includegraphics[width=\linewidth]{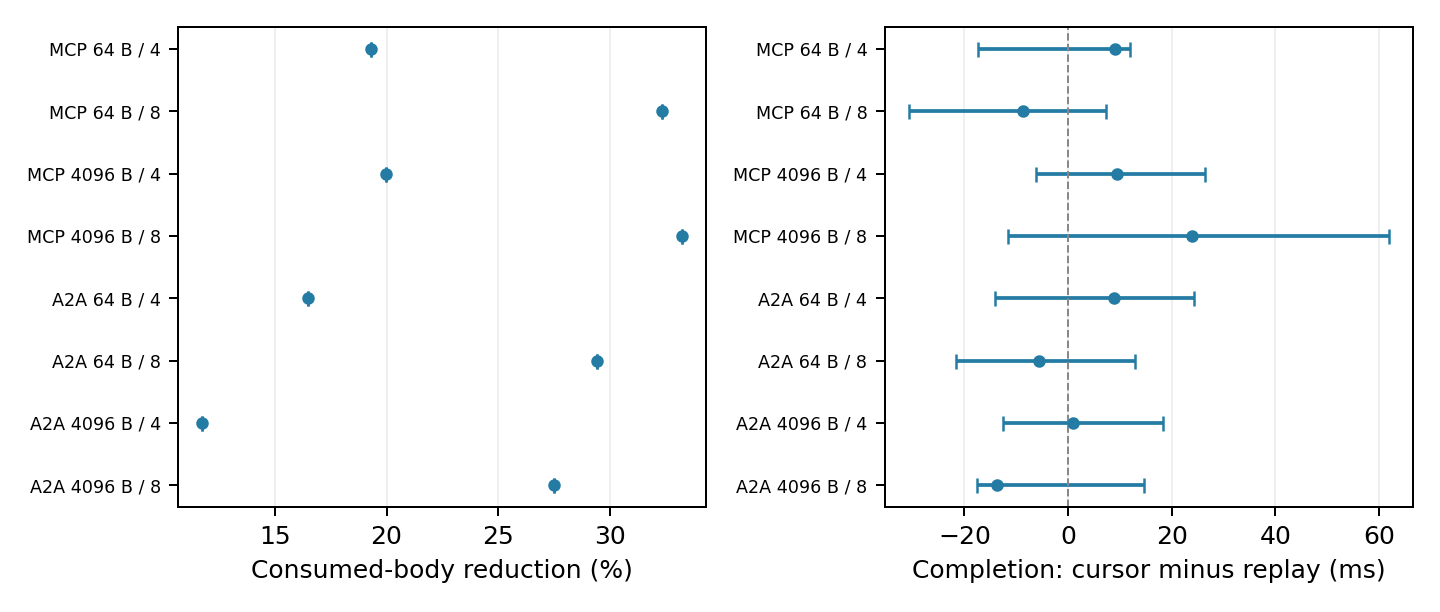}
\caption{MCP/A2A ingress through the OpenWrt adapter. Left: median paired SSE-body reduction. Right: median paired completion-time difference and exploratory 95\% intervals from 5,000 pair-bootstrap resamples, with ten pairs per cell.}
\label{fig:native-adapter}
\end{figure}

All 160 tasks reconstruct their exact output with 2,560 unique stage records, and every disconnect precedes the final stage. Continuation removes all duplicate events and reduces consumed SSE bodies by 11.7--33.3\% across the eight cells. For 4,096-byte text interrupted after eight events, MCP body consumption falls from 103,346 to 68,962 bytes and A2A from 100,096 to 72,550 bytes. A2A's initial events contain task/status information, so equal cut positions represent different business payloads across protocols; the controlled effects are within each protocol.

All total-completion intervals include zero, while the consumed-body reduction repeats across the workload cells. Both paths still await the paced producer, so less replayed output need not shorten task completion. Body counts exclude HTTP, TCP, SSH, and unread buffering. These measurements describe adapter conversion and in-memory history retention through sockets. They use the application-specific MCP stream described above; standard-client framing and explicit retained-stream recovery are evaluated separately.

\subsection{Asynchronous task identity across peer disconnection}
\label{sec:peer-task-recovery}
Asynchronous delegation separates task lifetime from the connection used to observe it. We evaluate this property with the Qwen-controlled expense/inventory worker on two logical nodes in one OpenWrt VM. The caller retains a private task handle, and every worker performs four independently recorded model decisions.

The initial SDK encodes admission information in private HTTP headers that the peer IPC path does not carry. Its two direct calls return handles, but two peer calls fail admission observation despite correct completed worker outputs. A revised SDK carries the authenticated start and receipt information in a private control envelope while leaving the routing daemons unchanged. Both caller and destination use this revision; the application payload and routing constraints are preserved without synchronous fallback or automatic resubmission.

\begin{table}[!ht]
\centering\small
\caption{Asynchronous admission across both directions. Each worker executes four model decisions. Correct worker output and a usable caller-held task handle are counted separately.}
\label{tab:peer-async-admission}
\begin{tabular}{@{}llrrr@{}}\toprule
SDK & Path & Handles & Correct workers & Calls \\
\midrule
Original & Direct & 2/2 & 2/2 & 8 \\
Original & Peer & 0/2 & 2/2 & 8 \\
Candidate & Direct & 2/2 & 2/2 & 8 \\
Candidate & Peer & 2/2 & 2/2 & 8 \\
\bottomrule\end{tabular}
\end{table}

All four revised admission probes return handles before worker completion and subsequently retrieve correct results. Each SDK version makes sixteen model calls. These version-specific outcomes distinguish executed business work from usable asynchronous admission.

\paragraph{Observation while the peer is unavailable.}
A separate four-arm schedule compares peer and direct access in both directions. Workers pause after read and calculate decisions; the destination peer daemon is stopped while the worker and process-local task store remain alive. Both peer observations fail during the outage, while direct controls continue observing the same workers. Releasing the gate lets all four workers finish their remaining decisions and artifact writes before the peer returns.

\begin{figure}[!ht]
\centering\includegraphics[width=\linewidth]{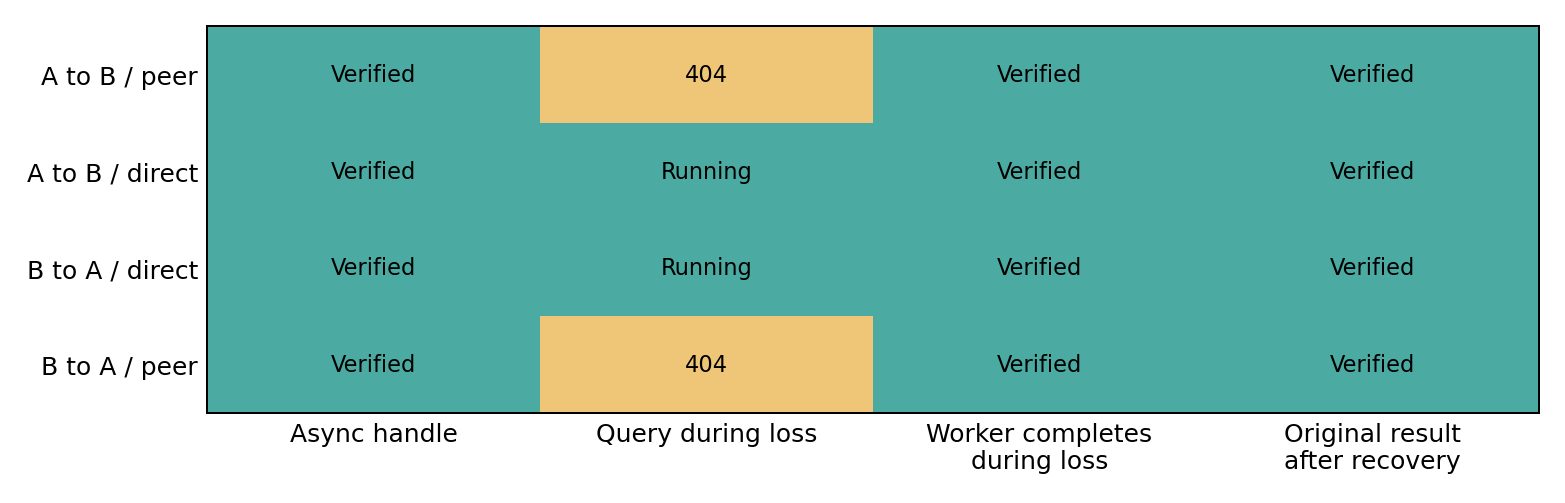}
\caption{Original-handle observation during peer interruption. Amber marks unavailable observation while the worker continues; green marks the validated result.}
\label{fig:peer-task-recovery}
\end{figure}

After daemon restart and stable route availability, all four original handles retrieve the correct reports. Each task creates one store entry and four model requests, with no second admission during recovery. The worker and its process-local task record survive the transport interruption, allowing the original handle to resume observation of that same execution. Worker or store loss is not injected here. The caller already possesses its handle before the interruption; the next experiments vary record lifetime and receipt delivery separately.

\subsection{Admission backpressure and task-record lifetime}
\label{sec:task-retention}
A task handle remains useful only while its record exists. We compare two capacity policies using a two-slot store, identical model-driven workers, and the same asynchronous admission protocol. The initial policy prunes at capacity on both admission and reads, potentially deleting an accepted task's identity while its handler continues. The retention-preserving policy separates expiry from admission: reads keep unexpired records, and a full store rejects new work before execution.

Eight counterbalanced scenarios cross the two policies with lookup at capacity, overflow admission, completed-task expiry, and active-task expiry. Stores restart empty between scenarios. Capacity cases use a 120-second lifetime and hold two workers after their first two model decisions. Lookup queries the first handle; overflow attempts a third task. Independent artifact and Gateway records account for every admitted worker, including those whose control record disappears.

\begin{table}[!ht]
\centering\small
\caption{Capacity scenarios. A correct worker artifact and a usable completed
task handle are separate outcomes. The candidate's refused third task is not
counted as a completed task.}
\label{tab:task-capacity}
\begin{tabular}{@{}llrrrr@{}}\toprule
Condition & Store & Admitted & Correct & Handles & Calls \\
\midrule
Lookup & Previous & 2 & 2 & 1 & 8 \\
Lookup & Candidate & 2 & 2 & 2 & 8 \\
Overflow & Previous & 3 & 3 & 1 & 12 \\
Overflow & Candidate & 2 & 2 & 2 & 8 \\
\bottomrule\end{tabular}
\end{table}

The retention-preserving policy keeps both accepted tasks and rejects the third admission with no model call. The initial policy admits additional work by evicting an existing record, and later reads can trigger further eviction. The policies change the admitted workload: the new policy preserves accepted identities by rejecting overflow before execution, and the rejected request remains a noncompletion.

Expiry scenarios use a natural thirty-second creation-time lifetime. All four handles are usable five seconds before expiry and absent at least one second afterward. Both completed and still-active records expire, while all handlers eventually produce correct artifacts. Across the eight scenarios, thirteen admitted workers complete with 52 verified model calls; one request is rejected before execution.

Capacity and lifetime govern task visibility in different ways. Admission backpressure preserves existing records at saturation, while the creation-time expiry still removes both completed and active records at their stated deadline. Workers can finish after the handle has expired. The tested store is process-local and its lifetime is not refreshed by reads; applications therefore need capacity for their outstanding tasks and a retention interval long enough to observe completion.

\subsection{Application reconciliation after a lost admission receipt}
\label{sec:admission-receipt}
A caller may lose the receipt that identifies work already admitted. We test this case with an application-level identity guard in the report Agent. Atomic local task-directory creation assigns ownership; matching replays are rejected while work is active and return the cached report after completion. The SDK's private control record remains a separate identity.

Eight scenarios cross two peer directions with intact receipt, lost receipt without replay, one active replay, and one completed replay. Inputs, source, SDK, daemons, and model configuration are fixed. A sixteen-slot store with a 600-second lifetime keeps capacity and expiry outside the intervention. A proxy consumes the real admission response and closes the caller connection without returning it. Workers pause after two model decisions so active replay occurs before the output write; completed replay uses the unchanged original request.

\begin{table}[!ht]
\centering\small
\caption{Two directions per receipt condition. ``Results'' counts caller-held
handles returning the completed application result. Control records include
lost and rejected-replay records; they are not distinct business executions.}
\label{tab:admission-receipt}
\begin{tabular}{@{}lrrrrr@{}}\toprule
Condition & Workers & Correct & Results & Records & Calls \\
\midrule
Intact & 2 & 2 & 2 & 2 & 8 \\
Lost, no replay & 2 & 2 & 0 & 2 & 8 \\
Lost, active replay & 2 & 2 & 0 & 4 & 8 \\
Lost, completed replay & 2 & 2 & 2 & 4 & 8 \\
\bottomrule\end{tabular}
\end{table}

All eight original workers produce correct artifacts with four model calls each, totaling 32 Gateway-linked requests. Six first receipts are lost. Intact callers retain their original handles; no-replay callers remain without usable handles despite completed work. Each active replay creates a new control handle whose task is rejected by the application, without recovering the original handle. Each completed replay creates a new control record and returns the original cached report with unchanged artifact hash and modification time.

The twelve control identities represent eight executed business tasks. In both completed-replay cases, the application's identity guard returns a cached report without another model workflow. Active replays instead leave callers without the original task's usable handle. These observations separate application-result recovery from SDK-handle recovery under retained local state. The guard is enabled throughout this experiment, so the measured outcome is its replay behavior, not a guard-enabled/disabled cost difference; durable handle discovery is not tested.

\subsection{Queued continuation during caller absence}
\label{sec:queued-turn}
A queued turn can remove the caller's continued presence from the execution dependency. Four balanced pairs use 256- and 2,048-byte inputs and the same two-turn handler. Once its bounded steering window closes, the next input is available to both callers. The queue arm submits it while turn one remains active; the explicit-continuation arm holds it locally until the caller returns. The second turn computes a digest and returns the byte count and pair nonce.

The caller closes its session for twenty seconds without reading or writing the Run. After reauthentication, the first display is captured before any manual continuation, and both Runs are then observed to completion. Fresh Run identities prevent cross-arm output reuse. The experiment uses actual Cloud/SDK execution with no model or device calls.

\begin{table}[!ht]
\centering\small
\caption{Four matched pairs. Readiness means the independently verified
second-turn result already exists at client return, before any manual resume.}
\label{tab:queued-turn}
\begin{tabular}{@{}lrrrr@{}}\toprule
Submission mode & Runs & Ready at return & Final correct & Invocations \\
\midrule
Worker-dispatched queue & 4 & 4/4 & 4/4 & 8 \\
Explicit continuation & 4 & 0/4 & 4/4 & 8 \\
\bottomrule\end{tabular}
\end{table}

All eight workflows eventually produce the correct second-turn result. In all four queue arms, the second turn is admitted and completed during caller absence; no manual arm is ready before explicit continuation. Each Run has exactly two successful invocations, with first-turn completion preceding second-turn admission. The queued record links to that second turn, while manual arms contain no queue record. All tariffs and charges are zero.

Queueing changes the submission dependency: the second request is already present when the first turn finishes, allowing execution during caller absence. The manual path waits for another online client action. Readiness at return measures this difference in submission time, with permissions, funding, successful preceding turns, and the queued contents held fixed.

\subsection{Persisted outputs, replies, and steering}
\label{sec:measured-interaction}
Thirty hosted invocations exercise three interaction patterns: digest output, a solicited yes/no reply, and one cooperative steering instruction. Ten Runs per pattern execute in shuffled order. Independent validation checks the payload digest and nonce, returned reply, or applied steering content. The caller uses the normal reply and steering interfaces, and the actual worker, MCP path, and SDK produce the results.

\begin{table}[htbp]
\centering\small
\caption{Hosted output, reply, and steering scenarios, with ten Runs per pattern.}
\label{tab:measured-interactions}
\begin{tabular}{@{}lrrr@{}}
\toprule
Scenario & Successful & Failed & Median time (s) \\
\midrule
Hash output & 10/10 & 0 & 7.515 \\
Solicited reply & 10/10 & 0 & 8.915 \\
Cooperative steering & 10/10 & 0 & 10.950 \\
\bottomrule
\end{tabular}
\end{table}
\begin{figure}[htbp]
\centering
\includegraphics[width=.92\linewidth]{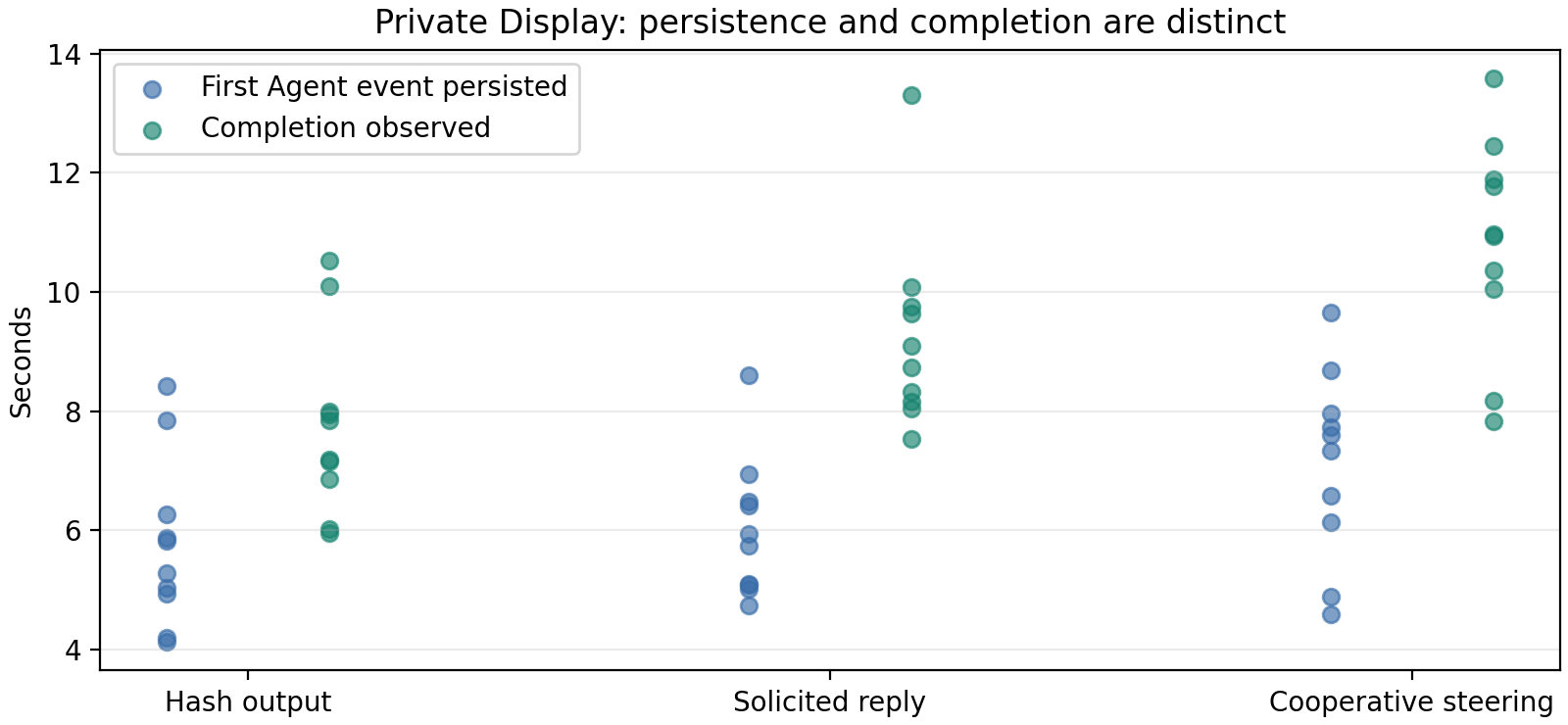}
\caption{Server-side event persistence and client-observed workflow completion for ten Runs per interaction pattern.}
\end{figure}

The first measured Agent event is persistence of the display title, following the earlier platform start event. Ordered snapshots and sequence numbers associate the reply or steering input with the final output. Event timing ends at server persistence, while API completion timing includes caller observation. These thirty free-tariff invocations characterize three interaction paths without a comparison implementation. Section~\ref{sec:measured-billing} adds nonzero reservation and settlement to the execution lifecycle.

\subsection{Reservation, reporting, and settlement under execution outcomes}
\label{sec:measured-billing}
The accounting experiment tests whether the charge follows the admitted terms and actual execution outcome. Seven scenarios use a PostgreSQL-backed Cloud application, its worker execution path, and a separate MCP/SDK Agent. The handler computes a digest, optionally reports a charge, and either publishes the result or raises a specified failure. Admission, authenticated callbacks, and settlement execute through their normal interfaces. Opening accounts, deployment metadata, and synthetic balances are controlled fixtures.

\begin{table}[htbp]
\centering\small
\caption{HTTP--worker--SDK accounting scenarios. Each accepted call
reserves 0.050000 synthetic USD; all reservations are released at termination.}
\label{tab:actual-billing}
\begin{tabularx}{\linewidth}{@{}>{\raggedright\arraybackslash}Xlr@{}}
\toprule
Scenario & Terminal outcome & Charge (USD) \\
\midrule
Fixed-price successful digest & Completed & 0.050000 \\
Reported-price successful digest & Completed & 0.032100 \\
Successful digest, missing report & Completed & 0.050000 \\
Report followed by handler failure & Failed & 0.000000 \\
Same report submitted twice & Completed & 0.032100 \\
Over-cap report attempted through SDK & Failed & 0.000000 \\
Caller cancels before worker claim & Task cancelled & 0.000000 \\
\bottomrule
\end{tabularx}
\end{table}

Four workflows complete with the expected nonce and digest; the other scenarios exercise failure and cancellation. All seven produce the expected accounting records. Starting from 10.000000 synthetic USD, the consumer pays 0.164200 and finishes with 9.835800. The provider receives 0.131360 and the platform receives 0.032840. At each settlement, exact six-decimal arithmetic checks
\begin{equation}
 B_c=10-\sum_i c_i,\qquad B_p=0.8\sum_i c_i,\qquad
 c_i=c_i^{\mathrm{provider}}+c_i^{\mathrm{platform}},\qquad L_c=0,
\end{equation}
where $c_i$ is the charge and $L_c$ is the remaining locked balance.

Successful execution without a final cost report uses the authorized maximum. An excessive report is rejected by the SDK, and queued cancellation releases the reservation. Cancellation terminates the task and records a failed Run with a cancellation reason; these states express the same unsuccessful business outcome at different lifecycle levels. A separate component experiment covers insufficient funds, concurrent reservations, conflicting reports, and settlement replay using predetermined completion outcomes. The component checks form a separate population from the seven executed workflows.

\begin{figure}[htbp]
\centering
\includegraphics[width=\linewidth]{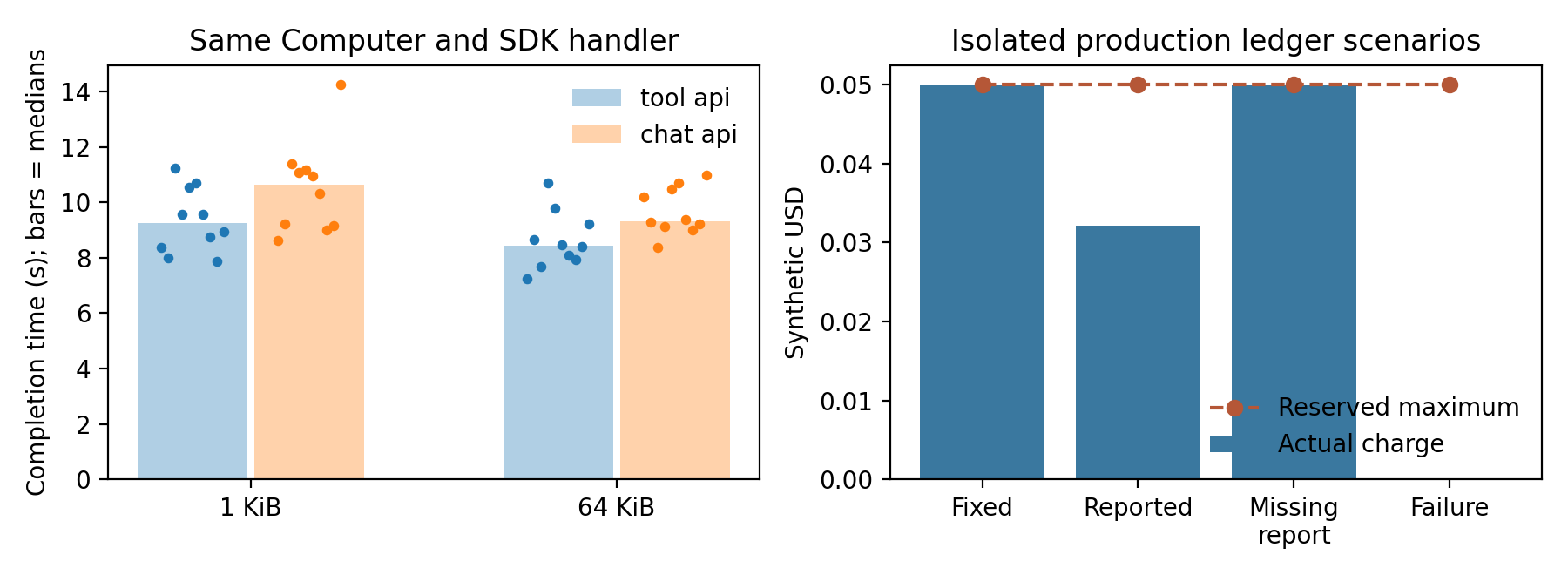}
\caption{Left: same-Computer admission-path durations. Right: reserved and charged synthetic amounts in the separate PostgreSQL component experiment.}
\label{fig:extensions}
\end{figure}

The results establish agreement among admitted pricing, execution, and ledger allocation for the specified scenarios. The amounts are controlled synthetic balances used to evaluate accounting consistency; no commercial payment is made. Section~\ref{sec:measured-cancellation} examines the separate question of residual work after cancellation.

\subsection{Workflow access through browser and programmatic clients}
\label{sec:web-api-measured}
Eight paired workflows compare the Web and API entry paths using the same caller, organization, hosted Agent, input bytes, output validator, and free tariff. Four pairs compute a digest from a 1-KiB input; four obtain a yes/no confirmation. The product order is balanced, and every arm creates a fresh invocation and executes independently.

The browser records admission, terminal-state observation, visible output, and confirmation through its ordinary interface. The API caller observes the corresponding states by polling every 250 ms. Web observation combines events with its normal five-second active polling. Browser timestamps refer to DOM observations rather than painted frames. In the confirmation workload, the browser controller reacts after approximately 2.8 s, whereas the API replies after its next poll. These client behaviors are part of the measured interaction paths.

\begin{table}[htbp]\centering\small
\caption{Matched invocation, four pairs per task family. All 16 attempts
complete with exact results and zero configured charge. Times are marginal
medians in seconds; client and server medians have different boundaries.}
\label{tab:web-api-measured}
\begin{tabular}{llrrrr}\toprule
Task & Client & Exact & Client s & Server s & Reply reaction s\\\midrule
Hash & WEB & 4/4 & 11.160 & 6.002 & 0.000 \\
Hash & API & 4/4 & 9.175 & 8.688 & 0.000 \\
Confirm & WEB & 4/4 & 19.814 & 15.780 & 2.809 \\
Confirm & API & 4/4 & 11.230 & 10.530 & 0.000 \\
\bottomrule\end{tabular}\end{table}
\begin{figure}[htbp]\centering
\includegraphics[width=\linewidth]{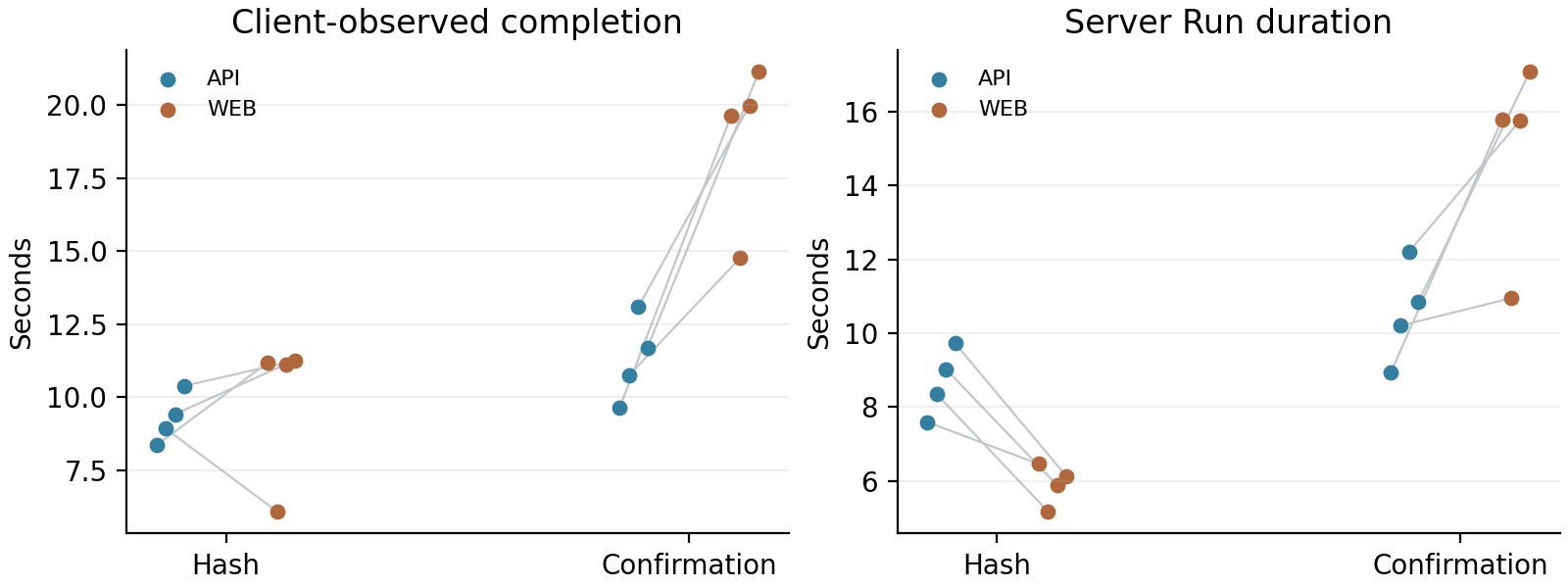}
\caption{Paired API and Web durations. Client time includes observation and confirmation response; server time includes scheduling and interaction but excludes browser presentation.}
\label{fig:web-api-timing}\end{figure}

All eight Web workflows and all eight API workflows produce the expected outcomes. The timing distributions include each client's observation and reply behavior, measuring completion through the Web or polling interface. Unique invocation records and separate captures associate every displayed result with the corresponding execution. Earlier preparation observations affected by reused task identifiers are separated from these pairs.

\begin{figure}[htbp]\centering
\includegraphics[width=\linewidth,trim={0 730bp 0 0},clip]{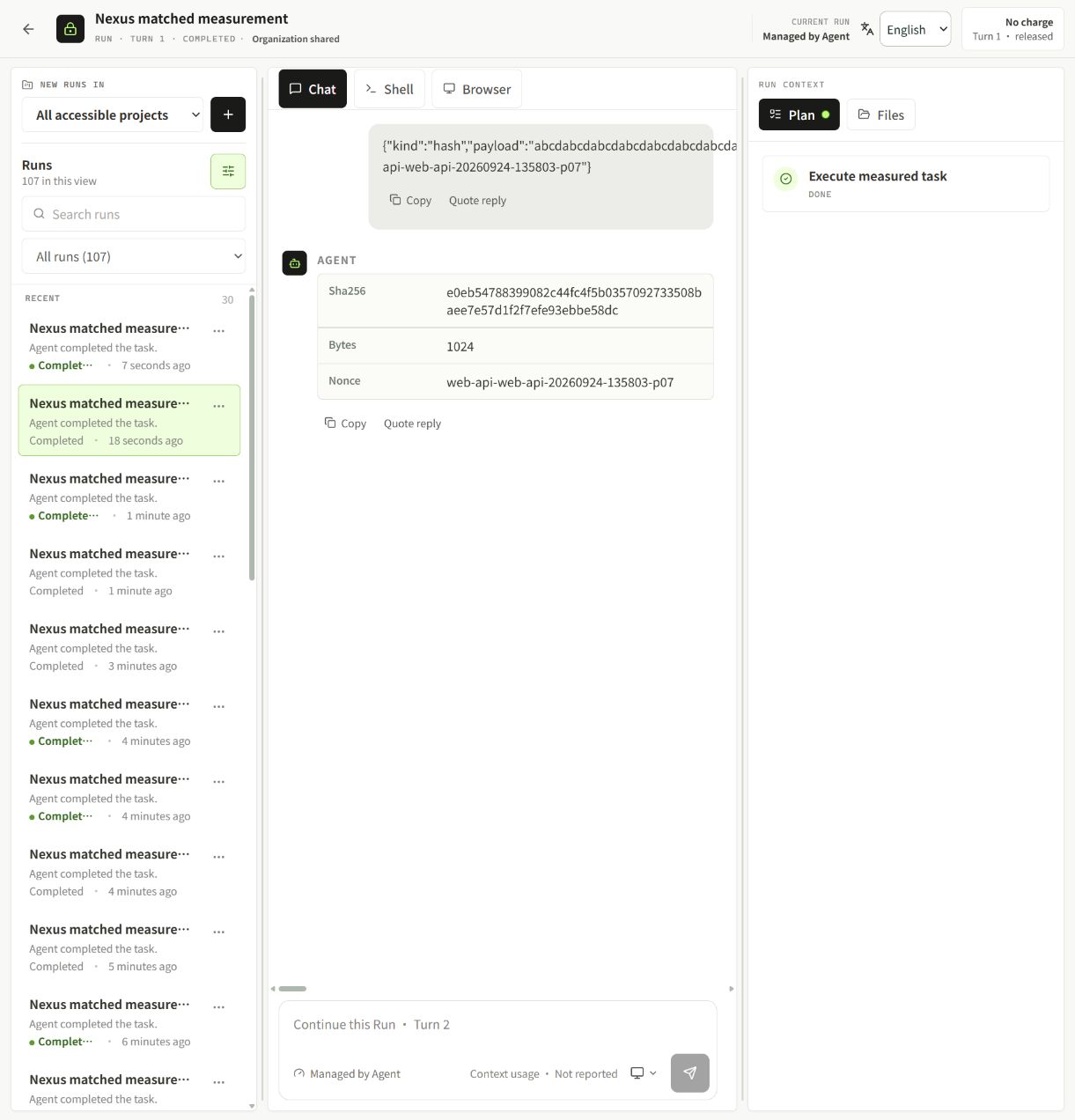}
\caption{Browser view of a completed 1,024-byte digest workflow, including its result, execution plan, and zero-charge settlement.}
\label{fig:web-api-screen}\end{figure}

The two interfaces therefore expose the same tested business operations and private outputs. Nonzero prices introduce an additional consistency requirement, evaluated next.

\subsection{Settlement consistency across Web and API invocation}
\label{sec:web-billing-measured}
Ten paired invocations test whether browser and programmatic access preserve the same admitted tariff and final charge. Each pair shares the caller, consumer organization, published cross-organization Agent, message, and price. The four two-pair families have balanced order; duplicate reporting and excessive reporting each contribute one additional pair. Every arm independently executes through the ordinary admission interface.

A PostgreSQL-backed Cloud application and its workers invoke an actual local MCP/SDK Agent. Deployment metadata and opening balances are fixtures: the consumer starts with 10 synthetic USD and the publisher with zero, and balances evolve without resets. Independent six-decimal calculations reconcile accepted reports, reservations, usage, ledger allocation, and browser-visible charges. No model request or commercial payment occurs.

\begin{table}[htbp]\centering\small
\caption{Paired nonzero settlement. Expected failures are part of the
fixed protocol. Each listed amount is per independently executed Run, in
synthetic USD; both clients agree.}
\label{tab:web-billing-measured}
\begin{tabular}{lrlrl}\toprule
Scenario & Pairs & Run status & Charge & Billing state\\\midrule
missing report & 2 & completed & 0.050000 & missing \\
reported failure & 2 & failed & 0.000000 & released \\
fixed success & 2 & completed & 0.050000 & captured \\
reported success & 2 & completed & 0.032100 & captured \\
over cap & 1 & failed & 0.000000 & released \\
duplicate report & 1 & completed & 0.032100 & captured \\
\bottomrule\end{tabular}\end{table}

Fourteen invocations complete and six fail as specified. All twenty produce the expected accounting outcome. A fixed-price success charges 0.05, and a reported-price success charges 0.0321. An identical repeated report creates one usage allocation. Successful execution without a report charges the authorized maximum of 0.05. Handler failure and an excessive report release the reservation without a charge; the latter is rejected by the SDK before the report callback, identifying SDK validation as the rejection point in this scenario.

The total debit is 0.592600, allocated as 0.474080 to the publisher and 0.118520 to the platform. The consumer finishes with 9.407400 and no locked balance. All twenty initial reservations of 0.05 are observed. Each positive charge has one consumer debit and one publisher credit, yielding 28 wallet entries; the six failures have zero-allocation usage records and no wallet transfer.

\begin{figure}[htbp]\centering
\includegraphics[width=.85\linewidth]{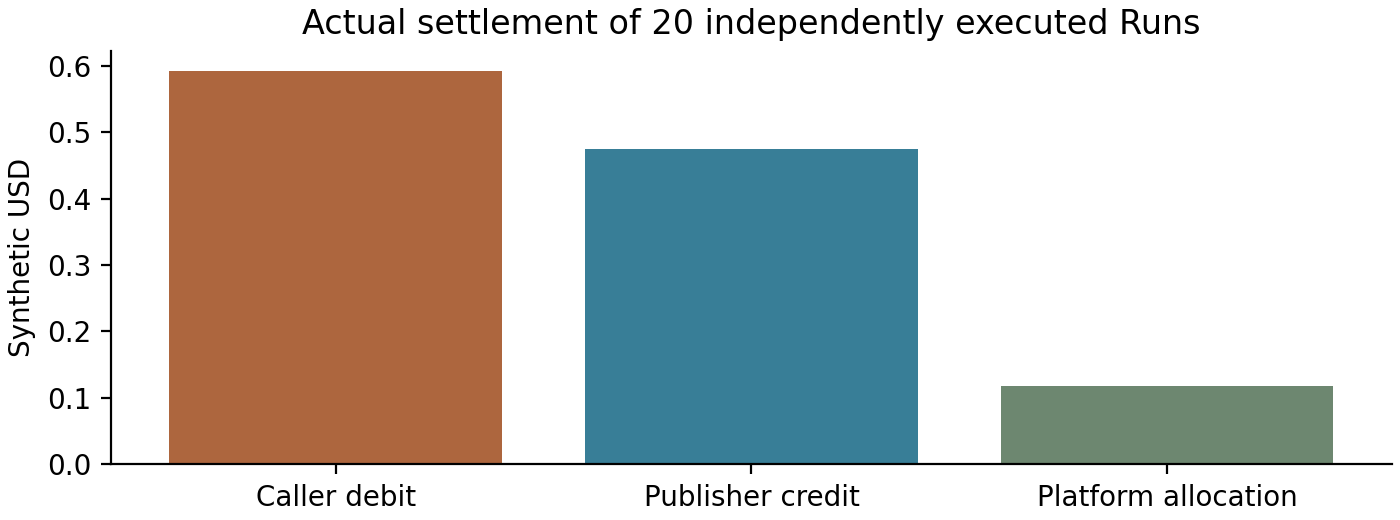}
\caption{Allocation of synthetic experimental funds. Publisher and platform shares sum to the consumer debit.}
\end{figure}
\begin{figure}[htbp]\centering
\includegraphics[width=\linewidth]{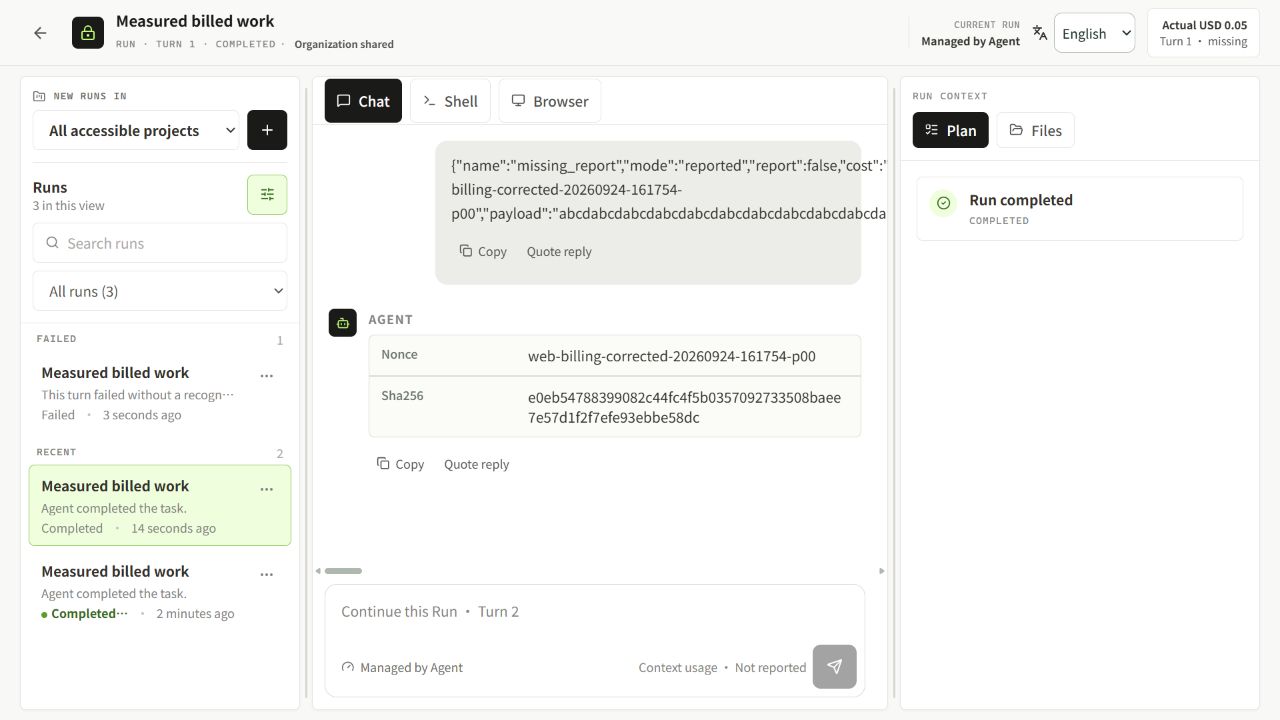}
\caption{Browser view of successful execution without a final cost report. The \texttt{missing} report status coexists with the settled authorized charge of 0.05 synthetic USD.}
\end{figure}

The Web summary matches settlement in the evaluated version, including the authorized-maximum charge when no report is supplied. An earlier presentation version labels that fallback as uncharged despite the server debit; its observation remains separate because the display changed while the settlement rule did not. The paired accounting results agree across the two entry paths, with rendering and observation time measured in the preceding workflow-access experiment.

\subsection{Protocol interoperability and externally visible completion}
\label{sec:mcp-client}
A service can execute a tool without delivering a result that a standard client recognizes. To examine this distinction, an unmodified Python MCP client invokes the OpenWrt adapter and a FastMCP service with the same staged handler. Eight balanced pairs cross 64- and 4,096-character outputs, with four pairs per size. The client, dependencies, connection timeouts, and producer behavior are fixed; each arm starts a fresh client session. Independent producer records identify execution even when the client rejects the response.

Both endpoints complete initialization and tool discovery in all eight arms. FastMCP delivers all eight results and 128 progress events. The initial OpenWrt adapter delivers no client-recognized result. Its first call returns HTTP 200 and executes all sixteen stages, but its application-specific event framing does not reach the client's JSON-RPC callbacks. The remaining seven calls return HTTP 409 before a new producer starts. Reuse of a session-local RPC identifier is consistent with these conflicts, although the retained error information does not independently establish every conflict's cause.

\begin{table}[htbp]
\centering\small
\caption{Standard-client tool invocation across eight fresh sessions per path. Producer completion is measured independently of client-visible completion.}
\label{tab:mcp-client}
\begin{tabular}{@{}lrrrr@{}}
\toprule
Path & Initialize/list & Client success & Progress callbacks & Producer complete \\
\midrule
FastMCP transport & 8/8 & 8/8 & 128 & 8/8 \\
Native adapter & 8/8 & 0/8 & 0 & 1/8 \\
\bottomrule
\end{tabular}
\end{table}

Both endpoints initialize and discover tools under versions accepted by the client, but only FastMCP delivers client-recognized results in this experiment. Producer records show why these are separate outcomes: the first adapter call executes the handler without delivering a usable protocol response. Section~\ref{sec:mcp-correction} repeats the comparison with standard message framing and per-invocation identity separation; the initial adapter outcomes remain a distinct implementation group.

\subsection{Standard tool responses and explicit retained-stream recovery}
\label{sec:mcp-correction}
The revised adapter separates standard MCP responses from application-specific retained-stream recovery. Standard calls emit JSON-RPC responses and notifications in SSE message frames, associate progress with the caller's progress token, and preserve the response's RPC identifier. Each call receives a fresh internal invocation identity, preventing session-local RPC identifiers from selecting another call's state. This endpoint is stateless: a new POST starts a new invocation, and standard GET-based resumption is unavailable.

The same eight balanced client pairs from Section~\ref{sec:mcp-client} are repeated with unchanged handlers, dependency versions, and timeouts. Both the standard adapter and FastMCP deliver eight complete outputs and 128 progress notifications. A subsequent build-controlled repetition of the original adapter again yields no completed client call: one times out after producer execution and seven encounter conflicts. This build-controlled repetition reproduces the client-visible outcome of the initial adapter. Its sequential execution is analyzed as an interoperability check, not a latency contrast.

\begin{table}[htbp]
\centering\small
\caption{Standard-client outcomes for the initial and revised adapter configurations, with separate initial, repeated, and build-controlled experiments.}
\label{tab:mcp-correction}
\begin{tabular}{@{}llrrr@{}}
\toprule
Campaign & Path & Client success & Progress & Producer complete \\
\midrule
Original & Direct FastMCP & 8/8 & 128 & 8/8 \\
Original & Installed adapter & 0/8 & 0 & 1/8 \\
Rebuilt original & Direct FastMCP & 8/8 & 128 & 8/8 \\
Rebuilt original & Control adapter & 0/8 & 0 & 1/8 \\
Corrected & Direct FastMCP & 8/8 & 128 & 8/8 \\
Corrected & Candidate adapter & 8/8 & 128 & 8/8 \\
\bottomrule
\end{tabular}
\end{table}

Retained-stream recovery uses a separate, explicitly selected transport. A caller supplies the stable task identity and last consumed event, allowing reconnection to the original producer while its history remains available. Standard MCP calls and this recovery transport therefore have different resumption contracts.

In the custom-transport experiment, eighty pairs compare cursor-based continuation with full-history replay. All 160 calls reconstruct the correct outputs with one producer execution per call, totaling 2,560 produced stages. Cursor-based continuation delivers no repeated events; full replay delivers 480 repeated events across its eighty calls before client-side deduplication. Standard clients receive complete tool responses through the stateless endpoint; callers selecting the retained-stream interface additionally control continuation through the original task identity and cursor. These remain separate transport contracts.

\par\bigskip
\section{Resource Reuse and Deployment Conditions}
\label{app:resource-evidence}
File references and model-usage identities connect resources to their originating invocation. The following experiments study that association, reuse across turns and collections, policy-controlled routing, and the deployment conditions under which resources remain available or reduce expenditure.

\subsection{Content-preserving continuation of private file transfers}
\label{sec:run-file}
Retaining transfer progress can reduce repeated communication without changing the final file. Four pairs cross 2- and 4-MiB binary payloads with upload and download interruption. The two arms share the caller, Agent, permissions, SDK, handler, content, and storage service. They differ only in whether recovery retains the upload identity or partial download. Order is balanced by fault type.

For uploads, the connection is closed after the first 1-MiB chunk has been accepted, and the responses to all three attempts of that chunk are discarded. The server recognizes the repeated chunk; a metadata request reports its accepted offset. Continuation retains that upload and sends the remaining suffix, whereas restart creates a new upload. For downloads, the injector terminates the response after delivering 256 KiB. Continuation requests the remaining range, while restart downloads from zero. Both paths verify the final length and digest before publication.

Let $N$ denote file length, $C=1$ MiB the upload chunk, and $D=256$ KiB the downloaded prefix. The three interrupted PUTs transmit $3C$ request-body bytes. Continuation adds $N-C$, giving $N+2C$; restart adds $N$, giving $N+3C$. Download totals are $N$ with continuation and $N+D$ with restart. The request and read counters agree with these predictions in every pair.

\begin{table}[htbp]
\centering\small
\caption{File-transfer outcomes. File size, PUT bodies, and delivered data are in MiB; end-to-end time is in seconds. All eight outputs match their expected content.}
\label{tab:run-file}
\begin{tabular}{@{}lrlrrr@{}}
\toprule
Fault & Size & Policy & PUT bodies & Download & Time \\
\midrule
Download & 2 & restart & 2.00 & 2.25 & 8.921 \\
Download & 2 & resume & 2.00 & 2.00 & 8.216 \\
Download & 4 & restart & 4.00 & 4.25 & 7.607 \\
Download & 4 & resume & 4.00 & 4.00 & 7.548 \\
Upload & 2 & restart & 5.00 & 2.00 & 10.107 \\
Upload & 2 & resume & 4.00 & 2.00 & 9.677 \\
Upload & 4 & restart & 7.00 & 4.00 & 9.302 \\
Upload & 4 & resume & 6.00 & 4.00 & 8.199 \\
\bottomrule
\end{tabular}
\end{table}
\begin{figure}[htbp]
\centering
\includegraphics[width=\linewidth]{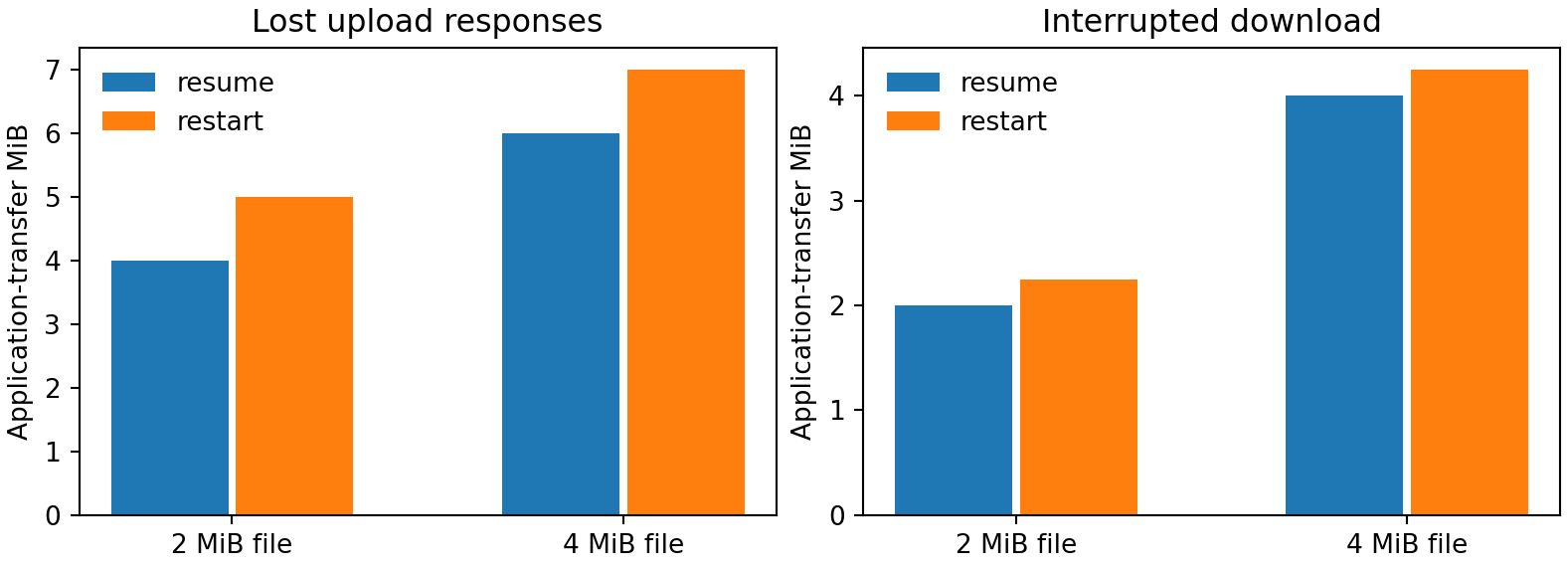}
\caption{Observed bytes in the affected transfer direction for all four pairs.
Continuation preserves the already accepted or delivered prefix. The plotted
values are recomputed from the retained request and read records.}
\label{fig:run-file}
\end{figure}

All eight Runs produce exact files. Continuing each upload avoids 1 MiB of repeated PUT bodies, and continuing each download avoids 256 KiB of repeated delivery. The resumed downloads return the expected partial-content range. Independent reads of the final stored objects and caller-visible artifacts confirm the same lengths and digests. Ten transfer records comprise eight final files and two abandoned upload drafts; published artifacts and creation identities remain associated with the originating Runs.

The counters include request bodies and delivered file content, excluding control messages, TLS framing, and the independent verification download. Continuation succeeds because the accepted upload state or partial download survives the connection interruption and the source bytes remain unchanged. Process and storage loss are not introduced by this fault model.

\subsection{Private-file transformation and policy-controlled reuse}
\label{sec:file-adapter}
Eight pairs compare SDK file operations with explicit HTTP code for an identical transformation workflow. The adapters share hosting, authentication, journal, storage, publication, and business logic. The workload crosses 64-KiB and 1-MiB inputs, text and binary formats, and two repetitions. Each size/format cell has opposite arm orders. Fresh inputs contain identical bytes within a pair and remain bound to their own Runs.

The Agent enumerates and downloads the input, verifies its digest, transforms text to uppercase or binary by XOR with \texttt{0x5a}, uploads the result, reads it back, and publishes a private reference. The caller independently downloads that result and reconstructs its expected digest. Both adapters use the same deployed image and scanner policy. No model or device operation contributes to the transformation.

\begin{table}[htbp]
\centering\small
\caption{Scheduled private-file workflow observations. Completion includes an
independent output digest; reuse is evaluated under the scanner policy.
Expected denotes the predeclared complete workflow oracle (including binary
reuse denial); Reused counts accepted references. Time is the median
invocation-completion time in seconds for two observations
per row, excluding caller preparation and post-completion validation/scan.}
\label{tab:file-adapter}
\begin{tabular}{@{}lrlrrrr@{}}
\toprule
Format & KiB & Adapter & Exact & Expected & Reused & Time \\
\midrule
Binary & 64 & SDK & 2/2 & 2/2 & 0/2 & 10.455 \\
Binary & 64 & HTTP & 2/2 & 2/2 & 0/2 & 9.592 \\
Text & 64 & SDK & 2/2 & 2/2 & 2/2 & 8.578 \\
Text & 64 & HTTP & 2/2 & 2/2 & 2/2 & 9.509 \\
Binary & 1024 & SDK & 2/2 & 2/2 & 0/2 & 9.640 \\
Binary & 1024 & HTTP & 2/2 & 2/2 & 0/2 & 9.385 \\
Text & 1024 & SDK & 2/2 & 1/2 & 1/2 & 8.464 \\
Text & 1024 & HTTP & 2/2 & 1/2 & 1/2 & 9.217 \\
\bottomrule
\end{tabular}
\end{table}

All sixteen transformations produce exact output bytes. Each handler uploads $N$ file-body bytes and downloads $2N$: one input read and one output-verification read. Independent storage and lifecycle records identify sixteen immutable outputs, thirty-two distinct stored objects, and thirty-two successful journal operations. The transformation holds its output in memory; the evaluated sizes are 64 KiB and 1 MiB, with no bounded-memory claim for larger inputs.

Transformation success and permission to reuse the output are separate outcomes. All initial reference selections are refused before scanning. Six of eight text outputs subsequently pass scanning, and repeated selection returns the same reference to the immutable object. All eight binary outputs remain ineligible under the supported-format policy. The remaining text pair unexpectedly matches a phone-number rule in the generated digest text. Both adapters receive the same refusal, leaving fourteen of sixteen complete workflows consistent with their planned outcomes despite sixteen correct transformations. These two false positives remain part of the result.

The SDK resource adapter uses nine nonblank, noncomment lines compared with 104 for explicit HTTP, excluding shared infrastructure and transformation code. Both adapters produce identical transformations and policy decisions within each pair, including the two unexpected text refusals. The code count measures application-side protocol work; developer time is not measured, and whole-workflow timing remains descriptive for this small workload. Section~\ref{sec:run-file} evaluates continuation after a transfer interruption.

\subsection{Repeatability of transformation and reuse decisions}
\label{sec:file-adapter-replication}
The private-file experiment is repeated with the exact original bytes, formats, sizes, and eight-pair order. Fresh files and Runs prevent reuse of earlier execution results. Retaining the same generated text also retains the scanner-triggering inputs, allowing the original acceptance and refusal outcomes to be repeated.

The Agent image, handler, SDK, scanner, and output validator remain fixed. Cloud runtime changes between experiments are recorded separately, so the repeat compares functional outcomes rather than cross-version timing. An earlier repetition attempt encounters an offline bound Runtime: all sixteen admissions are rejected before a file transformation begins. After reconnecting that Runtime, the complete paired schedule executes.

\begin{table}[htbp]\centering\small
\caption{Exact-input file-workflow replication. Full outcomes include the
original scan/reference policy, not only correct transformation. Unsupported
binary rejection is an expected control. Each cell has four Runs.}
\begin{tabular}{lrrrr}\toprule
& \multicolumn{2}{c}{Exact outputs} & \multicolumn{2}{c}{Full outcomes}\\
Input & Initial & Repeat & Initial & Repeat\\\midrule
64 KiB text & 4/4 & 4/4 & 4/4 & 4/4 \\
1 MiB text & 4/4 & 4/4 & 2/4 & 2/4 \\
64 KiB binary & 4/4 & 4/4 & 4/4 & 4/4 \\
1 MiB binary & 4/4 & 4/4 & 4/4 & 4/4 \\
\bottomrule\end{tabular}\end{table}

The completed repetition again yields sixteen exact transformations and fourteen complete expected workflows. Every arm agrees with its original transformation, scan, and reuse outcome, including the two unexpected text refusals. Independent object reads and lifecycle records confirm the file contents and lineage. The same two text inputs trigger the scanner in both experiments, localizing the repeated refusal to this input-policy combination. Broader scanner accuracy is not measured.

\subsection{Immutable output references across conversation turns}
\label{sec:file-reuse}
An approved output reference can become the next turn's input without requiring the caller to upload the same content again. Four paired workflows use 64-KiB and 1-MiB ASCII inputs, with two repetitions per size and balanced arm order. Both arms execute the same hosted handler with the same caller, permissions, and transformation. The first turn converts lowercase text to uppercase and scans the output. The second either selects the approved immutable reference or uploads identical uppercase bytes already held by the caller.

The experiment changes only second-turn attachment preparation. Both arms incur the first upload and scan; neither is assigned a hypothetical extra download. Each new turn consumes its admitted attachment rather than an arbitrary file from the Run-wide input list. Eight Runs execute sixteen turns, with a 240-second observation bound per turn.

\begin{table}[htbp]
\centering\small
\caption{Continuation after output preparation. PUT bytes are summed
caller upload bodies for second-turn admission; times are medians in seconds.
Each row contains two Runs. Application execution still transfers files between
the Agent and Cloud in both arms.}
\label{tab:file-reuse}
\begin{tabular}{rlrrrr}
\toprule
KiB & Input source & Exact & PUT KiB & Prepare s & Continue s\\
\midrule
64 & Reference & 2/2 & 0 & 0.226 & 9.705 \\
64 & Upload & 2/2 & 128 & 1.580 & 8.524 \\
1024 & Reference & 2/2 & 0 & 0.178 & 9.162 \\
1024 & Upload & 2/2 & 2048 & 0.757 & 10.161 \\
\bottomrule
\end{tabular}
\end{table}

All eight Runs complete both turns with exact, independently downloaded outputs. Reference selection sends no second-attachment PUT body, while reupload sends 2,228,224 bytes across its four Runs. The resulting 2.125-MiB reduction concerns repeated caller uploads; metadata exchange and Agent-side file operations remain necessary.

Each reference points to the first output's immutable stored object and introduces no upload parts. Reupload instead creates a distinct input object with identical bytes. Reference and reupload Runs therefore contain three and four distinct file objects, respectively, including both outputs. Turn indices and source lineage preserve the origin of the reused content. These are logical object counts, not allocated physical-disk measurements.

Approved immutable references support later-turn execution under the same scan and authorization policy. Across the two input sizes, upload-body savings are consistent while the faster continuation path varies. Reuse therefore changes attachment preparation and object lineage without changing task output. Eligibility still follows the format and scan outcomes measured in Section~\ref{sec:file-adapter}.

\subsection{Content lineage and immutable collection versions}
\label{sec:asset-archive}
Eight pairs compare server-side capture of an Agent output with caller download followed by multipart import. The sources are the eight second-turn artifacts from Section~\ref{sec:file-reuse}: four 64-KiB and four 1-MiB files, with one controlled content digest per size. The source order is fixed and arm order alternates within each size. This experiment operates on completed outputs and adds no model or device execution.

Both paths share the authenticated caller, tenant, source bytes, filename, private-collection policy, and scan preparation. Capture records Agent, Run, and artifact lineage; import records ordinary user-upload provenance. Each arm verifies the archived content, creates a one-file version, appends a marker to the live collection, and downloads the earlier version again.

\begin{table}[t]\centering\small
\begin{tabular}{rrrrrr}\toprule
KiB & Capture & Reupload & Capture (s) & Reupload (s) & Avoided KiB \\
\midrule
64 & 4/4 & 4/4 & 0.250 & 0.305 & 128 \\
1024 & 4/4 & 4/4 & 0.296 & 0.715 & 2048 \\
\bottomrule\end{tabular}
\caption{Archive comparison. Counts are verified operations; times are
means over four operations. Reupload time includes source download and import.
Avoided KiB is caller artifact-body traffic per pair, excluding metadata,
multipart framing and equal verification traffic.}
\label{tab:asset-archive}\end{table}
\begin{figure}[t]\centering
\includegraphics[width=\linewidth]{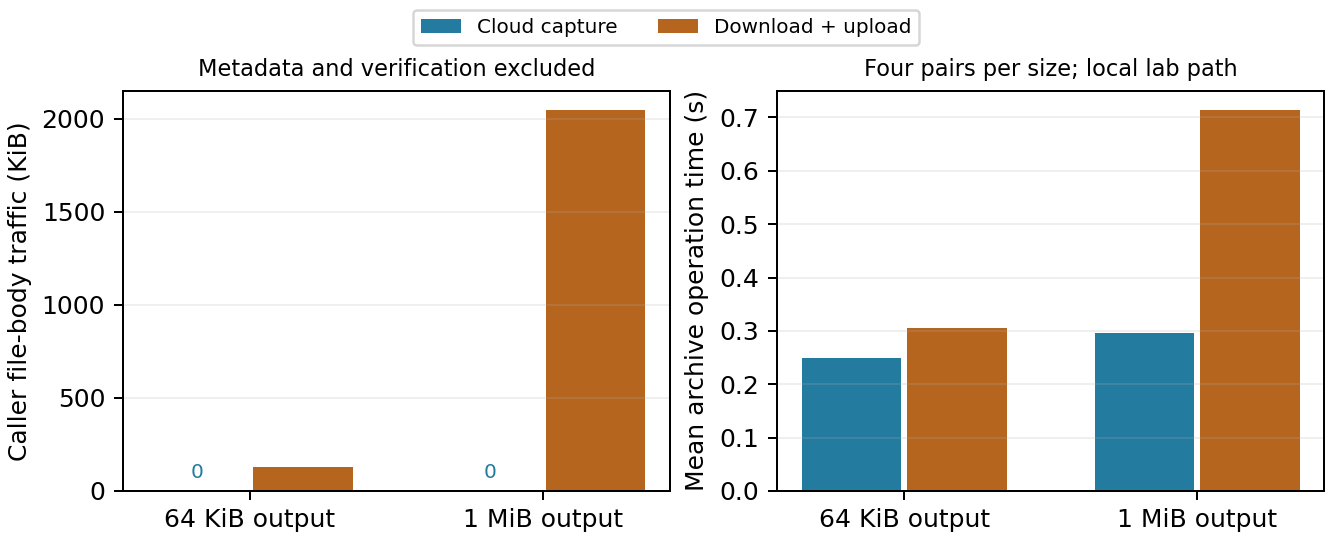}
\caption{Caller artifact-body traffic and archive duration. The metric excludes metadata, shared scan preparation, version administration, and verification reads.}
\label{fig:asset-archive}\end{figure}

All sixteen archive operations preserve the expected bytes, provenance, and version membership. After the append, each live collection contains two files while its earlier version retains one. Server-side capture avoids 8,912,896 caller artifact-body bytes across the eight pairs, or 8.5 MiB. Both paths materialize dataset content on the server, so the communication reduction is distinct from physical-storage deduplication.

A captured private output retains its execution lineage inside a versioned collection. Both tested Cloud interfaces preserve the earlier one-file version when the live collection receives an append. The intervention is capture versus import of completed artifacts; overwrite recovery and a dataset-specific SDK are not exercised.

\subsection{Route selection and invocation-scoped model usage}
\label{sec:model-routing}
A successful model response establishes neither the selector used by the gateway nor the Run to which its usage belongs. A four-arm experiment separates these responsibilities: direct pool selection versus explicit execution-router selection, crossed with invocation-attribution headers versus their omission. All arms use the same hosted Agent, gateway endpoint, model alias, single Qwen/Qwen3-8B source, permissions, prompts, and generation settings. Pool selection still traverses the gateway.

Four JSON tasks require addition, sorting, uppercasing, and list length. Every task runs once in each arm, giving sixteen fresh model calls. Cyclic ordering balances arm position across tasks. Temperature is zero, thinking is disabled, output is limited to 128 tokens, and inference is not automatically retried. HTTP and Run observation bounds are 90 and 180 seconds. The tasks expose selection and attribution behavior rather than broad reasoning ability.

\begin{table}[htbp]
\centering\small
\caption{Single-source Qwen requests, with four common tasks per row. Lineage matches source and usage records; Usage counts Run-attributed records. Tokens are reported input/output totals, and time is median model-request duration excluding admission and polling.}
\label{tab:model-routing}
\begin{tabular}{@{}llrrrrr@{}}
\toprule
Selector & Headers & Calls & Lineage & Usage & Tokens & Time \\
\midrule
Pool & NONE & 4/4 & 4/4 & 0 & 118/33 & 1.599 \\
Pool & SDK & 4/4 & 4/4 & 4 & 118/33 & 1.627 \\
Router & NONE & 4/4 & 4/4 & 0 & 118/33 & 1.656 \\
Router & SDK & 4/4 & 4/4 & 4 & 118/33 & 1.643 \\
\bottomrule
\end{tabular}
\end{table}

All sixteen requests produce the expected JSON and upstream usage. Request identities link each response to one gateway record with the expected source, pool, and router identity. The experiment records 472 input and 132 output tokens. All eight header-bearing Runs acquire one authenticated usage record linked to the corresponding Run and event; none of the eight omission controls acquires that Run-level record, although gateway usage remains available in every case.

Invocation attribution changes from zero of eight to eight of eight while source selection remains unchanged. Each arm reports the same aggregate 118 input and 33 output tokens across its four tasks. The added information is the authenticated link from gateway usage to its Run, not fewer model requests or tokens. All calls use one source and a zero configured tariff; provider failover, adversarial credentials, and commercial billing are not exercised by this selection-and-attribution comparison.

\subsection{Conformance to versioned source-selection policies}
\label{sec:multisource-routing}
Two logical model sources expose whether pool policy changes reach actual request selection. Both sources reference the same free Qwen/Qwen3-8B runtime and provider account, giving two policy candidates within one upstream failure domain. Four JSON tasks are each submitted under five policies: A-first order, A disabled, zero weight for A, rollback to A-first after an intermediate B-first revision, and fallback order A=0/B=1. Task order is shuffled and arm positions rotate across tasks. All requests share temperature zero, a 128-token bound, and disabled thinking.

The expected source is determined from the applied policy independently of inference correctness. Request records connect each response with its source, pool, policy revision, usage, and fallback count. Stale-revision writes are rejected without changing the policy and are not counted as additional inference requests.

\begin{table}[htbp]\centering\scriptsize
\begin{tabular}{rlrrrrrrr}\toprule
C & Policy & Sent/planned & Match & Differ & Unknown & Success & Correct & Mean (s) \\
\midrule
1 & A first & 4/4 & 4 & 0 & 0 & 4 & 4 & 2.114 \\
1 & Withdraw A & 4/4 & 4 & 0 & 0 & 4 & 4 & 5.869 \\
1 & Weight A=0 & 4/4 & 4 & 0 & 0 & 4 & 4 & 2.851 \\
1 & Rollback & 4/4 & 4 & 0 & 0 & 4 & 4 & 3.850 \\
1 & Order A=0 & 4/4 & 0 & 4 & 0 & 4 & 4 & 3.613 \\
2 & A first & 4/4 & 4 & 0 & 0 & 4 & 4 & 2.118 \\
2 & Withdraw A & 4/4 & 4 & 0 & 0 & 4 & 4 & 2.335 \\
2 & Weight A=0 & 4/4 & 4 & 0 & 0 & 4 & 4 & 3.386 \\
2 & Rollback & 4/4 & 4 & 0 & 0 & 4 & 4 & 2.380 \\
2 & Order A=0 & 4/4 & 4 & 0 & 0 & 4 & 4 & 2.528 \\
\bottomrule\end{tabular}
\caption{Policy-controlled selection. Match, Differ, and Unknown describe source choice; Success describes inference and Correct describes task output. Mean duration is conditional on a successful response. Experiments 1 and 2 use the initial and zero-preserving implementations.}
\label{tab:multisource-routing}\end{table}
\begin{figure}[htbp]\centering
\includegraphics[width=\linewidth]{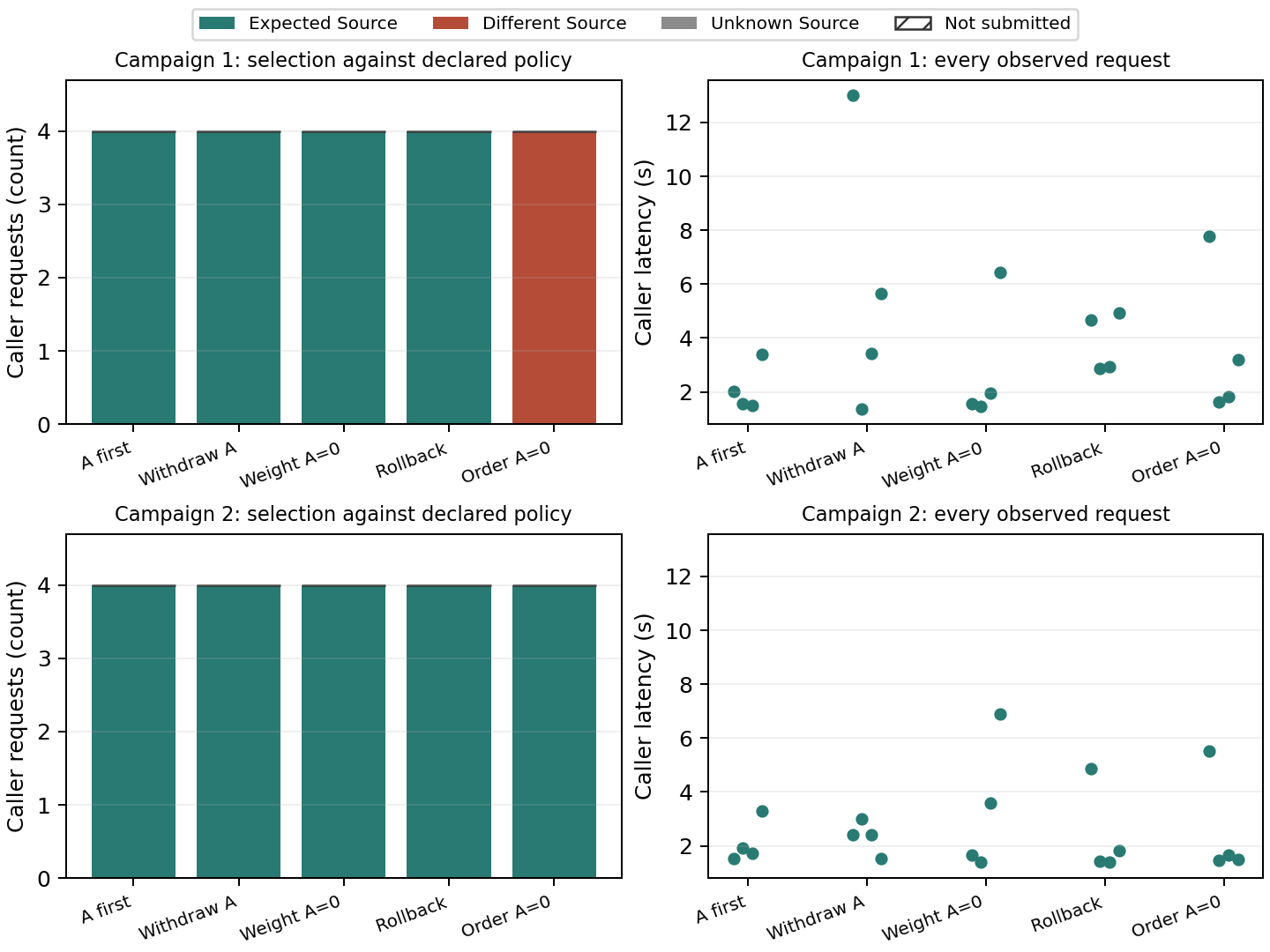}
\caption{Source selection and request duration for each policy. Circles indicate successful inference; crosses indicate unsuccessful or unresolved requests.}
\label{fig:multisource-routing}\end{figure}

Both complete campaigns return twenty correct JSON outputs. The initial implementation selects the expected source in sixteen cases but chooses B in all four zero-order cases. Its handling of a valid zero as a missing order causes this disagreement. The version that preserves zero repeats the same task and policy schedule and selects the expected source in all twenty cases. The change brings actual source selection into agreement with the declared order, including a valid zero-valued fallback order. The tested intervention is policy conformance in the implementation.

An earlier schedule stops after four of twenty planned requests when its health guard observes a degraded shared offer; sixteen requests remain undispatched. The two complete campaigns use the same subsequent eligibility rule and observe a healthy offer before every request. The preliminary five-request pilot has five successful responses and four policy matches; the stopped schedule has four successful responses and three matches. Neither group is pooled with the complete experiments.

Both logical candidates use the same upstream account and tariff, placing them in one provider failure domain. The observed outcomes concern exclusion, ordering, and revision selection within that domain. Independent-provider outages, weighted traffic shares, and price variation are not part of the workload.

\subsection{Workflow availability after a command-idle interval}
\label{sec:full-agent-idle}
A finite inactivity experiment examines whether a deployed Agent remains callable after two hours without Agent or device commands. Native OpenWrt and Cloud Docker use the same Agent source, Python, SDK, Qwen route, Computer Runtime, Android emulator, and one-CPU/512-MiB process ceilings. Each placement executes a baseline workflow and a fresh post-idle workflow. Baseline order is native then Docker, and the post-idle order is reversed.

Each workflow performs model selection, a file read and digest, Android Settings observation, confirmation, a verified write, and a model report. Normal heartbeat, route renewal, and health traffic continue during the interval. The shared Cloud also archives previously completed files without invoking either idle Agent. Caller authentication is renewed before each new invocation, so the experiment concerns new-work availability rather than survival of an old token or interrupted Run.

The shared command-idle interval is 7,200.00 s. Intermediate observations identify 60 distinct route-renewal timestamps and stable sampled runtime identities. Task and device records contain no intervening command. Availability observations have one-minute resolution, leaving between-sample transients unobserved.

\begin{table}[t]
\centering\small
\begin{tabular}{lrrrr}
\toprule
Deployment & Baseline (s) & Post-idle (s) & Verified & Gap (s) \\
\midrule
OpenWrt native & 19.30 & 20.33 & 2/2 & 7251.9 \\
Cloud Docker & 18.32 & 26.44 & 2/2 & 7203.1 \\
\bottomrule
\end{tabular}
\caption{Complete workflows before and after inactivity. Gap spans baseline terminal observation to post-idle submission for each placement, including the other placement's scheduled execution. There is one workflow per cell.}
\label{tab:full-agent-idle}
\end{table}
\begin{figure}[t]\centering
\includegraphics[width=\linewidth]{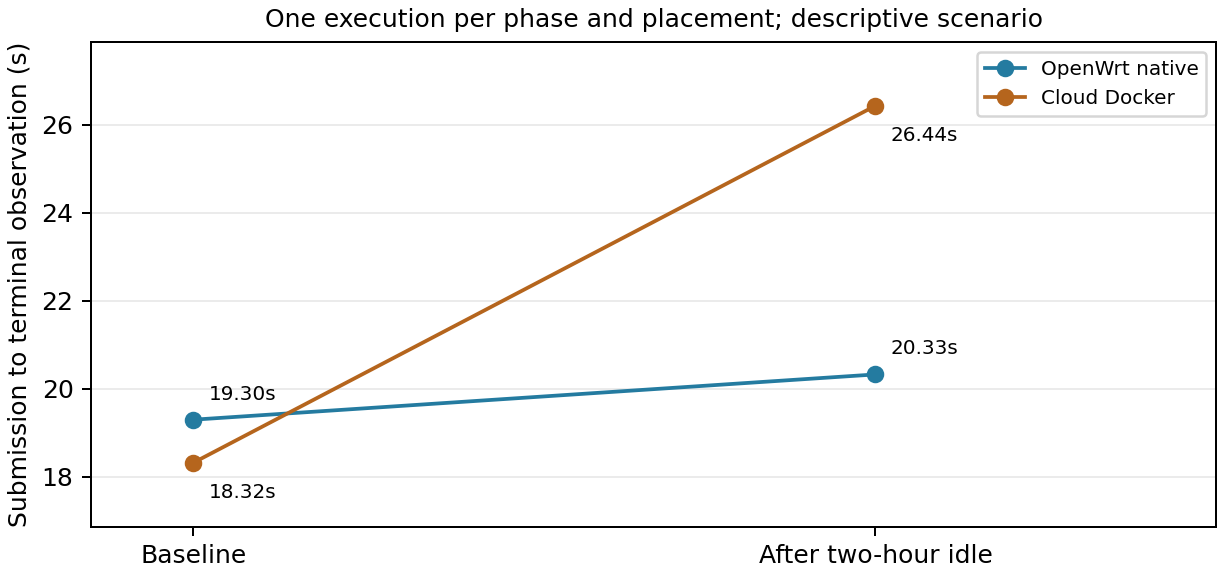}
\caption{Completion times before and after the command-idle interval. Each point represents a complete workflow with model requests and device operations.}
\label{fig:full-agent-idle}\end{figure}

All four workflows produce their expected outputs and all eight model calls appear in Gateway records. Fresh workflows complete after the specified idle interval while normal maintenance remains active. Their durations include model waiting and scripted interaction; four executions characterize this finite interval rather than a population uptime or latency distribution.

\subsection{Conditions for converting capacity relief into expenditure reduction}
\label{sec:cost-evidence-bridge}
Lower Cloud-side capacity requirements reduce expenditure only when provisioned resources can change and the additional edge cost remains smaller than the released Cloud cost. The capacity results provide the workload-specific starting point. The 1-KiB task passes at a 0.25-vCPU ceiling in both configurations. For the 16-MiB task, both experiments pass all batches at 0.25 vCPU with edge processing and 2 vCPU with Cloud processing. The corresponding ceiling reduction is 87.5\%; a ceiling is an execution constraint rather than a purchased allocation.

Let $q_C$ and $q_E$ be the tested passing Cloud ceilings, and assume a linear CPU tariff $p$ over equal duration $T$. Define the baseline component cost as $C_0=q_CTp$. Let $\gamma\in[0,1]$ denote the fraction of the potential ceiling reduction converted into lower billable allocation, and let $x=\Delta C_E/C_0$ denote the additional edge cost relative to $C_0$. With other components held fixed,
\[
 \frac{\Delta C_{\mathrm{net}}}{C_0}
 =\gamma\left(1-\frac{q_E}{q_C}\right)-x.
\]
Positive savings require $x<\gamma(1-q_E/q_C)$. For the large-input task, full allocation conversion permits additional edge cost below $0.875C_0$, while half conversion permits $0.4375C_0$. An unchanged allocation has $\gamma=0$. The small-input task provides no ceiling-reduction allowance. If the CPU component accounts for fraction $f$ of the original total bill, the corresponding modeled total-bill reduction is $f[\gamma(1-q_E/q_C)-x]$.

\begin{figure}[htbp]\centering
\includegraphics[width=\linewidth]{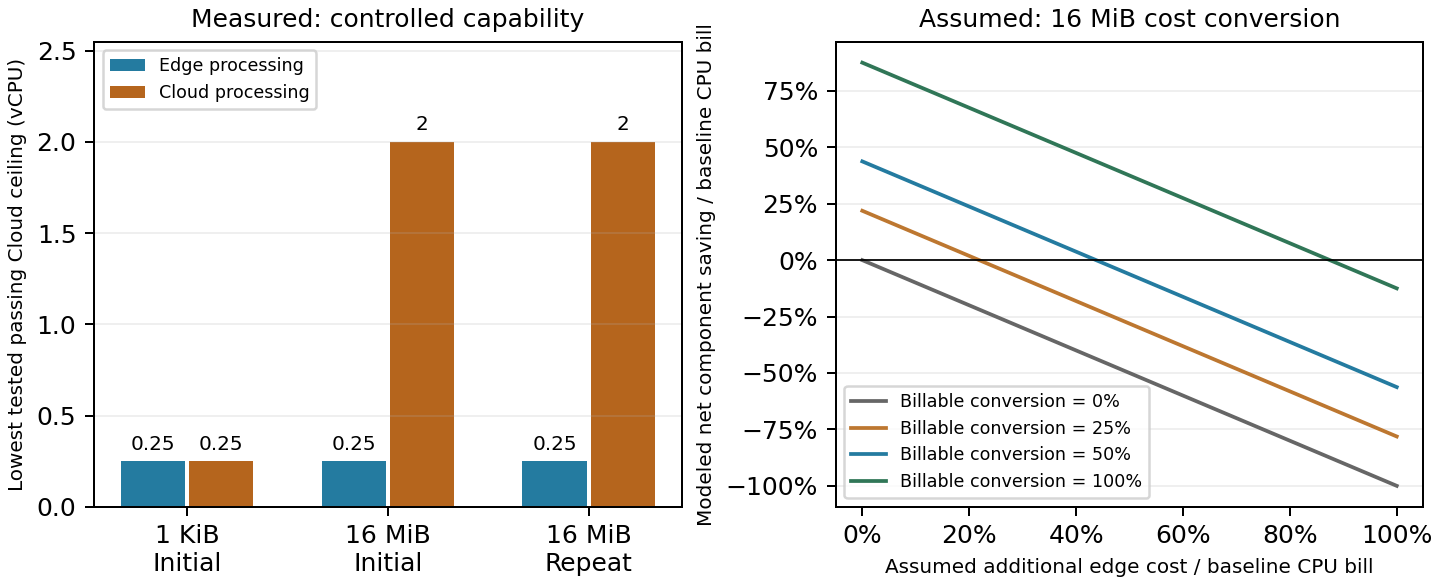}
\caption{Cloud capacity and conditional expenditure reduction. Left: tested passing ceilings in the initial and repeated experiments. Right: analytical sensitivity to allocation conversion and additional edge cost for the 16-MiB workload; these coordinates are assumptions rather than measured charges.}
\label{fig:cost-evidence-bridge}\end{figure}

The sensitivity parameters $\gamma$, $x$, $f$, and $p$ are explicit assumptions. The measured input is the capability workload's tested feasible CPU ceiling, reused from the capacity experiments. The sixteen complete-Agent pairs have equal model request counts and token totals, but no CPU-ceiling sweep, so they supply a different resource observation. This analysis connects capacity relief to expenditure only when released capacity changes the billable allocation and exceeds the additional edge cost; commercial savings remain unmeasured.

\FloatBarrier

\end{document}